\documentclass[12pt,a4paper,english]{article}
\pdfoutput=1

\usepackage[a4paper]{geometry}
\usepackage{babel}
\usepackage{textcomp}
\usepackage{amsmath}
\usepackage{amssymb}
\usepackage[nosort]{cite}
\usepackage{bbm}

\numberwithin{equation}{section}

\newcommand{\eq}[1]{\begin{equation}
                     \begin{split} #1 \end{split}
                     \end{equation}}

\allowdisplaybreaks

\begin{document}

%%%%%%%%%%%%%%%%%%%%%%%%%%%%%%%%%%%%%%%%%%%%%%%
%%%%%%%%%%%%%%%%%%%%%%%%%%%%%%%%%%%%%%%%%%%%%%%
%%%%%%%%%%%%%%%%%%%%%%%%%%%%%%%%%%%%%%%%%%%%%%%
%%%%%%%%%%%%%%%%%%%%%%%%%%%%%%%%%%%%%%%%%%%%%%%
%%%%%%%%%%%%%%%%%%%%%%%%%%%%%%%%%%%%%%%%%%%%%%%
%%%%%%%%%%%%%%%%%%%%%%%%%%%%%%%%%%%%%%%%%%%%%%%

\thispagestyle{empty}

\begin{flushright}
  {\small
  LMU-ASC 78/17
  }
\end{flushright}

\vspace*{2.5cm}

%%%%%%%%%%%%%%%%%%%%%%%%%%%%%%%%%%%%%%%%%%%%%%
%%%%%%%%%%%%%%%%%%%%%%%%%%%%%%%%%%%%%%%%%%%%%%

\begin{center}
{\LARGE
Dimensional Reductions of DFT and \\[4pt]  Mirror Symmetry for Calabi-Yau \\[4pt]  Three-folds
and $K3\times T^{2}$ \\[4pt] 
}
\end{center}

%%%%%%%%%%%%%%%%%%%%%%%%%%%%%%%%%%%%%%%%%%%%%%
%%%%%%%%%%%%%%%%%%%%%%%%%%%%%%%%%%%%%%%%%%%%%%

\vspace{0.6cm}

\begin{center}
  Philip Betzler, Erik Plauschinn
\end{center}

%%%%%%%%%%%%%%%%%%%%%%%%%%%%%%%%%%%%%%%%%%%%%%
%%%%%%%%%%%%%%%%%%%%%%%%%%%%%%%%%%%%%%%%%%%%%%

\vspace{0.6cm}

\begin{center} 
\emph{Arnold-Sommerfeld-Center for Theoretical Physics \\
Ludwig-Maximilians-Universit\"at M\"unchen \\ Theresienstra\ss e 37, 80333 M\"unchen \\ Germany} \\
\end{center} 

\vspace{1.5cm}

%%%%%%%%%%%%%%%%%%%%%%%%%%%%%%%%%%%%%%%%%%%%%%
%%%%%%%%%%%%%%%%%%%%%%%%%%%%%%%%%%%%%%%%%%%%%%

\begin{abstract}
\noindent
We perform dimensional reductions of type IIA and type IIB double field theory in the flux formulation 
on Calabi-Yau three-folds and on $K3\times T^2$. In addition to geometric and 
non-geometric three-index fluxes and Ramond-Ramond fluxes, we include generalized dilaton fluxes.
We relate our results to the scalar potentials of corresponding four-dimensional gauged supergravity 
theories, and we verify the expected behavior under mirror symmetry. 
For Calabi-Yau three-folds we extend this analysis to the full bosonic action including kinetic terms. 
\end{abstract}

\clearpage

%%%%%%%%%%%%%%%%%%%%%%%%%%%%%%%%%%%%%%%%%%%%%%%
%%%%%%%%%%%%%%%%%%%%%%%%%%%%%%%%%%%%%%%%%%%%%%%
%%%%%%%%%%%%%%%%%%%%%%%%%%%%%%%%%%%%%%%%%%%%%%%
%%%%%%%%%%%%%%%%%%%%%%%%%%%%%%%%%%%%%%%%%%%%%%%
%%%%%%%%%%%%%%%%%%%%%%%%%%%%%%%%%%%%%%%%%%%%%%%
%%%%%%%%%%%%%%%%%%%%%%%%%%%%%%%%%%%%%%%%%%%%%%%
%%%%%%%%%%%%%%%%%%%%%%%%%%%%%%%%%%%%%%%%%%%%%%%
%%%%%%%%%%%%%%%%%%%%%%%%%%%%%%%%%%%%%%%%%%%%%%%

\tableofcontents

%%%%%%%%%%%%%%%%%%%%%%%%%%%%%%%%%%%%%%%%%%%%%%%
%%%%%%%%%%%%%%%%%%%%%%%%%%%%%%%%%%%%%%%%%%%%%%%
%%%%%%%%%%%%%%%%%%%%%%%%%%%%%%%%%%%%%%%%%%%%%%%
%%%%%%%%%%%%%%%%%%%%%%%%%%%%%%%%%%%%%%%%%%%%%%%
%%%%%%%%%%%%%%%%%%%%%%%%%%%%%%%%%%%%%%%%%%%%%%%
%%%%%%%%%%%%%%%%%%%%%%%%%%%%%%%%%%%%%%%%%%%%%%%

\section{Introduction}

One of the important problems in string phenomenology is  moduli stabilization.  Moduli are  massless 
scalar fields which arise when compactifying string theory and which are inconsistent with 
experimental observations. 
A way to address this issue is to turn
on background fluxes on the internal manifold (see, e.g. \cite{Grana:2005jc,Douglas:2006es,Denef:2007pq}
for  reviews on the topic). At string tree-level, this creates
a scalar potential that can stabilize the moduli parametrizing the
vacuum degeneracy. It was, however, found that successive application
of T-duality transformations to backgrounds with fluxes gives rise to geometrically
ill-defined objects \cite{Shelton:2005cf,Wecht:2007wu} which play
an essential role in obtaining full moduli stabilization. Constructing
phenomenologically realistic models from flux compactifications therefore
requires suitable frameworks allowing for a mathematical description
of such ``non-geometric'' backgrounds.

One natural approach is to relax the Calabi-Yau condition and only assume
the existence of a nowhere vanishing spinor on the compactification manifold. As a consequence,
Calabi-Yau manifolds are replaced by more general $SU(3)$ structure
manifolds, which had previously been shown to arise as mirror symmetry
duals of Calabi-Yau backgrounds with non-vanishing Neveu-Schwarz--Neveu-Schwarz (NS-NS) fluxes \cite{Louis:2002ny,Gurrieri2003,Gurrieri:2002iw}.
Focusing on type II theories and going one step further, this idea
can be generalized by assuming the existence of a pair of non-vanishing
spinors, one for each of the ten-dimensional supercharges. This is
the underlying idea of compactifications on $SU(3)\!\times\! SU(3)$
structure manifolds. Such compactifications have been extensively
studied in \cite{Louis:2002ny,Gurrieri:2002iw,Gurrieri2003,Kachru:2004jr,DAuria:2004kwe,DAuria:2004mje,Tomasiello:2005bp,House:2005yc,Grana:2005ny,Chuang:2005qd,KashaniPoor:2006si,KashaniPoor:2007tr,Grana:2006hr}.
Interestingly, the latter show a natural connection to Hitchin's generalized
geometry \cite{Hitchin:2004ut,Gualtieri:2003dx}, where in this picture
$SU(3)\!\times\! SU(3)$ appears as the structure group of the generalized
tangent bundle $TM^{6}\oplus T^{*}M^{6}$ of the internal manifold
$M^{6}$.

In this paper, we will go another step further and consider compactifications
of type II actions in the framework of double field theory (DFT) \cite{Siegel:1993xq,Siegel:1993th,Hull:2009mi,Hohm:2010pp,Hohm:2010jy}
(see also \cite{Aldazabal:2013sca,Hohm:2013bwa,Berman:2013eva}
for pedagogical reviews). In addition to the generalized tangent bundle,
in DFT spacetime itself is doubled, allowing for a description of
ten-dimensional supergravities in which T-duality becomes a manifest
symmetry. In particular, it has been shown that there exists a ``flux
formulation'' \cite{Geissbuhler:2013uka} of DFT in which geometric
as well as non-geometric background fluxes arise naturally as constituents
of the action and can locally be described as operators acting on
differential forms.

It was found that compactifications and Scherk-Schwarz reductions
of DFT yield the scalar potential of half-maximal gauged supergravity
in four dimensions \cite{Aldazabal:2011nj,Geissbuhler:2011mx,Grana:2012rr}.
More recently, a connection between Calabi-Yau compactifications of
DFT and the scalar potential of four-dimensional $\mathcal{N}=2$
gauged supergravity was derived explicitly \cite{Blumenhagen:2015lta}.
The purpose of the present paper is to add to the picture by generalizing
the considered setting of \cite{Blumenhagen:2015lta} to a wider class
compactification manifolds and non-vanishing dilaton fluxes. We furthermore
extend the formalism to dimensional reductions of the full DFT action
by including the kinetic terms. This will eventually enable us to
show how in DFT $\textrm{IIA}\leftrightarrow\textrm{IIB}$ Mirror
Symmetry is restored due to the simultaneous presence of geometric
and non-geometric fluxes. 

\bigskip
In this paper we discuss the technical details of our analysis in some length, and therefore want 
to briefly summarize the main results of our work. In particular, the paper is 
organized as follows:
\begin{itemize}

\item In section~\ref{sec:2}, we provide a brief review on the framework of DFT.
The section is concluded by a short presentation of the flux formulation
and related notions which will be important for this paper.

\item In section~\ref{sec:3}, we compactify the purely internal part of the type
IIA and IIB DFT action on a Calabi-Yau three-fold. In doing so, we
mainly rely on the elaborations of \cite{Blumenhagen:2015lta} and
generalize the setting by including additional generalized dilaton
fluxes and cohomologically trivial terms in order to reveal more general
structures underlying the calculation. Both results are related to
the scalar potential of four-dimensional $\mathcal{N}=2$ gauged supergravity
constructed in \cite{DAuria:2007axr}, and a first manifestation of
Mirror Symmetry is discussed.

\item In section~\ref{sec:4}, the discussion of section~\ref{sec:3} is repeated for the
compactification manifold $K3\times T^{2}$. The necessary mathematical
steps to generalize the Calabi-Yau setting are highlighted, and the
special geometric properties of $K3\times T^{2}$ are discussed in
detail. The resulting four-dimensional scalar potential is related
to the framework of \cite{DAuria:2007axr}, and a set of mirror mappings
is constructed. A DFT origin of the $\mathcal{N}=4$ gauged
supergravity scalar potential has already been elaborated in the previous
works \cite{Geissbuhler:2011mx,Aldazabal:2011nj} using Scherk-Schwarz
reductions, however,  here we follow a different approach by employing the formalism
of generalized Calabi-Yau geometry \cite{Hitchin:2004ut} and generalized
K3 surfaces \cite{Huybrechts:2003ak}, giving rise to a scalar potential
formulated in the language of $\mathcal{N}=2$ gauged supergravity.
While the result shows characteristic features of its $\mathcal{N}=4$
counterpart, its relation to those of \cite{Geissbuhler:2011mx,Aldazabal:2011nj}
seems to be nontrivial and will be investigated in future work.

\item In section~\ref{sec:5}, we extend the setting of section~\ref{sec:3} by including
the kinetic terms.  We use a generalized Kaluza-Klein ansatz \cite{Aldazabal:2011nj,Geissbuhler:2011mx,Hohm:2013nja}
and treat the NS-NS and Ramond-Ramond (R-R) sectors separately. For the former, we
will mostly rely on the results of section~\ref{sec:3} and on the standard literature
on Calabi-Yau compactifications of type II theories. The latter is
more involved and gives rise to democratic type II supergravities
with all known NS-NS fluxes (including the non-geometric ones) and
R-R fluxes turned on. We first reduce the ten-dimensional equations
of motion, following a similar pattern as done in \cite{Cassani:2008rb}
for manifolds with $SU(3)\!\times\! SU(3)$ structure. The resulting
four-dimensional equations of motion can then be shown to originate
from the four-dimensional $\mathcal{N}=2$ gauged supergravity action
constructed in \cite{DAuria:2007axr}, where a subset of the axions
appearing in the standard formulation is dualized to two-forms in
order to account for both electric and magnetic charges. This will
finally enable us to once more read off a set of mirror mappings between
the full reduced type IIA and IIB actions.

\item Section~\ref{sec:6} concludes the discussion by summarizing the
results and giving an outlook on open questions and possible future
developments.

\end{itemize}
Throughout this work, we consider a doubled analogue of the spacetime
manifold $M^{10}=M^{1,4}\times M^{6}$, where $M^{1,4}$ denotes a
four-dimensional Lorentzian manifold and $M^{6}$ is an arbitrary
Calabi-Yau three-fold or  $K3\times T^{2}$. Moreover, we will apply
the framework of special geometry in order to describe the complex
structure and K\"ahler class moduli spaces of $M^{6}$. Due to the large
number of distinct indices used in this paper, we provide an accessible
indexing system in appendix~\ref{sec:A}.

\section{\label{sec:2}Basics of Double Field Theory}

This section will provide a brief overview on the notions of DFT,
which form the basis of our upcoming considerations. For more details,
we would like to refer the reader to \cite{Aldazabal:2013sca,Hohm:2013bwa,Berman:2013eva}.

\subsection{Doubled Spacetime}

The basic idea of DFT is to enhance ordinary supergravity theories
with additional structures in a way that T-duality becomes a manifest
symmetry. Motivated by the insights from toroidal compactifications
of the bosonic string, one doubles the dimension of the $D$-dimensional
spacetime manifold $M$ by introducing additional \emph{winding coordinates}\textit{
}$\tilde{x}_{\hat{m}}$ conjugate to the winding number $\tilde{p}_{\hat{m}}$
(just as the normal spacetime coordinates $x^{\hat{m}}$ relate to
the momenta $p^{\hat{m}}$) and arrange them in doubled coordinates
\begin{equation}
X^{\hat{M}}=\left(\tilde{x}_{\hat{m}},x^{\hat{m}}\right),\; P_{\hat{M}}=\left(\tilde{p}^{\hat{m}},p_{\hat{m}}\right)\qquad\textrm{with}\qquad\hat{m}=1,\ldots D\;\textrm{and}\;\hat{M}=0,\ldots2D.
\end{equation}
The corresponding derivatives are denoted by 
\begin{equation}
\partial_{\hat{m}}=\frac{\partial}{\partial x^{\hat{m}}},\quad\tilde{\partial}^{\hat{m}}=\frac{\partial}{\partial\tilde{x}_{\hat{m}}}.
\end{equation}
The spacetime manifold is locally equipped with the \emph{generalized
tangent bundle} 
\begin{equation}
E=TM\oplus T^{*}M
\end{equation}
and the $O\left(D,D,\mathbb{R}\right)$ \emph{invariant structure}
\begin{equation}
\eta_{\hat{M}\hat{N}}=\left(\begin{array}{cc}
0 & \delta^{\hat{m}}{}_{\hat{n}}\\
\delta_{\hat{m}}{}^{\hat{n}} & 0
\end{array}\right)=\eta^{\hat{M}\hat{N}}\label{eq: DFT: O(D,D) invariant Structure}
\end{equation}
defining the standard inner product of doubled vectors and taking
the same role as the Minkowski metric in general relativity. The spacetime
metric $g_{\hat{m}\hat{n}}$ and the Kalb-Ramond field $B_{\hat{m}\hat{n}}$
are repackaged into the \emph{generalized metric} 
\begin{equation}
\mathcal{\hat{H}}_{\hat{M}\hat{N}}=\left(\begin{array}{cc}
\hat{g}^{\hat{m}\hat{n}} & -\hat{g}^{\hat{m}\hat{p}}\hat{B}_{\hat{p}\hat{n}}\\
\hat{B}_{\hat{m}\hat{p}}\hat{g}^{\hat{p}\hat{n}} & g_{\hat{m}\hat{n}}-\hat{B}_{\hat{m}p}\hat{g}^{\hat{p}\hat{q}}\hat{B}_{\hat{q}\hat{n}}
\end{array}\right),\label{eq: Generalized Metric}
\end{equation}
whose structure is closely related to the Buscher rules for T-duality
transformations \cite{Buscher:1987sk,Buscher:1987qj}. It defines
a function $\mathcal{\hat{H}}_{\hat{M}\hat{N}}\left(X\right)$ of
the doubled coordinates and parametrizes the coset space $\frac{O\left(D,D,\mathbb{R}\right)}{O\left(D,\mathbb{R}\right)\times O\left(D,\mathbb{R}\right)}$.
Similarly to general relativity, indices in DFT are raised and lowered
by the $O\left(D,D,\mathbb{R}\right)$ invariant metric $\eta_{\hat{M}\hat{N}}$
and $\eta^{\hat{M}\hat{N}}$, respectively. In particular, one obtains
the relation 
\begin{equation}
\mathcal{\hat{H}}^{\hat{M}\hat{N}}=\eta^{\hat{M}\hat{P}}\mathcal{\hat{H}}_{\hat{P}\hat{Q}}\eta^{\hat{Q}\hat{N}},\label{eq: DFT: Generalized Metric Constraint}
\end{equation}
implying the existence of a \emph{generalized vielbein} $\mathcal{\mathcal{\hat{E}}}^{\hat{A}}{}_{\hat{M}}$
satisfying 
\begin{equation}
\mathcal{\hat{H}}_{\hat{M}\hat{N}}=\mathcal{\mathcal{\hat{E}}}^{\hat{A}}{}_{\hat{M}}\mathcal{\mathcal{\hat{E}}}^{\hat{B}}{}_{\hat{N}}S_{\hat{A}\hat{B}}.\label{eq: Relation: Generalized metric - Vielbein}
\end{equation}
Here, $\hat{M},\hat{N}$ denote curved spacetime indices, and $\hat{A},\hat{B}$
are flat tangent space indices. One can thus choose 
\begin{equation}
S_{\hat{A}\hat{B}}=\left(\begin{array}{cc}
s^{\hat{a}\hat{b}} & 0\\
0 & s_{\hat{a}\hat{b}}
\end{array}\right),\label{eq: Relation: Generalized Metric - Vielbein 2}
\end{equation}
where $s_{\hat{a}\hat{b}}$ denotes the flat $D$-dimensional Minkowski
metric. Using the vielbein $\hat{e}^{\hat{a}}\,_{\hat{m}}$ defined
by the relation 
\begin{equation}
g_{\hat{m}\hat{n}}=\hat{e}^{\hat{a}}\,_{\hat{m}}s_{\hat{a}\hat{b}}\hat{e}^{\hat{b}}\,_{\hat{n}},
\end{equation}
$\mathcal{\mathcal{\hat{E}}}^{\hat{A}}{}_{\hat{M}}$ can be parametrized
as 
\begin{equation}
\mathcal{\mathcal{\hat{E}}}^{\hat{A}}{}_{\hat{M}}=\left(\begin{array}{cc}
\hat{e}_{\hat{a}}{}^{\hat{m}} & -\hat{e}_{\hat{a}}{}^{\hat{p}}\hat{B}_{\hat{p}\hat{m}}\\
0 & \hat{e}^{\hat{a}}{}_{\hat{m}}
\end{array}\right).\label{eq: Relation: Generalized Metric - Vielbein 3}
\end{equation}
An action for DFT is determined by requiring invariance of the theory
under local doubled diffeomorphisms 
\begin{equation}
X^{M}=\left(\tilde{x}_{\hat{m}},x^{\hat{m}}\right)\rightarrow\left(\tilde{x}_{\hat{m}}+\widetilde{\xi}\left(X^{\hat{M}}\right),x^{\hat{m}}+\xi\left(X^{\hat{M}}\right)\right)
\end{equation}
and global $O\left(D,D,\mathbb{R}\right)$ transformations. In conjunction
with the requirement of the algebra of infinitesimal diffeomorphisms
to be closed, the latter give rise to the so-called \emph{strong constraint}
\begin{equation}
\eta^{\hat{M}\hat{N}}\partial_{\hat{M}}\Phi\partial_{\hat{N}}\Psi=0,\label{eq: Strong Constraint}
\end{equation}
where both $\Phi$ and $\Psi$ denote arbitrary fields or gauge parameters.
One possible solution is given by setting $\widetilde{\partial}^{\hat{m}}=0$,
in which case the dual coordinates become unphysical and the theory
reduces to ordinary supergravity. This also reveals an interpretation
of T-duality transformations as rotations of a ``physical section''
in doubled spacetime.

\subsection{Flux Formulation of Double Field Theory\label{sub:2.2}}

There exist two physically equivalent formulations of DFT, differing
only by terms that are either total derivatives or vanish by the strong
constraint. For the purpose of this paper, working with the so-called
\emph{flux formulation} \cite{Hohm:2010xe,Aldazabal:2011nj,Geissbuhler:2011mx}
(see also \cite{Siegel:1993xq,Siegel:1993th} for early developments)
will be more convenient since it provides a natural (local) description
of geometric as well as non-geometric background fluxes.

\subsubsection{NS-NS Sector}

As starting point for the NS-NS sector, we consider the action \cite{Hohm:2010xe,Aldazabal:2011nj,Geissbuhler:2011mx}
\begin{equation}
\begin{array}{ccl}
S_{\textrm{NS-NS}} & {\displaystyle \!\!\!\!=\!\!\!\!} & {\displaystyle \frac{1}{2}\int_{M^{10}}\textrm{d}^{20}Xe^{-2d}\left[\mathcal{\hat{F}}_{\hat{M}\hat{N}\hat{P}}\mathcal{\hat{F}}_{\hat{M}'\hat{N}'\hat{P}'}\left(\frac{1}{4}\mathcal{H}^{\hat{M}\hat{M}'}\eta^{\hat{N}\hat{N}'}\eta^{\hat{P}\hat{P}'}-\frac{1}{12}\mathcal{H}^{\hat{M}\hat{M}'}\mathcal{H}^{\hat{N}\hat{N}'}\mathcal{H}^{\hat{P}\hat{P}'}\right.\right.}\\
 & \vphantom{^{X^{X^{X^{X^{X^{X}}}}}}} & {\displaystyle \left.-\frac{1}{6}\eta^{\hat{M}\hat{M}'}\eta^{\hat{N}\hat{N}'}\eta^{\hat{P}\hat{P}'}\right)\left.+\mathcal{\hat{F}}_{\hat{M}}\mathcal{\hat{F}}_{\hat{M}'}\left(\vphantom{\frac{1}{4}S^{AA'}}\eta^{\hat{M}\hat{M}'}-\mathcal{H}^{\hat{M}\hat{M}'}\right)\right],}
\end{array}\label{eq: NS-NS Action Flux Formulation}
\end{equation}
where the \emph{generalized dilaton} $d$ is defined by the relation
\begin{equation}
e^{-2d}=\sqrt{\hat{g}}e^{-2\phi}.
\end{equation}
When performing dimensional reduction, an obvious first step is to
rewrite the action in terms of objects representing four-dimensional
fields and assume all fields with external legs to be independent
of the internal coordinates. We will do this by applying a generalized
Kaluza-Klein ansatz for DFT \cite{Aldazabal:2011nj,Geissbuhler:2011mx,Hohm:2013nja},
for which we split the coordinates into external and internal parts
\begin{equation}
X^{\hat{M}}=\left(\tilde{x}_{\mu},x^{\mu},Y^{\check{I}}\right),\qquad X^{\hat{A}}=\left(\tilde{x}_{e},x^{e},Y^{\check{A}}\right),
\end{equation}
where we used the collective notation $Y^{\check{I}}=\left(\tilde{y}_{\check{i}},y^{\check{i}}\right)$
and $Y^{\check{A}}=\left(\tilde{y}_{\check{a}},y^{\check{a}}\right)$
for the latter. In order to preserve rigid $O(6,6,\mathbb{R})$ symmetry,
we impose the section condition only on the external coordinates,
therefore assuming also independence of all fields and gauge parameters
of the external dual coordinates $\tilde{x}_{\mu}$, while leaving
the dependence of purely internal fields on the doubled coordinates
$Y^{\check{I}},Y^{\check{A}}$ untouched.

For the ten-dimensional metric and Kalb-Ramond field, we employ the
splitting \cite{Aldazabal:2011nj}
\begin{equation}
\hat{g}_{\hat{m}\hat{n}}=\left(\begin{array}{cc}
{\displaystyle g_{\mu\nu}+g_{\check{k}\check{l}}A^{\check{k}}{}_{\mu}A^{\check{l}}{}_{\nu}\vphantom{\frac{1}{2}}} & {\displaystyle A^{\check{k}}{}_{\mu}g_{\check{k}\check{j}}\vphantom{\frac{1}{2}}}\\
{\displaystyle g_{\check{i}\check{k}}A^{\check{k}}{}_{\nu}\vphantom{\frac{1}{2}}} & {\displaystyle g_{\check{i}\check{j}}\vphantom{\frac{1}{2}}}
\end{array}\right),\qquad\hat{B}_{\hat{m}\hat{n}}=\left(\begin{array}{cc}
{\displaystyle B_{\mu\nu}\vphantom{\frac{1}{2}}} & {\displaystyle -B_{\mu\check{j}}\vphantom{\frac{1}{2}}}\\
{\displaystyle B_{\check{i}\nu}\vphantom{\frac{1}{2}}} & {\displaystyle B_{\check{i}\check{j}}\vphantom{\frac{1}{2}}}
\end{array}\right)
\end{equation}
and arrange the parts with mixed external and internal indices in
a generalized Kaluza-Klein vector 
\begin{equation}
\mathcal{A}^{\check{I}}{}_{\mu}=\left(\begin{array}{c}
{\displaystyle B_{\check{i}\mu}\vphantom{\frac{1}{2}}}\\
{\displaystyle -A^{\check{i}}{}_{\mu}\vphantom{\frac{1}{2}}}
\end{array}\right).\label{eq: generalized Kaluza-Klein vector}
\end{equation}
Inserting this ansatz into (\ref{eq: NS-NS Action Flux Formulation}),
the NS-NS contribution to the action can be reformulated as \cite{Aldazabal:2011nj,Geissbuhler:2011mx,Hohm:2013nja}
\begin{equation}
\begin{array}{cll}
S_{\textrm{NS-NS}} & {\displaystyle \!\!=\!\!\!\!\!} & {\displaystyle \frac{1}{2}\int_{M^{10}}\textrm{d}^{4}x\textrm{d}^{12}Y\,\sqrt{g^{\left(4\right)}}e^{-2\phi}\left[\vphantom{\frac{1}{4}}\right.}\\
 & \vphantom{^{X^{X^{X^{X^{X^{X}}}}}}} & {\displaystyle {\displaystyle \widetilde{R}^{\left(4\right)}+4g^{\mu\nu}D_{\mu}\phi D_{\nu}\phi-\frac{1}{4}g^{\mu\nu}g^{\rho\sigma}\mathcal{H}_{IJ}\widetilde{\mathcal{F}}^{I}{}_{\mu\rho}\widetilde{\mathcal{F}}^{J}{}_{\nu\sigma}}}\\
 & \vphantom{^{X^{X^{X^{X^{X^{X}}}}}}} & {\displaystyle -\frac{1}{12}g^{\mu\nu}g^{\rho\sigma}g^{\tau\lambda}\mathcal{\widetilde{H}}_{\mu\rho\tau}\mathcal{\widetilde{H}}_{\nu\sigma\lambda}+g^{\mu\nu}\frac{1}{8}D_{\mu}\mathcal{H}_{\check{I}\check{J}}D_{\nu}\mathcal{H}^{\check{I}\check{J}}}\\
 & \vphantom{^{X^{X^{X^{X^{X^{X}}}}}}} & +{\displaystyle \mathcal{F}_{\check{I}\check{J}\check{K}}\mathcal{F}_{\check{I}'\check{J}'\check{K}'}\left(-\frac{1}{12}\mathcal{H}^{\check{I}\check{I}'}\mathcal{H}^{\check{J}\check{J}'}\mathcal{H}^{\check{K}\check{K}'}+\frac{1}{4}\mathcal{H}^{\check{I}\check{I}'}\eta^{\check{J}\check{J}'}\eta^{\check{K}\check{K}'}-\frac{1}{6}\eta^{\check{I}\check{I}'}\eta^{\check{J}\check{J}'}\eta^{\check{K}\check{K}'}\right)}\\
 & \vphantom{^{X^{X^{X^{X^{X^{X}}}}}}} & \left.{\displaystyle +\mathcal{F}_{\check{I}}\mathcal{F}_{\check{I}'}\left(\vphantom{\frac{1}{4}S^{AA'}}\eta^{\check{I}\check{I}'}-\mathcal{H}^{\check{I}\check{I}'}\right)}\right]
\end{array}\label{eq: Full Action NSNS-sector}
\end{equation}
where we defined the field strengths 
\begin{equation}
\begin{array}{lcl}
\widetilde{\mathcal{F}}^{\check{I}}{}_{\mu\nu} & {\displaystyle \!\!\!\!=\!\!\!\!} & {\displaystyle 2\partial_{[\underline{\mu}}A^{\check{I}}{}_{\underline{\nu}]}-\mathcal{F}^{\check{I}}{}_{\check{J}\check{K}}\mathcal{A}^{\check{J}}{}_{\mu}\mathcal{A}^{\check{K}}{}_{\nu}+2\mathcal{F}_{\check{J}}\mathcal{A}^{\check{J}}{}_{[\underline{\mu}}\mathcal{A}^{\check{I}}{}_{\underline{\nu}]}-2\mathcal{F}^{\check{I}}B_{\mu\nu}},\\
\mathcal{\widetilde{H}}_{\mu\nu\rho} & {\displaystyle \!\!\!\!=\!\!\!\!} & {\displaystyle 3\partial_{[\underline{\mu}}B_{\underline{\nu\rho}]}-3\partial_{[\underline{\mu}}\mathcal{A}^{\check{K}}{}_{\underline{\nu}}\mathcal{A}{}_{\underline{\rho}]\check{K}}-6\mathcal{F}_{\check{K}}\mathcal{A}^{\check{K}}{}_{[\underline{\mu}}B_{\underline{\nu\rho}]}-\mathcal{F}_{\check{I}\check{J}\check{K}}\mathcal{A}^{\check{I}}{}_{\mu}\mathcal{A}^{\check{J}}{}_{\nu}\mathcal{A}^{\check{K}}{}_{\rho}\vphantom{^{X^{X^{X^{X^{X^{X}}}}}}}}
\end{array}\label{eq: NSNS Gauge Field Strengths}
\end{equation}
and the covariant derivatives 
\begin{equation}
\begin{array}{rcl}
D_{\mu}\mathcal{H}_{\check{I}\check{J}} & {\displaystyle \!\!\!\!=\!\!\!\!} & \partial_{\mu}\mathcal{H}_{\check{I}\check{J}}+\mathcal{A}^{\check{K}}{}_{\mu}\mathcal{F}{}_{\check{K}\check{J}}{}^{\check{L}}\mathcal{H}_{\check{I}\check{L}}-\mathcal{A}_{\mu\check{J}}\mathcal{H}_{\check{I}\check{K}}\mathcal{F}^{\check{K}}+\mathcal{F}_{\check{J}}\mathcal{H}_{\check{I}\check{K}}\mathcal{A}^{\check{K}}{}_{\mu},\\
D_{\mu}\phi & {\displaystyle \!\!\!\!=\!\!\!\!} & \partial_{\mu}\phi-\mathcal{F}_{\check{K}}\mathcal{A}^{\check{K}}{}_{\mu}.\vphantom{^{X^{X^{X^{X^{X^{X}}}}}}}
\end{array}\label{eq: NSNS Covariant Derivative}
\end{equation}
Using the \emph{generalized Weizenb\"ock connection}\textit{ 
\begin{equation}
\Omega_{\check{A}\check{B}\check{C}}=\mathcal{E}_{\check{A}}{}^{\check{I}}\left(\partial_{\check{I}}\mathcal{E}_{\check{B}}{}^{\check{J}}\right)\mathcal{E}_{\check{C}\check{J}}
\end{equation}
}the\textit{ }\emph{generalized fluxes}\textit{ $\mathcal{F}_{\check{A}}$
}and\textit{ }$\mathcal{F}_{\check{A}\check{B}\check{C}}$ with flat
indices can be written as 
\begin{equation}
\mathcal{F}_{\check{A}}=\Omega^{\check{B}}{}_{\check{B}\check{A}}+2\mathcal{E}_{\check{A}}{}^{\check{I}}\partial_{\check{I}}d\qquad\textrm{and}\qquad\mathcal{F}_{\check{A}\check{B}\check{C}}=3\Omega_{[\underline{\check{A}\check{B}\check{C}}]},\label{eq: Generalized NSNS-Fluxes}
\end{equation}
where the squared brackets denote the antisymmetrization operator
defined in appendix~\ref{sec:A}. It will be explained in subsection~\ref{sub:2.3.1} how
these are related to the generalized fluxes with curved indices.

\subsubsection{\label{sub:R-R-Sector-Basics}R-R Sector}

A similar analysis has been done for the R-R sector in \cite{Rocen:2010bk,Hohm:2011zr,Hohm:2011dv,Hohm:2011cp,Jeon:2012kd}.
Recalling that the fields transform as $O\left(10,10\right)$ spinors
by construction, we expand
\begin{equation}
\hat{\mathfrak{G}}=\sum_{n}\frac{1}{n!}\mathcal{\hat{\mathfrak{G}}}_{\hat{m}_{1}\ldots\hat{m}_{n}}^{\left(n\right)}\hat{e}_{\hat{a}_{1}}\,^{\hat{m}_{1}}\ldots e_{\hat{a}_{n}}\,^{\hat{m}_{n}}\Gamma^{\hat{a}_{1}\ldots\hat{a}_{n}}\left|0\right\rangle ,
\end{equation}
where $\Gamma^{\hat{a}_{1}\ldots\hat{a}_{n}}$ denotes the totally
antisymmetrized product of $n$ gamma-matrices. The R-R gauge potentials
can be combined into a spinor 
\begin{equation}
\hat{\mathcal{C}}=\begin{cases}
\sum_{n=0}^{4}\hat{C}_{2n+1} & \textrm{for type IIA theory}\\
\sum_{n=0}^{4}\hat{C}_{2n} & \textrm{for type IIB theory},
\end{cases}\label{eq:R-R-Sector-C-Poly-Form}
\end{equation}
which can be used to write 
\begin{equation}
\hat{\mathfrak{G}}=\begin{cases}
G_{0}+/\hspace{-0.25cm}\nabla\hat{\mathcal{C}} & \textrm{for type IIA theory}\\
/\hspace{-0.25cm}\nabla\hat{\mathcal{C}} & \textrm{for type IIB theory},
\end{cases}\label{eq:R-R-Sector-G-Poly-Form}
\end{equation}
with the \emph{generalized fluxed Dirac operator} 
\begin{equation}
/\hspace{-0.25cm}\nabla=\Gamma^{\hat{A}}E_{\hat{A}}{}^{\hat{M}}\partial_{\hat{M}}-\frac{1}{2}\Gamma^{\hat{A}}\mathcal{F}_{\hat{A}}-\frac{1}{6}\Gamma^{\hat{A}\hat{B}\hat{C}}\mathcal{F}_{\hat{A}\hat{B}\hat{C}}.\label{eq: DFT RR twsited Dirac operator}
\end{equation}
The zero-form R-R flux $G_{0}$ in the type IIA case arises as dual
of the background field strength of $\hat{C}_{9}$. A pseudo-action
for the R-R sector can be obtained by summing over all relevant components
of a particular theory,
\begin{equation}
S_{\textrm{R-R}}=\frac{1}{2}\int_{M^{10}}\textrm{d}^{4}x\textrm{d}^{12}Y\left({\displaystyle -\frac{1}{2}\hat{\mathfrak{G}}\wedge\star\hat{\mathfrak{G}}}\right).\label{eq: DFT: RR Action}
\end{equation}
Since all fields $\hat{C}_{n}$ of a certain theory appear explicitly,
this has to be supplemented by duality constraints. Denoting the ten-dimensional
$n$-form contributions by $\hat{\mathfrak{G}}_{n}$, these take the
form \cite{Bergshoeff:2001pv} 
\begin{equation}
\mathfrak{\hat{G}}_{n}=\left(-1\right)^{\left\lfloor \frac{n}{2}\right\rfloor }\star\mathfrak{\hat{G}}_{n},\label{eq: Duality Relations General}
\end{equation}
where the floor operator $\left\lfloor \cdot\right\rfloor $ gives
as output the least integer that is greater than or equal to the argument.

\subsection{Fluxes in Doubled Geometry\label{sub:2.3}}

This section will focus on the scalar potential component of (\ref{eq: Full Action NSNS-sector})
and introduce a DFT interpretation of the NS-NS fluxes. This has first
been investigated in \cite{Blumenhagen:2015lta}, and much of this
section will be based on this work.

\subsubsection{\label{sub:2.3.1}Fluxes as Fluctuations about the Calabi-Yau Background}

The main idea is to treat the generalized fluxes (\ref{eq: Generalized NSNS-Fluxes})
as manifestations of small deviations from the Calabi-Yau background,
arising from perturbations of the internal vielbeins 
\begin{equation}
\mathcal{E}^{\check{A}}{}_{\check{I}}=\overset{{\scriptscriptstyle \circ}}{\mathcal{E}}^{\check{A}}{}_{\check{I}}+\overline{\mathcal{E}}^{\check{A}}{}_{\check{I}}+\mathcal{O}\left(\overline{\mathcal{E}}^{2}\right),\label{eq: Vielbein Deviations}
\end{equation}
where $\overset{{\scriptscriptstyle \circ}}{\mathcal{E}}^{\check{A}}{}_{_{\check{I}}}$
describes the Calabi-Yau background and $\overline{\mathcal{E}}^{\check{A}}{}_{_{\check{I}}}$
the fluctuations. Inserting this expansion into the generalized fluxes
(\ref{eq: Generalized NSNS-Fluxes}), we can write 
\begin{equation}
\mathcal{F}_{\check{A}}=\overset{{\scriptscriptstyle \circ}}{\mathcal{F}}_{\check{A}}+\overline{\mathcal{F}}_{\check{A}}+\mathcal{O}\left(\overline{\mathcal{E}}^{2}\right),\qquad\mathcal{F}_{\check{A}\check{B}\check{C}}=\overset{{\scriptscriptstyle \circ}}{\mathcal{F}}_{\check{A}\check{B}\check{C}}+\overline{\mathcal{F}}_{\check{A}\check{B}\check{C}}+\mathcal{O}\left(\overline{\mathcal{E}}^{2}\right).\label{eq: Generalized Fluxes Deviations}
\end{equation}
As the notation implies, $\overset{{\scriptscriptstyle \circ}}{\mathcal{F}}_{\check{A}}$
and $\overset{{\scriptscriptstyle \circ}}{\mathcal{F}}_{\check{A}\check{B}\check{C}}$
depend only on $\overset{{\scriptscriptstyle \circ}}{\mathcal{E}}^{\check{A}}{}_{\check{I}}$
and do not contribute to the scalar potential since $\overset{{\scriptscriptstyle \circ}}{\mathcal{E}}^{\check{A}}{}_{\check{I}}$
satisfies the DFT equations of motion. By contrast, $\overline{\mathcal{F}}_{\check{A}}$
and $\overline{\mathcal{F}}_{\check{A}\check{B}\check{C}}$ depend
linearly on the fluctuations $\overline{\mathcal{E}}^{\check{A}}{}_{\check{I}}$
and therefore have to be taken into account.

In the following, we will use the background component $\overset{{\scriptscriptstyle \circ}}{\mathcal{E}}^{\check{A}}{}_{\check{I}}$
of the vielbein to switch between flat and curved indices (defining,
e.g. $\mathcal{\overline{F}}_{\check{I}\check{J}\check{K}}=\overset{{\scriptscriptstyle \circ}}{\mathcal{E}}^{\check{A}}{}_{\check{I}}\overset{{\scriptscriptstyle \circ}}{\mathcal{E}}^{\check{B}}{}_{\check{J}}\overset{{\scriptscriptstyle \circ}}{\mathcal{E}}^{\check{C}}{}_{\check{K}}\mathcal{\overline{F}}_{\check{A}\check{B}\check{C}}$).
For the case of constant expectation values, the three-indexed object
$\overline{\mathcal{F}}_{\check{I}\check{J}\check{K}}$ has been shown
to encode the known geometric and non-geometric NS-NS fluxes by 
\begin{equation}
\mathcal{\overline{F}}_{\check{i}\check{j}\check{k}}=H_{\check{i}\check{j}\check{k}},\quad\mathcal{\overline{F}}^{\check{i}}{}_{\check{j}\check{k}}=F^{\check{i}}{}_{\check{j}\check{k}},\quad\mathcal{\overline{F}}_{\check{i}}{}^{\check{j}\check{k}}=Q_{\check{i}}{}^{\check{j}\check{k}},\quad\mathcal{\overline{F}}^{\check{i}\check{j}\check{k}}=R^{\check{i}\check{j}\check{k}}.\label{eq: Relations Generalized Fluxes - (Non-)Geometric Fluxes}
\end{equation}
Similarly, we define for the trace-terms and generalized dilaton fluxes
(cf. the first relation of (\ref{eq: Generalized NSNS-Fluxes})) 
\begin{equation}
\overline{\mathcal{F}}_{\check{i}}=2Y_{\check{i}}+F^{\check{m}}{}_{\check{m}\check{i}},\quad\overline{\mathcal{F}}^{\check{i}}=2Z^{\check{i}}+Q_{\check{m}}{}^{\check{m}\check{i}}.\label{eq:One-Index-Fluxes}
\end{equation}
As was discussed in \cite{Blumenhagen:2013hva}, writing out the generalized
metric $\mathcal{H}$ in terms of the internal metric and Kalb-Ramond
field gives rise to certain combinations of the latter with the fluxes,
for which it is convenient to use the shorthand notation 
\begin{equation}
\begin{array}{ccl}
\mathfrak{H}_{\check{i}\check{j}\check{k}} & {\displaystyle \!\!\!\!=\!\!\!\!} & H_{\check{i}\check{j}\check{k}}+3F^{\check{m}}{}_{[\underline{\check{i}\check{j}}}B_{\check{m}\underline{\check{k}}]}+3Q_{[\underline{\check{i}}}{}^{\check{m}\check{n}}B_{\check{m}\underline{\check{j}}}B_{\check{n}\underline{\check{k}}]}+R^{\check{m}\check{n}\check{p}}B_{\check{m}[\underline{\check{i}}}B_{\check{n}\underline{\check{j}}}B_{\check{p}\underline{\check{k}}]},\\
\mathfrak{F}^{\check{i}}{}_{\check{j}\check{k}} & {\displaystyle \!\!\!\!=\!\!\!\!} & F^{\check{i}}{}_{\check{j}\check{k}}+2Q_{[\underline{\check{j}}}{}^{\check{m}\check{i}}B_{\check{m}]\underline{\check{k}}}+R^{\check{m}\check{n}\check{i}}B_{\check{m}[\underline{\check{j}}}B_{\check{n}\underline{\check{k}}]},\vphantom{^{X^{X^{X^{X^{X^{X}}}}}}}\\
\mathfrak{Q}_{\check{k}}{}^{\check{i}\check{j}} & {\displaystyle \!\!\!\!=\!\!\!\!} & Q_{\check{k}}{}^{\check{i}\check{j}}+R^{\check{m}\check{i}\check{j}}B_{\check{m}\check{k}},\vphantom{^{X^{X^{X^{X^{X^{X}}}}}}}\\
\mathfrak{R}^{\check{i}\check{j}\check{k}} & {\displaystyle \!\!\!\!=\!\!\!\!} & R^{\check{i}\check{j}\check{k}},\vphantom{^{X^{X^{X^{X^{X^{X}}}}}}}\\
\mathfrak{Y}_{\check{i}} & {\displaystyle \!\!\!\!=\!\!\!\!} & Y_{\check{i}}+Z^{\check{m}}B_{\check{m}\check{i}},\vphantom{^{X^{X^{X^{X^{X^{X}}}}}}}\\
\mathfrak{Z}^{\check{i}} & {\displaystyle \!\!\!\!=\!\!\!\!} & Z^{\check{i}}.\vphantom{^{X^{X^{X^{X^{X^{X}}}}}}}
\end{array}\label{eq: Fraktur fluxes}
\end{equation}

\subsubsection{Operator Interpretation of Fluxes}

It will be useful to interpret the geometric and non-geometric fluxes
as operators acting on differential forms. Employing a local basis
$\left(dx^{1},\ldots dx^{6}\right)$ and the contractions $\left(\iota_{1},\ldots\iota_{6}\right)$
satisfying $\iota_{\check{i}}dx^{\check{j}}=\delta_{\check{i}}{}^{\check{j}}$,
we define \cite{Aldazabal:2006up,Villadoro:2006ia,Shelton:2006fd}\\
\begin{equation*}
\begin{array}{crcl}
H\wedge: & \Omega^{p}\left(CY_{3}\right) & \longrightarrow & \Omega^{p+3}\left(CY_{3}\right)\\
 & \omega_{p} & \mapsto & {\displaystyle \frac{1}{3!}H_{\check{i}\check{j}\check{k}}\; dx^{\check{i}}\wedge dx^{\check{j}}\wedge dx^{\check{k}}\wedge\omega_{p}},\\
\\
F\circ: & \Omega^{p}\left(CY_{3}\right) & \longrightarrow & \Omega^{p+1}\left(CY_{3}\right)\\
 & \omega_{p} & \mapsto & {\displaystyle \frac{1}{2!}F^{\check{k}}{}_{\check{i}\check{j}}\; dx^{\check{i}}\wedge dx^{\check{j}}\wedge\iota_{\check{k}}\wedge\omega_{p}},\\
 \\
Q\bullet: & \Omega^{p}\left(CY_{3}\right) & \longrightarrow & \Omega^{p-1}\left(CY_{3}\right)\\
 & \omega_{p} & \mapsto & {\displaystyle \frac{1}{2!}Q_{\check{i}}{}^{\check{j}\check{k}}\; dx^{\check{i}}\wedge\iota_{\check{j}}\wedge\iota_{\check{k}}\wedge\omega_{p}},
\end{array}
\end{equation*}
\begin{align}
\nonumber
\begin{array}{crcl}
R\llcorner: & \Omega^{p}\left(CY_{3}\right) & \longrightarrow & \Omega^{p-3}\left(KCY_{3}\right)\\
 & \omega_{p} & \mapsto & {\displaystyle \frac{1}{3!}R^{\check{i}\check{j}\check{k}}\;\iota_{\check{i}}\wedge\iota_{\check{j}}\wedge\iota_{\check{k}}\,\wedge\omega_{p}},\\
\\
Y\wedge: & \Omega^{p}\left(CY_{3}\right) & \longrightarrow & \Omega^{p+1}\left(CY_{3}\right)\\
 & \omega_{p} & \mapsto & {\displaystyle \vphantom{\frac{1}{3!}}{\displaystyle Y_{\check{i}}\; dx^{\check{i}}\wedge\omega_{p}}},\\
\\
Z\blacktriangledown: & \Omega^{p}\left(CY_{3}\right) & \longrightarrow & \Omega^{p-1}\left(CY_{3}\right)\\
 & \omega_{p} & \mapsto & {\displaystyle \vphantom{\frac{1}{3!}}{\displaystyle Z^{\check{i}}\;\iota_{\check{i}}\wedge\omega_{p}},}
\end{array}\label{eq: Flux Operators}
\\[-15pt]
\end{align}
the last two of which denote the newly introduced generalized dilaton
fluxes first considered in a non-DFT context in \cite{DallAgata:2009wsi,Derendinger:2007xp}.
These operators can be combined with the exterior derivative $\hat{\textrm{d}}$
to define the \emph{twisted differential} 
\begin{equation}
\hat{\mathcal{D}}=\hat{\textrm{d}}-H\wedge-F\circ-Q\bullet-R\llcorner-Y\wedge-Z\blacktriangledown.\label{eq: Twisted Differential}
\end{equation}
Notice that the exterior derivative is that of the full ten-dimensional
spacetime manifold. In the following, we will often distinguish between
internal and external components, for which it makes sense to split
the exterior derivative as 
\begin{equation}
\hat{\textrm{d}}=\textrm{d}+\textrm{d}_{CY_{3}}
\end{equation}
and define a purely internal twisted differential $\mathcal{D}$ with
respect to $\textrm{d}_{CY_{3}}$. For later convenience, we can furthermore
define analogous operators for the Fraktur fluxes (\ref{eq: Fraktur fluxes}),
including the Fraktur twisted differential $\hat{\mathfrak{D}}$.
As shown for a simplified setting in \cite{Blumenhagen:2015lta},
requiring nilpotency $\mathcal{\hat{D}}^{2}=0$ of the twisted differential
(and similarly for $\hat{\mathfrak{D}}$) gives rise to the Bianchi
identities

\begin{equation}
\begin{array}{rcl}
0 & {\displaystyle \!\!\!\!=\!\!\!\!} & {\displaystyle H_{\check{m}\left[\underline{\check{i}\check{j}}\right.}F^{\check{m}}{}_{\left.\underline{\check{k}\check{l}}\right]}-\frac{2}{3}\partial_{\left[\underline{\check{i}}\right.}H_{\left.\underline{\check{j}\check{k}\check{l}}\right]}},\\
0 & {\displaystyle \!\!\!\!=\!\!\!\!} & {\displaystyle F^{\check{m}}{}_{\left[\underline{\check{j}\check{k}}\right.}F^{\check{l}}{}_{\left.\underline{\check{i}}\right]\check{m}}+H_{\check{m}\left[\underline{\check{i}\check{j}}\right.}Q_{\left.\underline{\check{k}}\right]}{}^{\check{m}\check{l}}+\partial_{\left[\underline{\check{k}}\right.}F^{\check{l}}{}_{\left.\underline{\check{i}\check{j}}\right]}},\vphantom{^{X^{X^{X^{X^{X^{X}}}}}}}\\
0 & {\displaystyle \!\!\!\!=\!\!\!\!} & {\displaystyle F^{\check{m}}{}_{\left[\underline{\check{i}\check{j}}\right]}Q_{\check{m}}{}^{\left[\underline{\check{k}\check{l}}\right]}-4F^{\left[\underline{\check{k}}\right.}{}_{\check{m}\left[\underline{\check{i}}\right.}Q_{\left.\underline{\check{j}}\right]}{}^{\left.\underline{\check{l}}\right]\check{m}}+H_{\check{m}\check{i}\check{j}}R^{\check{m}\check{k}l}-2\partial_{\left[\underline{\check{i}}\right.}Q_{\left.\underline{\check{j}}\right]}{}^{\check{k}\check{l}}},\vphantom{^{X^{X^{X^{X^{X^{X}}}}}}}\\
0 & {\displaystyle \!\!\!\!=\!\!\!\!} & {\displaystyle Q_{\check{m}}{}^{\left[\underline{\check{j}\check{k}}\right.}Q_{\check{l}}{}^{\left.\underline{\check{i}}\right]\check{m}}+R^{\check{m}\left[\underline{\check{i}\check{j}}\right.}F^{\left.\underline{\check{k}}\right]}{}_{\check{m}\check{l}}-\frac{1}{3}\partial_{\check{l}}R^{\check{i}\check{j}\check{k}}},\vphantom{^{X^{X^{X^{X^{X^{X}}}}}}}\\
\\
0 & {\displaystyle \!\!\!\!=\!\!\!\!} & {\displaystyle R^{\check{m}\left[\underline{\check{i}\check{j}}\right.}Q_{\check{m}}{}^{\left.\underline{\check{k}\check{l}}\right]}},\vphantom{^{X^{X^{X^{X^{X^{X}}}}}}}\\
0 & {\displaystyle \!\!\!\!=\!\!\!\!} & {\displaystyle R^{\check{m}\check{n}\left[\underline{\check{i}}\right.}F^{\left.\underline{\check{j}}\right]}{}_{\check{m}\check{n}}}-R^{\check{m}[\underline{\check{i}\check{j}}]}Y_{\check{m}}-Z^{\check{m}}Q_{\check{m}}{}^{[\underline{\check{i}\check{j}}]},\vphantom{^{X^{X^{X^{X^{X^{X}}}}}}}\\
0 & {\displaystyle \!\!\!\!=\!\!\!\!} & {\displaystyle R^{\check{i}\check{m}\check{n}}H_{\check{j}\check{m}\check{n}}-F^{\check{i}}{}_{\check{m}\check{n}}Q_{\check{j}}\,^{\check{m}\check{n}}}-2Q_{\check{j}}{}^{\check{m}\check{i}}Y_{\check{m}}+2Z^{\check{m}}F^{\check{i}}{}_{\check{m}\check{j}}-2\partial_{\check{j}}Z^{\check{i}},\vphantom{^{X^{X^{X^{X^{X^{X}}}}}}}\\
0 & {\displaystyle \!\!\!\!=\!\!\!\!} & {\displaystyle Q_{\left[\underline{\check{i}}\right.}{}^{\check{m}\check{n}}H_{\left.\underline{\check{j}}\right]\check{m}\check{n}}-F^{\check{m}}{}_{[\underline{\check{i}\check{j}}]}Y_{\check{m}}}-Z^{\check{m}}H_{\check{m}[\underline{\check{i}\check{j}}]}+2\partial_{[\underline{\check{i}}}Y_{\check{j}]},\vphantom{^{X^{X^{X^{X^{X^{X}}}}}}}\\
\\
0 & {\displaystyle \!\!\!\!=\!\!\!\!} & {\displaystyle 6R^{\check{m}\check{n}\check{p}}H_{\check{m}\check{n}\check{p}}+Z^{\check{m}}Y_{\check{m}}},\vphantom{^{X^{X^{X^{X^{X^{X}}}}}}}
\end{array}\label{eq:Bianchi-Identities-NSNS}
\end{equation}
where the derivative terms vanish in the setting discussed in this
paper and were included only for the sake of completeness. This form
of the Bianchi identities generalizes the result of \cite{Blumenhagen:2015lta}
and matches with the relations presented earlier in \cite{Geissbuhler:2013uka}
when taking into account the definitions (\ref{eq:One-Index-Fluxes})
and assuming independence of the dual coordinates.

Another central role will be played by the generalized primitivity
constraints 
\begin{equation}
H_{\check{i}\check{a}\check{\bar{a}}}g^{\check{a}\check{\bar{a}}}=0,\qquad F^{\check{i}}{}_{\check{a}\check{\bar{a}}}g^{\check{a}\check{\bar{a}}}=0,\qquad Q_{\check{i}}{}^{\check{a}\check{\bar{a}}}g_{\check{a}\check{\bar{a}}}=0,\qquad R^{\check{i}\check{a}\check{\bar{a}}}g_{\check{a}\check{\bar{a}}}=0,\label{eq:Generalized-Primitivity}
\end{equation}
which extend the corresponding condition for $H$ arising from supersymmetry
considerations in traditional approaches to flux compactifications.
Indeed, the first condition is equivalent to requiring the interior
product $H\lrcorner J$ of $H$ and the K\"ahler form $J$ to vanish.
Analogous formulations are possible for the remaining fluxes by taking
the interior product with $F\lrcorner$ to be with respect to the
subscript indices and defining analogous contraction-like operators
$Q\urcorner,R\urcorner$ for the superscript indices of the non-geometric
fluxes. The primitivity constraints can then be recast in the coordinate-independent
forms 
\begin{equation}
H\lrcorner J=0,\qquad F\lrcorner J=0,\qquad Q\urcorner J=0,\qquad R\urcorner J=0.\label{eq:Generalized-Primitivity-Operator}
\end{equation}
Notice that the interior product of non-geometric fluxes looks very
similar to the corresponding operators defined in (\ref{eq: Flux Operators}),
but contracts only as many indices as there are in the differential
form it acts on. This structure is motivated by that of the Hodge-star
operator (\ref{eq:Hodge-Star}), and the relations (\ref{eq:Generalized-Primitivity-Operator})
describe a generalization of the corresponding constraints used in
\cite{Blumenhagen:2015lta}. As we will see in the next section, this
slight relaxation is necessary in order to make the framework applicable
to more general settings of flux compactifications.

\subsubsection{Geometric Tools}

To conclude this section, let us briefly introduce the most essential
geometric objects which will become important in the following discussion.
A useful tool to handle the flux operators is the so-called the \emph{Mukai-pairing}\textit{
}of two differential forms $\eta$ and $\rho$. It is defined by 
\begin{equation}
\left\langle \eta,\rho\right\rangle =\left[\eta\wedge\lambda\left(\rho\right)\right]_{6},\label{eq:MukaiPairing}
\end{equation}
where $\left[\cdot\right]_{\textrm{6}}$ picks the six-form-component
and the involution $\lambda$ acts on an $n$-form $\rho$ as 
\begin{equation}
\lambda\left(\rho\right)=\left(-1\right)^{\left\lceil \frac{n}{2}\right\rceil }\rho.\label{eq:MukaiPairing-Involution}
\end{equation}
The operator $\left\lceil \cdot\right\rceil $ denotes the ceiling
function, giving as output the greatest integer that is less than
or equal to the argument. Furthermore, we denote the purely external
and internal components of Kalb-Ramond field $\hat{B}$ by 
\begin{equation}
B=\frac{1}{2!}B_{\mu\nu}\; dx^{\mu}\wedge dx^{\nu}\qquad\textrm{and}\qquad b=\frac{1}{2!}B_{\check{i}\check{j}}\; dx^{\check{i}}\wedge dx^{\check{j}},
\end{equation}
respectively, and define the $b$-twisted Hodge-star operator $\star_{b}$
by \cite{Jeschek:2004wy,Benmachiche:2006df,Cassani:2007pq}
\begin{equation}
\star_{b}\eta=e^{b}\wedge\star\lambda\left(e^{-b}\eta\right),\label{eq:b-twisted-hodge-star}
\end{equation}
which allows for a natural extension of the framework to the Fraktur
fluxes (\ref{eq: Fraktur fluxes}).

\section{The Scalar Potential on a Calabi-Yau Three-Fold\label{sec:3}}

We start our discussion by considering only the purely internal parts
of (\ref{eq: Full Action NSNS-sector}) and (\ref{eq: DFT: RR Action})
on a Calabi-Yau three-fold $CY_{3}$. A simplified version of the
type IIB setting was already discussed in \cite{Blumenhagen:2015lta},
and the following elaborations are to be considered as an extension
of this work. The aim of this section is to show that both the type
IIA and IIB case correctly give rise to the scalar potential of four-dimensional
$\mathcal{N}=2$ gauged supergravity. We furthermore illustrate how
the simultaneous presence of geometric and non-geometric fluxes allows
for preservation of $\textrm{IIA}\leftrightarrow\textrm{IIB}$ Mirror
Symmetry in DFT.

One important point to remark is that the original work \cite{Blumenhagen:2015lta}
builds upon the a priori assumptions of vanishing trace- and dilaton-terms
due to the lack of homological one-cycles in $CY_{3}$. We will relax
these assumptions here in order to keep the calculation as general
as possible and therefore allow a straightforward application to arbitrary
compactification manifolds. This in particular means that we will
take into account fluxes which cannot be supported on $CY_{3}$ as
well fields which become massive in four-dimensions for most of the
calculation and hold off setting them to zero until right before expanding
the action in terms of the cohomology bases. Besides revealing more
general structures underlying the framework, this is also done for
the sake of mathematical accuracy: While one can argue that proper
one-form fluxes such as $Y$ from (\ref{eq:One-Index-Fluxes}) cannot
exist on $CY_{3}$ due to the lack of homological one-cycles (and
similar for $H$ and $K3\times T^{2}$), the same argument cannot
be applied for expressions arising from $,F,Q,R$ or $Z$ as they
all involve dual indices. A natural generalization would be to extend
the argument to all expressions with effectively one or five free
indices (and particular combinations of holomorphic and antiholomorphic
indices), however, this would require a doubled geometry analogue
of the notions of differential geometric homology and cohomology.
To our knowledge, such a framework has not been worked out yet, and
we therefore try to go without cohomological arguments as long as
possible.

Since we do not have to distinguish between different components of
the internal manifold, we will drop the ``checks'' above internal
indices ($\check{I},\check{J},\ldots\rightarrow I,J,\ldots$) for
the rest of this section. We furthermore impose the strong constraint
on the underlying Calabi-Yau background and the field fluctuations,
assuming independence of the dual coordinates $\tilde{y}_{i}$. We
will, however, not do so for the fluxes and only apply the weaker
(quadratic) Bianchi identities (\ref{eq:Bianchi-Identities-NSNS}),
ensuring that the theory is capable of describing electric and magnetic
gaugings and does not merely reduce to ordinary type II supergravities.

\subsection{NS-NS Sector\label{sub:3.1}}

When substituting the expansions (\ref{eq: Generalized Fluxes Deviations})
into the purely internal terms of (\ref{eq: Full Action NSNS-sector}),
those terms involving only the objects $\overset{{\scriptscriptstyle \circ}}{\mathcal{F}}_{I}$
and $\overset{{\scriptscriptstyle \circ}}{\mathcal{F}}_{IJK}$ describe
the Calabi-Yau background and do not contribute to the scalar potential
since $\overset{{\scriptscriptstyle \circ}}{\mathcal{E}}^{A}{}_{I}$
satisfies the DFT equations of motion. Furthermore, mixings between
background values and fluctuations describe first order terms in the
expansion about the minimum of the scalar potential and can be neglected
as well. Considering the action up to second order in the deviations,
we are therefore left with 
\begin{equation}
\begin{array}{cll}
S_{\textrm{NS-NS, scalar}} & {\displaystyle \!\!=\!\!\!\!\!} & {\displaystyle \frac{1}{2}\int_{M^{10}}\textrm{d}^{4}x\textrm{d}^{12}Y\,\sqrt{g^{\left(4\right)}}e^{-2\phi}\left[{\displaystyle \mathcal{\overline{F}}_{IJK}\mathcal{\overline{F}}_{I'J'K'}\left(-\frac{1}{12}\mathcal{H}^{II'}\mathcal{H}^{JJ'}\mathcal{H}^{KK'}\right.}\right.}\\
 & \vphantom{^{X^{X^{X^{X^{X^{X}}}}}}} & {\displaystyle \left.\left.+\frac{1}{4}\mathcal{H}^{II'}\eta^{JJ'}\eta^{KK'}-\frac{1}{6}\eta^{II'}\eta^{JJ'}\eta^{KK'}\right)+\mathcal{\overline{F}}_{I}\mathcal{\overline{F}}_{I'}\left(\vphantom{\frac{1}{4}S^{AA'}}\eta^{II'}-\mathcal{H}^{II'}\right)\right].}
\end{array}
\end{equation}
Inserting the relations (\ref{eq: Relations Generalized Fluxes - (Non-)Geometric Fluxes})
and (\ref{eq:One-Index-Fluxes}), this can be rewritten in terms of
the geometric and non-geometric fluxes as
\begin{equation}
\begin{array}{cll}
S_{\textrm{NS-NS, scalar}} & {\displaystyle \!\!=\!\!\!\!\!} & {\displaystyle \frac{1}{2}\int_{M^{10}}\textrm{d}^{4}x\textrm{d}^{12}Y\,\sqrt{g^{\left(4\right)}}e^{-2\phi}\left[\vphantom{\frac{1}{4}}\right.}\\
 &  & -{\displaystyle \frac{1}{12}\left(\mathfrak{H}_{ijk}\mathfrak{H}_{i'j'k'}g^{ii'}g^{jj'}g^{kk'}+3\mathfrak{F}^{i}{}_{jk}\mathfrak{F}^{i'}{}_{j'k'}g_{ii'}g^{jj'}g^{kk'}\right.}\\
 & \vphantom{^{X^{X^{X^{X^{X^{X}}}}}}} & {\displaystyle \left.+3\mathfrak{Q}_{i}{}^{jk}\mathfrak{Q}_{i'}{}^{j'k'}g^{ii'}g_{jj'}g_{kk'}+\mathfrak{R}^{ijk}\mathfrak{R}^{i'j'k'}g_{ii'}g_{jj'}g_{kk'}\right)}\\
 & \vphantom{^{X^{X^{X^{X^{X^{X}}}}}}} & {\displaystyle -\frac{1}{2}\left(\mathfrak{F}^{m}{}_{ni}\mathfrak{F}^{n}{}_{mi'}g^{ii'}+\mathfrak{Q}_{m}{}^{ni}\mathfrak{Q}_{n}{}^{mi'}g_{ii'}{\displaystyle -\mathfrak{H}_{mni}\mathfrak{Q}_{i'}{}^{mn}g^{ii'}-\mathfrak{F}^{i}{}_{mn}\mathfrak{R}^{mni'}g_{ii'}}\right)}\\
 & \vphantom{^{X^{X^{X^{X^{X^{X}}}}}}} & {\displaystyle \vphantom{\frac{1}{4}}-\left(\vphantom{\mathfrak{Q}_{m'}{}^{m'i'}}\mathfrak{F}^{m}{}_{mi}+2\mathfrak{Y}_{i}\right)\left(\vphantom{\mathfrak{Q}_{m'}{}^{m'i'}}\mathfrak{F}^{m'}{}_{m'i'}+2\mathfrak{Y}_{i'}\right)g^{ii'}}\\
 & \vphantom{^{X^{X^{X^{X^{X^{X}}}}}}} & {\displaystyle \left.\vphantom{\frac{1}{4}}-\left(\vphantom{\mathfrak{Q}_{m'}{}^{m'i'}}\mathfrak{Q}_{m}{}^{mi}+2\mathfrak{Z}^{i}\right)\left(\mathfrak{Q}_{m'}{}^{m'i'}+2\mathfrak{Z}^{i'}\right)g_{ii'}\right]},
\end{array}\label{eq:NS-NS-Action-WrittenOut}
\end{equation}
where the topological terms involving only the $O\left(6,6,\mathbb{R}\right)$
invariant structure $\eta^{II'}$  cancel by the Bianchi identities
(\ref{eq:Bianchi-Identities-NSNS}). Now a key issue of this action
is that the (generally unknown) metric $g_{ij}$ of $CY_{3}$ appears
explicitly. In traditional Calabi-Yau compactifications, this can
be remedied by applying differential form notation and expanding the
fields in terms of the cohomology bases. While this framework is not
readily applicable to the setting of this paper, we can resolve this
problem by employing the operator interpretation (\ref{eq: Flux Operators})
in order to build a bridge to the special geometry of the Calabi-Yau
moduli spaces.

\subsubsection{Single Flux Settings}

As already demonstrated in \cite{Blumenhagen:2015lta}, it is convenient
to first assume vanishing internal $B$-field components and consider
only one flux turned on at a time. It can then easily be shown that
the constructed reformulation is still applicable in more general
settings.

\subsubsection*{Pure $H$-Flux}

Due to its differential form nature, the discussion of the pure $H$
-flux setting is particularly simple and requires only the tools of
standard differential geometry. The corresponding Lagrangian of (\ref{eq:NS-NS-Action-WrittenOut})
takes the form 
\begin{equation}
\mathcal{L}_{\textrm{NS-NS, scalar,}H}=\frac{e^{-2\phi}}{4}H_{ijk}H_{i'j'k'}g^{ii'}g^{jj'}g^{kk'}.
\end{equation}
It is obvious that this can be written as 
\begin{equation}
\mathcal{L}_{\textrm{NS-NS, scalar,}H}=-\frac{e^{-2\phi}}{2}H\wedge\star H,
\end{equation}
where we the three-form $H$ is related to the first operator of (\ref{eq: Flux Operators})
by formally defining $H:=H\wedge\mathbf{1}_{CY_{3}}$.

\subsubsection*{Pure $F$-Flux}

The NS-NS scalar potential Lagrangian in the pure $F$-flux scenario
reads
\begin{equation}
\mathcal{L}_{\textrm{NS-NS, scalar,}F}=-\frac{e^{-2\phi}}{4}\left(\vphantom{\frac{1}{2}}F^{i}{}_{jk}F^{i'}{}_{j'k'}g_{ii'}g^{jj'}g^{kk'}+2F^{m}{}_{ni}F^{n}{}_{mi'}g^{ii'}+4F^{m}{}_{mi}F^{m}{}_{mi'}g^{ii'}\right).\label{eq:NS-NS-Lagrangian-PureF}
\end{equation}
While the three-form interpretation of $H$ does not apply to $F$,
we can construct a similar object by letting the operator $F\circ$
act on the K\"ahler form $J$ of $CY_{3}$. We then obtain 
\begin{equation}
-\frac{1}{2}\left(\vphantom{\frac{1}{2}}F\circ J\right)\wedge\star\left(\vphantom{\frac{1}{2}}F\circ J\right)=\left[\frac{1}{4}F^{m}{}_{ij}F^{m'}{}_{i'j'}g_{mm'}g^{ii'}g^{jj'}-\frac{1}{2}F^{m}{}_{ij}F^{m'}{}_{i'j'}I^{j'}{}_{m}I^{j}{}_{m'}g^{ii'}\right]\star\mathbf{1}_{CY_{3}}\label{eq:NS-NS-Lagrangian-PureF-FJ}
\end{equation}
and find that only the first terms of (\ref{eq:NS-NS-Lagrangian-PureF})
and (\ref{eq:NS-NS-Lagrangian-PureF-FJ}) match, while the second
term 
\begin{equation}
\begin{array}{cl}
 & {\displaystyle -\frac{1}{2}F^{m}{}_{ij}F^{m'}{}_{i'j'}I^{j'}{}_{m}I^{j}{}_{m'}g^{ii'}}\\
= & {\displaystyle \left(\vphantom{\frac{1}{2}}F^{c}{}_{ab}F^{b}{}_{\bar{a}c}+F^{\bar{c}}{}_{a\bar{b}}F^{\bar{b}}{}_{\bar{a}\bar{c}}-F^{\bar{c}}{}_{ab}F^{b}{}_{\bar{a}\bar{c}}-F^{c}{}_{a\bar{b}}F^{\bar{b}}{}_{\bar{a}c}\right)g^{a\bar{a}}}\vphantom{^{X^{X^{X^{X^{X^{X}}}}}}}
\end{array}
\end{equation}
comes with reversed signs for the last two components. To see how
this can be compensated for, notice that appropriate contraction of
indices in the second Bianchi identity of (\ref{eq:Bianchi-Identities-NSNS})
yields (for vanishing $Q$-flux) the relation 
\begin{equation}
F^{k}{}_{a\bar{b}}F^{\bar{b}}{}_{\bar{a}k}+F^{k}{}_{\bar{b}\bar{a}}F^{\bar{b}}{}_{ak}+F^{k}{}_{\overline{a}a}F^{\bar{b}}{}_{\bar{b}k}=0.\label{eq:Bianchi-F}
\end{equation}
Multiplying this by $g^{a\bar{a}}$, we find after taking into account
the corresponding primitivity constraint of (\ref{eq:Generalized-Primitivity})
\begin{equation}
F^{c}{}_{a\bar{b}}F^{\bar{b}}{}_{\bar{a}c}g^{a\bar{a}}=F^{\bar{c}}{}_{ab}F^{b}{}_{\bar{a}\bar{c}}g^{a\bar{a}}
\end{equation}
Using this, adding the expression 
\eq{
\frac{1}{2}\left(\vphantom{\frac{1}{2}}\Omega\wedge F\circ J\right)\wedge\star\left(\vphantom{\frac{1}{2}}\overline{\Omega}\wedge F\circ J\right)=-2\left[F^{\bar{c}}{}_{ab}F^{c}{}_{\bar{a}\bar{b}}g_{c\bar{c}}g^{a\bar{a}}g^{b\bar{b}}-2F^{\bar{c}}{}_{ab}F^{b}{}_{\bar{a}\bar{c}}g^{a\bar{a}}\right]\star\mathbf{1}_{CY_{3}}
}
involving the holomorphic three-form $\Omega$ of $CY_{3}$ gives
the correct second term of (\ref{eq:NS-NS-Lagrangian-PureF-FJ}),
but also comes with an additional contribution that has to be canceled.
We once more resolve this by adding 
\begin{equation}
-\frac{1}{2}\left(\vphantom{\frac{1}{2}}F\circ\Omega\right)\wedge\star\left(\vphantom{\frac{1}{2}}F\circ\overline{\Omega}\right)=\left[2F^{\bar{c}}{}_{ab}F^{c}{}_{\bar{a}\bar{b}}g_{c\bar{c}}g^{a\bar{a}}g^{b\bar{b}}+\frac{1}{2}F^{m}{}_{mi}F^{m}{}_{mi'}g^{ii'}\right]\star\mathbf{1}_{CY_{3}}.
\end{equation}
Finally, the missing trace-term can be obtained by substituting the
primitivity constraint (cf. (\ref{eq:Generalized-Primitivity})) into
the only remaining non-trivial expression related the Calabi-Yau structure
forms, 
\begin{equation}
-\frac{1}{2}\left(F\circ\frac{1}{2}J^{2}\right)\wedge\star\left(F\circ\frac{1}{2}J^{2}\right)=\left[\frac{1}{2}F^{m}{}_{mi}F^{m}{}_{mi'}g^{ii'}\right]\star\mathbf{1}_{CY_{3}},
\end{equation}
and we find in total 
\begin{equation}
\begin{array}{ccl}
{\displaystyle \mathcal{L}_{\textrm{NS-NS, scalar,}F}} & {\displaystyle \!\!\!\!=\!\!\!\!} & {\displaystyle -\frac{e^{-2\phi}}{2}\left[\left(\vphantom{\frac{1}{2}}F\circ J\right)\wedge\star\left(\vphantom{\frac{1}{2}}F\circ J\right)+\left(F\circ\frac{1}{2}J^{2}\right)\wedge\star\left(F\circ\frac{1}{2}J^{2}\right)\right.}\\
\vphantom{^{X^{X^{X^{X^{X^{X}}}}}}} &  & {\displaystyle \hphantom{-\frac{e^{-2\phi}}{2}||}\left.+\left(\vphantom{\frac{1}{2}}F\circ\Omega\right)\wedge\star\left(\vphantom{\frac{1}{2}}F\circ\overline{\Omega}\right)-\left(\vphantom{\vphantom{\frac{1}{2}}}\Omega\wedge F\circ J\right)\wedge\star\left(\vphantom{\vphantom{\frac{1}{2}}}\overline{\Omega}\wedge F\circ J\right)\right]}.
\end{array}\label{eq:NS-NS-Lagrangian-PureF-Reformulated}
\end{equation}
Notice that this poses a slight generalization of the corresponding
expression found in \cite{Blumenhagen:2015lta} due to the presence
of additional trace-terms of $F$. In particular, the reformulation
only works when employing only the relaxed primitivity constraints
(\ref{eq:Generalized-Primitivity}), (\ref{eq:Generalized-Primitivity-Operator}).

\subsubsection*{Pure $Q$-Flux}

The analysis of the pure $Q$-flux setting follows a very similar
pattern as for the $F$-flux, and we will only sketch the basic idea
here. By proceeding completely analogously to the $F$-flux case,
one can show that the Lagrangian can be reformulated as 
\eq{
&\mathcal{L}_{\textrm{NS-NS, scalar,}Q}  =  {\displaystyle -\frac{e^{-2\phi}}{2}\left[\left(Q\bullet\frac{1}{2}J^{2}\right)\wedge\star\left(Q\bullet\frac{1}{2}J^{2}\right)+\left(Q\bullet\frac{1}{3!}J^{3}\right)\wedge\star\left(Q\bullet\frac{1}{3!}J^{3}\right)\right.}
\\
&\hspace{115pt}{\displaystyle \left.+\left(\vphantom{\frac{1}{2}}Q\bullet\Omega\right)\wedge\star\left(\vphantom{\frac{1}{2}}Q\bullet\overline{\Omega}\right)-\left(\vphantom{\vphantom{\frac{1}{2}}}\Omega\wedge Q\bullet\frac{1}{2}J^{2}\right)\wedge\star\left(\vphantom{\vphantom{\frac{1}{2}}}\overline{\Omega}\wedge Q\bullet\frac{1}{2}J^{2}\right)\right]},
\label{eq:NS-NS-Lagrangian-PureQ}
}
where the only nontrivial step is to take into account the relation
\begin{equation}
Q_{k}{}^{a\bar{b}}Q_{\bar{b}}{}^{\bar{a}k}+Q_{k}{}^{\bar{b}\bar{a}}Q_{\bar{b}}{}^{ak}+Q_{k}{}^{\bar{a}a}Q_{\bar{b}}{}^{\bar{b}k}=0\label{eq:Bianchi-Q}
\end{equation}
obtained by appropriately contracting the fourth Bianchi identity
of (\ref{eq:Bianchi-Identities-NSNS}), which can eventually be recast
in the form 
\begin{equation}
g_{a\bar{a}}Q_{\bar{b}}{}^{ac}Q_{c}{}^{\bar{a}\bar{b}}=g_{a\bar{a}}Q_{b}{}^{a\bar{c}}Q_{\bar{c}}{}^{\bar{a}b}
\end{equation}
and used to identify certain contributions arising from the first
and third term of (\ref{eq:NS-NS-Lagrangian-PureQ}). Again, the result
describes a slight generalization of the one found in \cite{Blumenhagen:2015lta},
and matching for the trace-terms requires one to use the relaxes primitivity
constraints (\ref{eq:Generalized-Primitivity}), (\ref{eq:Generalized-Primitivity-Operator}).

\subsubsection*{Pure $R$-Flux}

Similarly to the symmetry between the pure $F$- and $Q$-flux settings,
the reformulation of pure $R$-flux case shows a strong resemblance
of the pure $H$-flux setting, and it seems natural to consider the
term $R\llcorner\frac{1}{3!}J^{3}$. This expression can be handled
best by exploiting the relation 
\begin{equation}
\frac{1}{3!}J^{3}=\star\mathbf{1}^{(6)}=\frac{\sqrt{g}}{6!}\varepsilon_{i_{1}\ldots i_{6}}dx^{i_{1}}\wedge\ldots\wedge dx^{i_{6}},
\end{equation}
to show that 
\begin{equation}
R\llcorner\left(\frac{1}{3!}J^{3}\right)=-\frac{\sqrt{g}}{3\text{!3!}}R^{ijk}\varepsilon_{ijklmn}dx^{l}\wedge dx^{m}\wedge dx^{n}.
\end{equation}
Inserting the relation (\ref{eq:Levi-Civita-Delta-Relation}) for
$D=3$ and $p=3$, we then find 
\begin{equation}
\mathcal{L}_{\textrm{NS-NS, scalar,}R}=-\frac{e^{2\phi}}{2}\left(R\llcorner\frac{1}{3!}J^{3}\right)\wedge\star\left(R\llcorner\frac{1}{3!}J^{3}\right).
\end{equation}

\subsubsection*{Pure $Y$- and $Z$-Flux}

While the nature of the generalized dilaton fluxes $Y$ and $Z$ differs
from that of their (three-indexed) geometric and non-geometric counterparts,
including them into the framework presented here requires only minor
modifications. The idea is again to consider all possible combinations
of flux operators with the holomorphic three-form $\Omega$ or powers
of the K\"ahler-form $J$. Direct computation of the corresponding expressions
then shows that the Lagrangian (\ref{eq:NS-NS-Action-WrittenOut})
for the (combined) pure $Y$- and $Z$-flux settings can be rewritten
as 
\begin{equation}
\begin{array}{ccl}
\mathcal{L}_{\textrm{NS-NS, scalar,}Y} & {\displaystyle \!\!\!\!=\!\!\!\!} & {\displaystyle -\frac{e^{-2\phi}}{2}\left[\left(\vphantom{\frac{1}{2}}Y\wedge\mathbf{1}_{CY_{3}}\right)\wedge\star\left(\vphantom{\frac{1}{2}}Y\wedge\mathbf{1}_{CY_{3}}\right)+\left(\vphantom{\frac{1}{2}}Y\wedge J\right)\wedge\star\left(\vphantom{\frac{1}{2}}Y\wedge J\right)\right.}\\
\vphantom{^{X^{X^{X^{X^{X^{X}}}}}}} &  & {\displaystyle \hphantom{-\frac{e^{-2\phi}}{2}||}\left.+\left(\vphantom{\frac{1}{2}}Y\wedge\frac{1}{2}J^{2}\right)\wedge\star\left(\vphantom{\frac{1}{2}}Y\wedge\frac{1}{2}J^{2}\right)+\left(\vphantom{\vphantom{\frac{1}{2}}}Y\wedge\Omega\right)\wedge\star\left(\vphantom{\vphantom{\frac{1}{2}}}Y\wedge\overline{\Omega}\right)\right]}
\end{array}
\end{equation}
and 
\begin{equation}
\begin{array}{ccl}
\mathcal{L}_{\textrm{NS-NS, scalar,}Z} & {\displaystyle \!\!\!\!=\!\!\!\!} & {\displaystyle -\frac{e^{-2\phi}}{2}\left[\left(\vphantom{\frac{1}{2}}Z\blacktriangledown J\right)\wedge\star\left(\vphantom{\frac{1}{2}}Z\blacktriangledown J\right)+\left(\vphantom{\frac{1}{2}}Z\blacktriangledown\frac{1}{2}J^{2}\right)\wedge\star\left(\vphantom{\frac{1}{2}}Z\blacktriangledown\frac{1}{2}J^{2}\right)\right.}\\
\vphantom{^{X^{X^{X^{X^{X^{X}}}}}}} &  & {\displaystyle \hphantom{-\frac{e^{-2\phi}}{2}||}\left.+\left(\vphantom{\frac{1}{2}}Z\blacktriangledown\star\mathbf{1}_{CY_{3}}\right)\wedge\star\left(\vphantom{\frac{1}{2}}Z\blacktriangledown\mathbf{1}_{CY_{3}}\right)+\left(\vphantom{\vphantom{\frac{1}{2}}}Y\wedge\Omega\right)\wedge\star\left(\vphantom{\vphantom{\frac{1}{2}}}Y\wedge\overline{\Omega}\right)\right]},
\end{array}
\end{equation}
respectively. Notice that, although there do exist corresponding non-trivial
expressions, we did not include any mixings between $J$ and $\Omega$.
The reason for this discrepancy will become clear when considering
more general settings in the next subsection.

\subsubsection{Generalization}

\subsubsection*{$H$-,$F$-,$Q$- and $R$-Fluxes}

Before turning to the most general setting, it makes sense to first
consider the case of all three-indexed fluxes $H,F,Q,R$ being present
and vanishing one-indexed fluxes $Y$ and $Z$. It was shown in \cite{Blumenhagen:2015lta}
that the Lagrangian (\ref{eq:NS-NS-Action-WrittenOut}) can then be
written as 
\eq{
&\star\mathcal{L}_{\textrm{NS-NS, scalar, \ensuremath{HFQR}}} = {\displaystyle -e^{-2\phi}\left[\frac{1}{2}\chi\wedge\star\overline{\chi}+\frac{1}{2}\Psi\wedge\star\overline{\Psi}\right.}
\\
&\hspace{110pt} {\displaystyle \hphantom{-e^{-2\phi}||}\left.-\frac{1}{4}\left(\vphantom{\overline{\Omega}}\Omega\wedge\chi\right)\wedge\star\left(\overline{\Omega}\wedge\overline{\chi}\right)-\frac{1}{4}\left(\vphantom{\overline{\Omega}}\Omega\wedge\overline{\chi}\right)\wedge\star\left(\overline{\Omega}\wedge\chi\right)\right]},
\label{eq:NS-NS-Scalar-Potential-General-Formulation}
}
where 
\begin{equation}
\chi=\mathcal{D}e^{iJ},\qquad\Psi=\mathcal{D}\Omega\label{eq:eq:NS-NS-Scalar-Potential-General-Formulation-Definitions}
\end{equation}
and the twisted differential $\mathcal{D}$ defined in (\ref{eq: Twisted Differential})
(with vanishing $Y$- and $Z$-components). Taking into account the
generalized primitivity constraints (\ref{eq:Generalized-Primitivity}),
it is easy to check that this formula correctly reproduces the single
flux settings. Concerning the mixings between different fluxes, a
minimal requirement for matching with the original Lagrangian (\ref{eq:NS-NS-Action-WrittenOut})
is that all mixings between different fluxes except for the $HQ$-
and $FR$-combinations vanish. Since the only nontrivial contributions
of (\ref{eq:NS-NS-Scalar-Potential-General-Formulation}) to the integral
over $CY_{3}$ are the ones proportional to its volume form $\star\mathbf{1}_{CY_{3}}$,
the relevant combinations of differential forms to check are those
where both constituents share the same degree. This in particular
excludes all components of the poly-form $\Psi$. Furthermore, those
terms arising from quadratic combinations of $\chi$ involving precisely
one even and one odd power of $iJ$ cancel due to the complex conjugation
operator reversing the signs only for imaginary differential forms.
A simple computation shows that the remaining terms of (\ref{eq:NS-NS-Scalar-Potential-General-Formulation})
are the desired $HQ$- and $FR$-combinations, which read 
\begin{equation}
\begin{array}{ccl}
T_{HQ} & {\displaystyle \!\!\!\!=\!\!\!\!} & {\displaystyle -H\wedge\star\left(Q\bullet\frac{1}{2}J^{2}\right)+\textrm{Re}\left(\vphantom{\frac{1}{2}}\Omega\wedge H\right)\wedge\star\left(\overline{\Omega}\wedge Q\bullet\frac{1}{2}J^{2}\right)},\\
\vphantom{^{X^{X^{X^{X^{X^{X}}}}}}}T_{FR} & {\displaystyle \!\!\!\!=\!\!\!\!} & {\displaystyle -F\circ J\wedge\star\left(R\llcorner\frac{1}{3!}J^{3}\right)+\textrm{Re}\left(\vphantom{\frac{1}{2}}\Omega\wedge F\circ J\right)\wedge\star\left(\overline{\Omega}\wedge R\llcorner\frac{1}{3!}J^{3}\right).}
\end{array}
\end{equation}
To show that these correctly reproduce the mixing terms of (\ref{eq:NS-NS-Action-WrittenOut}),
one can again follow a similar pattern as in the single flux settings,
and we refer the reader to the original work \cite{Blumenhagen:2015lta}
for detailed calculations. The most important step here is to once
more make use of the second and fourth Bianchi identities of (\ref{eq:Bianchi-Identities-NSNS})
in order to relate the above expressions to the original action, which
will in particular offset additional contributions arising from modifications
of the relations (\ref{eq:Bianchi-F}) and (\ref{eq:Bianchi-Q}) we
used in the pure $F$- and $Q$-flux settings.

\subsubsection*{Including the $Y$- and $Z$-Fluxes}

When trying to incorporate the generalized dilaton fluxes $Y$ and
$Z$ into the framework, one immediate problem is that the relation
(\ref{eq:NS-NS-Scalar-Potential-General-Formulation}) does not even
hold for the single flux settings. This is due to the appearance of
additional mixings between $e^{iJ}$ and $\Omega$ arising from the
expressions in the second line, which cancel half of the desired terms
and leave an overall mismatch by a factor of $\frac{1}{2}$. We resolve
this by slightly modifying the expression in such a way that only
the $Y$- and $Z$- terms are affected: Using the Mukai-pairing defined
in (\ref{eq:MukaiPairing}), we find the more general Lagrangian
\begin{equation}
\mathcal{L}_{\textrm{NS-NS, scalar}}={\displaystyle -e^{-2\phi}\left[\frac{1}{2}\left\Vert \left\langle \vphantom{\overline{\Omega}}\chi,\star\overline{\chi}\right\rangle \right\Vert +\frac{1}{2}\left\Vert \left\langle \vphantom{\overline{\Omega}}\Psi,\star\overline{\Psi}\right\rangle \right\Vert -\frac{1}{4}\left\Vert \left\langle \vphantom{\overline{\Omega}}\chi,\Omega\right\rangle \right\Vert ^{2}-\frac{1}{4}\left\Vert \left\langle \vphantom{\overline{\Omega}}\chi,\overline{\Omega}\right\rangle \right\Vert ^{2}\right]},\label{eq:NS-NS-Scalar-Potential-InnerProduct-Formulation}
\end{equation}
where the norm $\left\Vert \cdot\right\Vert $ is with respect to
the scalar product (\ref{eq:Scalar-Product-Differential-Forms}) and
$\chi$ and $\Psi$ are defined as in (\ref{eq:eq:NS-NS-Scalar-Potential-General-Formulation-Definitions}),
the twisted differential taking its general form (\ref{eq: Twisted Differential}).
It is easy to check by direct computation and use of the primitivity
constraints (\ref{eq:Generalized-Primitivity}) that (\ref{eq:NS-NS-Scalar-Potential-InnerProduct-Formulation})
reduces to the previously described special cases when setting the
corresponding subsets of fluxes to zero. Of the newly appearing mixing
terms, the non-vanishing ones are precisely the $FY$- and $QZ$-combinations,
which correctly give rise to the trace-dilaton-mixings found in the
last two lines of (\ref{eq:NS-NS-Action-WrittenOut}).

Notice that this formulation of the scalar potential shows a stronger
resemblance of its generalized geometry counterpart found in \cite{Cassani:2008rb}
for compactifications of type II supergravities on manifolds with
general $SU(3)\!\times\! SU(3)$ structures.

\subsubsection{Including the Kalb-Ramond Field}

In a final step, the above results are once more generalized to the
setting of a non-vanishing internal Kalb-Ramond field $b$. As can
be inferred from the structure of the Lagrangian (\ref{eq:NS-NS-Action-WrittenOut}),
this can be achieved by simply replacing 
\begin{equation}
H\rightarrow\mathfrak{H},\quad F\rightarrow\mathfrak{F},\quad Q\rightarrow\mathfrak{Q},\quad R\rightarrow\mathfrak{R},\quad Y\rightarrow\mathfrak{Y},\quad Z\rightarrow\mathfrak{Z}
\end{equation}
 and, thus, for the twisted differential 
\begin{equation}
\mathcal{D}\rightarrow\mathfrak{D}=\mbox{d}-\mathfrak{H}\wedge-\mathfrak{F}\circ-\mathfrak{Q}\bullet-\mathfrak{R\llcorner}-\mathfrak{Y}\wedge-\mathfrak{Z}\blacktriangledown.
\end{equation}
Mathematically, the K\"ahler and complex structures of Calabi-Yau manifolds
with non-vanishing $b$-field are described by the modified poly-forms
\begin{equation}
e^{i\mathfrak{J}}\rightarrow e^{b+iJ},\qquad\Omega\rightarrow e^{b}\Omega.\label{eq:Modified-Calabi-Yau-Structure-Forms}
\end{equation}
At a later point, it will be convenient to absorb the factor $e^{b}$
into the twisted differential. We therefore consider the relation
\cite{Blumenhagen:2015lta} 
\begin{equation}
\mathfrak{D}=e^{-b}\mathcal{D}e^{b}-\frac{1}{2}\left(\mathfrak{Q}_{i}{}^{mn}B_{mn}dx^{i}+\mathfrak{R}^{imn}B_{mn}\iota_{i}\right),\label{eq:FrakturTwsited-Differential-incomplete}
\end{equation}
which can be derived by direct computation and using closure of
$b$. Imposing primitivity constraints analogous to (\ref{eq:Generalized-Primitivity})
for the Fraktur fluxes and the modified Calabi-Yau structure forms
(\ref{eq:Modified-Calabi-Yau-Structure-Forms}), 
\[
\mathfrak{Q}\urcorner\mathfrak{J}=0,\qquad\qquad\mathfrak{R}\urcorner\mathfrak{J}=0,
\]
we furthermore obtain the  relations 
\begin{equation}
\begin{array}{rcl}
Q_{i}{}^{mn}B_{mn}+iR^{mnp}B_{im}J_{np}+R^{mnp}B_{im}B_{np} & {\displaystyle \!\!\!\!=\!\!\!\!} & 0,\\
\vphantom{^{X^{X^{X^{X^{X^{X}}}}}}}R^{mnp}B_{np}+iR^{mnp}J_{np} & {\displaystyle \!\!\!\!=\!\!\!\!} & 0,
\end{array}
\end{equation}
showing that the terms in the brackets of (\ref{eq:FrakturTwsited-Differential-incomplete})
vanish and, in fact, 
\begin{equation}
\mathfrak{D}=e^{-b}\mathcal{D}e^{b}.\label{eq:FrakturTwisted-Differential-complete}
\end{equation}
We thus find for the NS-NS scalar potential in the most general case
\begin{equation}
\mathcal{L}_{\textrm{NS-NS, scalar}}={\displaystyle -e^{-2\phi}\left[\frac{1}{2}\left\Vert \left\langle \vphantom{\overline{\Omega}}\chi,\star\overline{\chi}\right\rangle \right\Vert +\frac{1}{2}\left\Vert \left\langle \vphantom{\overline{\Omega}}\Psi,\star\overline{\Psi}\right\rangle \right\Vert -\frac{1}{4}\left\Vert \left\langle \vphantom{\overline{\Omega}}\chi,\Omega\right\rangle \right\Vert ^{2}-\frac{1}{4}\left\Vert \left\langle \vphantom{\overline{\Omega}}\chi,\overline{\Omega}\right\rangle \right\Vert ^{2}\right]}\label{eq:NS-NS-Scalar-Potential-Complete}
\end{equation}
 with
\begin{equation}
\chi=e^{-b}\mathcal{D}e^{b+iJ},\qquad\Psi=e^{-b}\mathcal{D}\left(e^{b}\Omega\right).\label{eq:NS-NS-Scalar-Potential-Complete-Definitions}
\end{equation}

\subsection{R-R Sector\label{sub:3.2}}

Reformulating the scalar potential contribution of the R-R action
(\ref{eq: DFT: RR Action}) is much more straightforward as one encounters
only differential form terms. We will do this separately for the type
IIA and IIB cases.

\subsubsection{Type IIA Theory}

Starting from the purely internal component of (\ref{eq: DFT: RR Action})
and substituting the definitions (\ref{eq:R-R-Sector-G-Poly-Form})
and (\ref{eq:R-R-Sector-C-Poly-Form}), we find for the internal components
of the poly-form $\hat{\mathfrak{G}}^{{\scriptscriptstyle \left(\mathrm{IIA}\right)}}$
\begin{equation}
\begin{array}{ccl}
\mathfrak{G}_{0}^{{\scriptscriptstyle \left(\mathrm{IIA}\right)}} & {\displaystyle \!\!\!\!=\!\!\!\!} & {\displaystyle G_{0}-\mathfrak{Q}\bullet C_{1}-\mathfrak{R}\llcorner C_{3}-\mathfrak{Z}\blacktriangledown C_{1},\vphantom{\frac{1}{2}}}\\
\vphantom{^{X^{X^{X^{X^{X^{X}}}}}}}\mathfrak{G}_{2}^{{\scriptscriptstyle \left(\mathrm{IIA}\right)}} & {\displaystyle \!\!\!\!=\!\!\!\!} & {\displaystyle G_{2}-B\wedge G_{0}-\mathfrak{F}\circ C_{1}-\mathfrak{Q}\bullet C_{3}-\mathfrak{R}\llcorner C_{5}-\mathfrak{Y}\wedge C_{1}-\mathfrak{Z}\blacktriangledown C_{3},\vphantom{\frac{1}{2}}}\\
\vphantom{^{X^{X^{X^{X^{X^{X}}}}}}}\mathfrak{G}_{4}^{{\scriptscriptstyle \left(\mathrm{IIA}\right)}} & {\displaystyle \!\!\!\!=\!\!\!\!} & {\displaystyle G_{4}-B\wedge G_{2}+\frac{1}{2}B^{2}\wedge G_{0}}-\mathfrak{H}\wedge C_{1}-\mathfrak{F}\circ C_{3}-\mathfrak{Q}\bullet C_{5}{\displaystyle -\mathfrak{Y}\wedge C_{3}-\mathfrak{Z}\blacktriangledown C_{5}\vphantom{,\frac{1}{2}}}\\
\vphantom{^{X^{X^{X^{X^{X^{X}}}}}}}\mathfrak{G}_{6}^{{\scriptscriptstyle \left(\mathrm{IIA}\right)}} & {\displaystyle \!\!\!\!=\!\!\!\!} & {\displaystyle G_{6}-B\wedge G_{4}+\frac{1}{2}B^{2}\wedge G_{2}-\frac{1}{3!}B^{3}\wedge G_{0}}-\mathfrak{H}\wedge C_{3}-\mathfrak{F}\circ C_{5}-\mathfrak{Y}\wedge C_{5},
\end{array}
\end{equation}
immediately revealing that the Lagrangian takes the form 
\begin{equation}
\star\mathcal{L}_{\textrm{R-R}}^{{\scriptscriptstyle \left(\mathrm{IIA}\right)}}=-\frac{1}{2}\mathfrak{G}^{{\scriptscriptstyle \left(\mathrm{IIA}\right)}}\wedge\star\mathfrak{G}^{{\scriptscriptstyle \left(\mathrm{IIA}\right)}}.\label{eq:RR-IIA-ScalarPotential}
\end{equation}
Here, $\mathfrak{G}^{{\scriptscriptstyle \left(\mathrm{IIA}\right)}}$
denotes the purely internal part of $\hat{\mathfrak{G}}^{{\scriptscriptstyle \left(\mathrm{IIA}\right)}}$
given by 
\begin{equation}
\begin{array}{ccc}
\mathfrak{G}^{{\scriptscriptstyle \left(\mathrm{IIA}\right)}} & = & e^{-B}\mathcal{G}^{{\scriptscriptstyle \left(\mathrm{IIA}\right)}}+e^{-B}\mathcal{D}\left(e^{B}\mathcal{C}^{{\scriptscriptstyle \left(\mathrm{IIA}\right)}}\right),\end{array}\label{eq:RR-IIA-Poly-Form}
\end{equation}
with 
\begin{equation}
\begin{array}{lcl}
\mathcal{C}^{{\scriptscriptstyle \left(\mathrm{IIA}\right)}} & {\displaystyle \!\!\!\!=\!\!\!\!} & C_{1}+C_{3}+C_{5}+C_{7}+C_{9},\\
\mathcal{G}^{{\scriptscriptstyle \left(\mathrm{IIA}\right)}} & {\displaystyle \!\!\!\!=\!\!\!\!} & G_{0}+G_{2}+G_{4}+G_{6}\vphantom{^{X^{X^{X^{X^{X^{X}}}}}}}
\end{array}
\end{equation}
comprising the purely internal components of the $C_{2n+1}$-fields
(including those which become massive in the process of compactification)
and the background R-R fluxes $G_{2n}$. Notice that the former are
to be understood as fluctuations $\overline{C}_{2n+1}$, and one can
equivalently write (\ref{eq:RR-IIA-Poly-Form}) as $\mathfrak{G}^{{\scriptscriptstyle \left(\mathrm{IIA}\right)}}=G_{0}+e^{-B}\mathcal{D}\left[e^{B}\left( \overset{{\scriptscriptstyle \circ}}{\mathcal{C}}{}^{{\scriptscriptstyle \left(\mathrm{IIA}\right)}}+\overline{\mathcal{C}}{}^{{\scriptscriptstyle \left(\mathrm{IIA}\right)}}\right)\right]$.
The former formulation will, however, be more convenient since it
allows one to treat all R-R fluxes on equal footing and obtain the
same structure for the type IIA und IIB settings.

\subsubsection{Type IIB Theory}

The analysis of the type IIB setting is completely analogous to the
type IIA case, and one eventually arrives at 
\begin{equation}
\star\mathcal{L}_{\textrm{R-R}}^{{\scriptscriptstyle \left(\mathrm{IIB}\right)}}=-\frac{1}{2}\mathfrak{G}^{{\scriptscriptstyle \left(\mathrm{IIB}\right)}}\wedge\star\mathfrak{G}^{{\scriptscriptstyle \left(\mathrm{IIB}\right)}}\label{eq:RR-IIB-ScalarPotential}
\end{equation}
with 
\begin{equation}
\begin{array}{ccc}
\mathfrak{G}^{{\scriptscriptstyle \left(\mathrm{IIA}\right)}} & = & e^{-B}\mathcal{G}^{{\scriptscriptstyle \left(\mathrm{IIB}\right)}}+e^{-B}\mathcal{D}\left(e^{B}\mathcal{C}^{{\scriptscriptstyle \left(\mathrm{IIB}\right)}}\right)\end{array}\label{eq:RR-IIB-Poly-Form}
\end{equation}
and 
\begin{equation}
\begin{array}{lcl}
\mathcal{G}^{{\scriptscriptstyle \left(\mathrm{IIB}\right)}} & {\displaystyle \!\!\!\!=\!\!\!\!} & G_{1}+G_{3}+G_{5},\\
\mathcal{\hat{C}}^{{\scriptscriptstyle \left(\mathrm{IIB}\right)}} & {\displaystyle \!\!\!\!=\!\!\!\!} & \hat{C}_{0}+\hat{C}_{2}+\hat{C}_{4}+\hat{C}_{6}+\hat{C}_{8}.\vphantom{^{X^{X^{X^{X^{X^{X}}}}}}}
\end{array}
\end{equation}
Notice that the cohomologically trivial R-R fluxes $G_{1}$ and $G_{5}$
cannot be supported on $CY_{3}$ and were included only to keep the
structure as general as possible.

\subsection{Dimensional Reduction\label{sub:3.3}}

The reformulated scalar potential described in (\ref{eq:NS-NS-Scalar-Potential-Complete}),
(\ref{eq:RR-IIA-ScalarPotential}) and (\ref{eq:RR-IIB-ScalarPotential})
depends only on the K\"ahler form and the holomorphic three-form of
$CY_{3}$ and can be evaluated by utilizing the framework of special
geometry for the Calabi-Yau moduli spaces.

\subsubsection{Special Geometry of Calabi-Yau Three-Folds\label{sub:3.3.1}}

Since we are interested only in those fields which do not acquire
mass in the course of the compactification, we would like to follow
the standard procedure of Calabi-Yau compactifications and expand
the appearing fields in terms of the cohomology bases of $CY_{3}$.
In the setting discussed here, this additionally requires a way to
describe the action of the flux operators (\ref{eq: Flux Operators})
on the field expansions. We therefore start by reviewing the topological
properties of Calabi-Yau manifolds and proceed by constructing a framework
that incorporates the flux operators of DFT.

\subsubsection*{Even Cohomology }

The nontrivial even cohomology groups are precisely $H^{n,n}\left(CY_{3}\right)$
with $n=0,1,2,3$. We denote the corresponding bases by 
\begin{equation}
\begin{array}{ll}
\left\{ \vphantom{\frac{\sqrt{g_{CY_{3}}}}{\mathcal{K}}}\mathbf{1}^{\left(6\right)}\right\} \in H^{0,0}\left(CY_{3}\right),\\
\\
\left\{ \vphantom{\frac{\sqrt{g_{CY_{3}}}}{\mathcal{K}}}\omega_{\mathsf{i}}\right\} \in H^{1,1}\left(CY_{3}\right),\\
 & \qquad\textrm{with }\mathsf{i}=1,\ldots h^{1,1}\\
\left\{ \vphantom{\frac{\sqrt{g_{CY_{3}}}}{\mathcal{K}}}\widetilde{\omega}^{\mathsf{i}}\right\} \in H^{2,2}\left(CY_{3}\right),\\
\\
\left\{ \frac{\sqrt{g_{CY_{3}}}}{\mathcal{K}}\star\mathbf{1}^{\left(6\right)}\right\} \in H^{3,3}\left(CY_{3}\right),
\end{array}\label{eq: Cohomology Basis even}
\end{equation}
where $\mathcal{K}$ is the volume of $CY_{\text{3}}$. For later
convenience, it makes sense to set $\omega_{0}=\star\mathbf{1}^{\left(6\right)}$
and $\widetilde{\omega}^{0}=\mathbf{1}^{\left(6\right)}$, allowing
us to use the collective notation 
\begin{equation}
\begin{array}{lc}
\omega_{\mathsf{I}}=\left(\begin{array}{cc}
\vphantom{\widetilde{\omega}^{0}}\omega_{0}, & \omega_{\mathsf{i}}\end{array}\right),\\
 & \qquad\textrm{with }\mathsf{I}=0,\ldots h^{1,1}\\
\widetilde{\omega}^{\mathsf{I}}=\left(\begin{array}{cc}
\widetilde{\omega}^{0}, & \widetilde{\omega}^{\mathsf{i}}\end{array}\right).
\end{array}\label{eq: Collective Notation for Even Cohomology}
\end{equation}
This structure is motivated by the action of the involution operator
(\ref{eq:MukaiPairing-Involution}). We choose the two bases such
that the normalization condition 
\begin{equation}
\int_{CY_{3}}\omega_{\mathsf{I}}\wedge\widetilde{\omega}^{\mathsf{J}}=\delta_{\mathsf{I}}{}^{\mathsf{J}}
\end{equation}
holds. For the K\"ahler form $J$ of $CY_{3}$ and the Kalb-Ramond field
$\hat{B}$, we use the expansions 
\begin{equation}
J=v^{\mathsf{i}}\omega_{\mathsf{i}}\qquad\textrm{and}\qquad\hat{B}=B+b=B+b^{\mathsf{i}}\omega_{\mathsf{i}},\label{eq: Expansion K=0000E4hler Form and B-Field}
\end{equation}
where $B$ denotes the external component of $\hat{B}$ living in
$M^{1,4}$ and $b$ its internal counterpart. The internal expansion
coefficients $b^{\mathsf{i}}$ can be combined with $v^{\mathsf{i}}$
to define the complexified K\"ahler form 
\begin{equation}
\mathfrak{J}=\left(b^{\mathsf{i}}+iv^{\mathsf{i}}\right)\omega_{\mathsf{i}}=:t^{\mathsf{i}}\omega_{\mathsf{i}}.\label{eq: Expansion Complexified K=0000E4hler Form}
\end{equation}
We furthermore introduce the shorthand notation 
\begin{equation}
\begin{array}{cclcl}
\mathcal{K}_{\mathsf{ijk}} & {\displaystyle \!\!\!\!=\!\!\!\!} & {\displaystyle \int_{CY_{\text{3}}}\omega_{\mathsf{i}}\wedge\omega_{\mathsf{j}}\wedge\omega_{\mathsf{k}}},\\
\\
\mathcal{K}_{\mathsf{ij}} & {\displaystyle \!\!\!\!=\!\!\!\!} & {\displaystyle \int_{CY_{\text{3}}}\omega_{\mathsf{i}}\wedge\omega_{\mathsf{j}}\wedge J} & {\displaystyle \!\!\!\!=\!\!\!\!} & {\displaystyle \mathcal{K}_{\mathsf{ijk}}v^{\mathsf{k}}},\\
\\
\mathcal{K}_{\mathsf{i}} & {\displaystyle \!\!\!\!=\!\!\!\!} & {\displaystyle \int_{CY_{\text{3}}}\omega_{\mathsf{i}}\wedge J\wedge J} & {\displaystyle \!\!\!\!=\!\!\!\!} & {\displaystyle \mathcal{K}_{\mathsf{ijk}}v^{\mathsf{j}}v^{\mathsf{k}}},\\
\\
\mathcal{K} & {\displaystyle \!\!\!\!=\!\!\!\!} & {\displaystyle \frac{1}{3!}\int_{CY_{\text{3}}}J\wedge J\wedge J} & {\displaystyle \!\!\!\!=\!\!\!\!} & {\displaystyle \frac{1}{\text{6}}\mathcal{K}_{\mathsf{ijk}}v^{\mathsf{i}}v^{\mathsf{j}}v^{\mathsf{k}},}
\end{array}
\end{equation}
where the $\mathcal{K}_{\mathsf{ijk}}$, $\mathcal{K}_{\mathsf{ij}}$
and $\mathcal{K}_{\mathsf{i}}$ are called intersection numbers. Using
this, one can eventually expand the first poly-form of (\ref{eq:NS-NS-Scalar-Potential-Complete-Definitions})
in terms of the complexified K\"ahler class moduli 
\begin{equation}
e^{B+iJ}=e^{\mathfrak{J}}=\tilde{\omega}^{0}+t^{\mathsf{i}}\omega_{\mathsf{i}}+\frac{1}{2!}\left(\mathcal{K}_{\mathsf{ijk}}t^{\mathsf{i}}t^{\mathsf{j}}\right)\tilde{\omega}^{\mathsf{k}}+\frac{1}{3!}\left(\mathcal{K}_{\mathsf{ijk}}t^{\mathsf{i}}t^{\mathsf{j}}t^{\mathsf{k}}\right)\omega_{0},\label{eq: DFT: chi expansion-1}
\end{equation}
where all powers of order $\geq4$ vanish on $CY_{3}$.

\subsubsection*{Odd Cohomology}

The nontrivial odd cohomology groups are given by $H^{3,0}\left(CY_{3}\right)$,
$H^{2,1}\left(CY_{3}\right)$,$H^{1,2}\left(CY_{3}\right)$ and $H^{0,3}\left(CY_{3}\right)$.
For these we introduce the collective basis 
\begin{equation}
\left\{ \alpha_{\mathsf{A}},\beta^{\mathsf{A}}\right\} \in H^{3}\left(CY_{3}\right)\qquad\textrm{with }\mathsf{A}=0,\ldots h^{1,2},\label{eq: Cohomology Basis odd}
\end{equation}
which can be normalized to satisfy 
\begin{equation}
{\displaystyle \int_{CY_{3}}\alpha_{\mathsf{A}}\wedge\beta^{\mathsf{B}}}=\delta_{\mathsf{A}}{}^{\mathsf{B}}.
\end{equation}
The complex structure moduli are encoded by the holomorphic three-form
$\Omega$ of $CY_{3}$, which we expand in terms of the periods $X^{\mathsf{A}}$
and $F_{\mathsf{A}}$ as 
\begin{equation}
\Omega=X^{\mathsf{A}}\alpha_{\mathsf{A}}-F_{\mathsf{A}}\beta^{\mathsf{A}}.\label{eq: Holomorphic Three-Form Expansion}
\end{equation}
Notice that there is a minus sign in front of the $\beta^{\mathsf{A}}$.
Throughout this paper, we will apply this convention to all odd cohomology
expansions of fields, while the signs are exchanged for field strengths.
The periods $F_{\mathsf{A}}$ are functions of $X^{\mathsf{A}}$ and
can be determined from a holomorphic prepotential $F$ by $F_{\mathsf{A}}=\frac{\partial F}{\partial X^{\mathsf{A}}}$.
Defining $F_{\mathsf{AB}}=\frac{\partial F_{\mathsf{A}}}{\partial X^{\mathsf{B}}}$,
one can write the period matrix $\mathcal{M}_{\mathsf{AB}}$ as 
\begin{equation}
\mathcal{M}_{\mathsf{AB}}=F_{\mathsf{AB}}+2i\frac{\textrm{Im}\left(F_{\mathsf{AC}}\right)X^{\mathsf{C}}\textrm{Im}\left(F_{\mathsf{BD}}\right)X^{\mathsf{D}}}{X^{\mathsf{E}}\textrm{Im}\left(F_{\mathsf{EF}}\right)X^{\mathsf{F}}},
\end{equation}
which is related to the cohomology bases (\ref{eq: Cohomology Basis odd})
by
\begin{equation}
\begin{array}{l}
{\displaystyle \int_{CY_{\text{3}}}\alpha_{\mathsf{A}}\wedge\star\alpha_{\mathsf{B}}=-\left[\left(\textrm{Im}\mathcal{M}\right)+\left(\textrm{Re}\mathcal{M}\right)\left(\textrm{Im}\mathcal{M}\right)^{-1}\left(\textrm{Re}\mathcal{M}\right)\right]_{\mathsf{AB}}},\\
\\
{\displaystyle {\displaystyle \int_{CY_{\text{3}}}\alpha_{\mathsf{A}}\wedge\star\beta^{\mathsf{B}}=-\left[\left(\textrm{Re}\mathcal{M}\right)\left(\textrm{Im}\mathcal{M}\right)^{-1}\right]_{\mathsf{A}}{}^{\mathsf{B}}},}\\
\\
{\displaystyle \int_{CY_{\text{3}}}\beta^{\mathsf{A}}\wedge\star\beta^{\mathsf{B}}=-\left[\textrm{Im}\mathcal{M}^{-1}\right]^{A\mathsf{B}}.}
\end{array}\label{eq: Gauge Coupling Matrix M (odd)}
\end{equation}

\subsubsection*{Gauge Coupling Matrices}

Denoting some arbitrary poly-form field $A$ which can be expanded
in terms of the nontrivial cohomology bases of $CY_{3}$ by 
\begin{equation}
A=A^{\mathsf{I}}\omega_{\mathsf{I}}+A_{\mathsf{I}}\tilde{\omega}^{\mathsf{I}}+A^{\mathsf{A}}\alpha_{\mathsf{A}}-A_{\mathsf{A}}\beta^{\mathsf{A}},
\end{equation}
one can define a collective notation by 
\begin{equation}
A^{\mathbb{I}}=\left(\begin{array}{cc}
A^{\mathsf{I}}, & A_{\mathsf{I}}\end{array}\right)^{T}\qquad\textrm{and}\qquad A^{\mathbb{A}}=\left(\begin{array}{cc}
A^{\mathsf{A}}, & -A_{\mathsf{A}}\end{array}\right)^{T}.\label{eq: Even-More-Collective Notation f=0000FCr Moduli Spaces}
\end{equation}
Again, notice that we will use reversed signs for the third cohomology
group in case of field strengths. Similarly, we define the collective
cohomology bases 
\begin{equation}
\Sigma_{\mathbb{I}}=\left(\begin{array}{cc}
\omega_{\mathsf{I}}, & \tilde{\omega}^{\mathsf{I}}\end{array}\right)\qquad\textrm{and}\qquad\Xi_{\mathbb{A}}=\left(\begin{array}{cc}
\alpha_{\mathsf{A}}, & \beta^{\mathsf{A}}\end{array}\right)\label{eq: Cohomology Bases Collective Notation}
\end{equation}
and the matrix 
\begin{equation}
\mathbb{M_{AB}}{\displaystyle =}{\displaystyle \int_{CY_{3}}\left(\begin{array}{cc}
\,\,-\left\langle \vphantom{\beta^{\mathsf{B}}}\alpha_{\mathsf{A}},\star_{b}\alpha_{\mathsf{B}}\right\rangle \, & \,\left\langle \alpha_{\mathsf{A}},\star_{b}\beta^{\mathsf{B}}\right\rangle \,\,\\
\vphantom{^{X^{X^{X^{X^{X^{X}}}}}}}\,\,\left\langle \beta^{\mathsf{A}},\star_{b}\alpha_{\mathsf{B}}\right\rangle \, & \,-\left\langle \beta^{\mathsf{A}},\star_{b}\beta^{\mathsf{B}}\right\rangle \,\,
\end{array}\right)},\label{eq:CollectiveMatrix-M}
\end{equation}
which can be expressed in terms of the period matrix (\ref{eq: Gauge Coupling Matrix M (odd)})
as
\begin{equation}
\mathbb{M}={\displaystyle \left(\begin{array}{cc}
\mathbbm{1} & -\textrm{Re}\mathcal{M}\\
0 & \mathbbm{1}
\end{array}\right)}{\displaystyle \left(\begin{array}{cc}
\textrm{Im}\mathcal{M} & 0\\
0 & \textrm{Im}\mathcal{M}^{-1}
\end{array}\right)}{\displaystyle \left(\begin{array}{cc}
\mathbbm{1} & 0\\
-\textrm{Re}\mathcal{M} & \mathbbm{1}
\end{array}\right)}.\label{eq: Gauge Coupling Matrices: M}
\end{equation}
For later convenience, we parametrize the even cohomology analogue
\begin{equation}
\mathbb{N_{IJ}}=\int_{CY_{3}}\left(\begin{array}{cc}
\,\,\left\langle \omega_{\mathsf{I}},\star_{b}\omega_{\mathsf{J}}\vphantom{\tilde{\omega}^{\mathsf{J}}}\right\rangle \, & \,\left\langle \omega_{\mathsf{I}},\star_{b}\tilde{\omega}^{\mathsf{J}}\right\rangle \,\,\\
\vphantom{^{X^{X^{X^{X^{X^{X}}}}}}}\,\,\left\langle \tilde{\omega}^{\mathsf{I}},\star_{b}\omega_{\mathsf{J}}\right\rangle \, & \,\left\langle \tilde{\omega}^{\mathsf{I}},\star_{b}\tilde{\omega}^{\mathsf{J}}\right\rangle \,\,
\end{array}\right)\label{eq:CollectiveMatrix-N}
\end{equation}
as 
\begin{equation}
\begin{array}{lll}
\mathbb{N} & = & {\displaystyle \left(\begin{array}{cc}
\mathbbm{1} & -\textrm{Re}\mathcal{N}\\
0 & \mathbbm{1}
\end{array}\right)}{\displaystyle \left(\begin{array}{cc}
\textrm{Im}\mathcal{N} & 0\\
0 & \textrm{Im}\mathcal{N}^{-1}
\end{array}\right)}{\displaystyle \left(\begin{array}{cc}
\mathbbm{1} & 0\\
-\textrm{Re}\mathcal{N} & \mathbbm{1}
\end{array}\right)},\end{array}\label{eq: Gauge Coupling Matrices: N}
\end{equation}
where $\mathcal{N}_{\mathsf{IJ}}$ denotes the corresponding period
matrix of the special K\"ahler manifold spanned by the complexified
K\"ahler class moduli. A detailed discussion of its structure can be
found in \cite{Gurrieri:2003st}.

Using the notation (\ref{eq: Collective Notation for Even Cohomology}),
one can also see that the Mukai-pairing (\ref{eq:MukaiPairing}) induces
a symplectic structure by 
\begin{equation}
\int_{CY_{3}}\left\langle \Sigma_{\mathbb{I}},\Sigma_{\mathbb{J}}\right\rangle =(S_{even})_{\mathbb{IJ}}=\left(\begin{array}{cc}
0 & \mathbbm{1}\\
-\mathbbm{1} & 0
\end{array}\right)\in Sp\left(2h^{1,1}+2,\mathbb{R}\right)
\end{equation}
and 
\begin{equation}
\int_{CY_{3}}\left\langle \Xi_{\mathbb{A}},\Xi_{\mathbb{B}}\right\rangle =(S_{odd})_{\mathbb{IJ}}=\left(\begin{array}{cc}
0 & \mathbbm{1}\\
-\mathbbm{1} & 0
\end{array}\right)\in Sp\left(2h^{1,2}+2,\mathbb{R}\right).
\end{equation}
For simplicity, we will omit the subscripts ``even'' and ``odd''
from now on. The dimension can, however, easily be inferred from the
context or read off from the indices when using component notation.

\subsubsection{Fluxes and Cohomology Bases}

In the previous subsections, we treated the fluxes as operators in
a local basis. We now want to find a way to express how they relate
to the cohomology basis elements (\ref{eq: Cohomology Basis even})
and (\ref{eq: Gauge Coupling Matrix M (odd)}). For the $H$-flux,
it is clear that one can write 
\begin{equation}
H=-\tilde{h}^{\mathsf{A}}\alpha_{\mathsf{A}}+h_{\mathsf{A}}\beta^{\mathsf{A}}
\end{equation}
since it acts as a wedge product with a three-form. While there is
no such obvious relation for the remaining fluxes, one can extract
useful structures by letting them act on the basis elements. Following
the idea of \cite{Grana:2006hr}, we define 
\begin{equation}
\begin{array}{l}
\mathcal{D}\alpha_{\mathsf{A}}=O_{\mathsf{A}}{}^{\mathsf{I}}\omega_{\mathsf{I}}+\mathcal{O}_{\mathsf{AI}}\tilde{\omega}^{\mathsf{I}},\hphantom{\hspace{0.15cm}-}\qquad\mathcal{D}\beta^{\mathsf{A}}=\tilde{P}^{\mathsf{AI}}\omega_{\mathsf{I}}+\tilde{P}^{\mathsf{A}}{}_{\mathsf{I}}\tilde{\omega}^{\mathsf{I}},\\
\mathcal{D}\omega_{\mathsf{I}}=-\tilde{P}^{\mathsf{A}}{}_{\mathsf{I}}\alpha_{\mathsf{A}}+O_{\mathsf{AI}}\beta^{\mathsf{A}},\qquad\mathcal{D}\tilde{\omega}^{\mathsf{I}}=\tilde{P}^{\mathsf{AI}}\alpha_{\mathsf{A}}-O_{\mathsf{A}}{}^{\mathsf{I}}\beta^{\mathsf{A}},\vphantom{^{X^{X^{X^{X^{X^{X}}}}}}}
\end{array}\label{eq: Fluxes - Cohomology Bases}
\end{equation}
where the components with $\mathsf{I}\neq0$ encode the contributions
of both the one- and three-indexed fluxes, e.g. by 
\eq{
\left(F\circ+Y\wedge\right)\omega_{\mathsf{i}}=\left(\tilde{f}^{\mathsf{A}}{}_{\mathsf{i}}+\tilde{y}^{\mathsf{A}}{}_{\mathsf{i}}\right)\alpha_{\mathsf{A}}-\left(\vphantom{\tilde{f}^{\mathsf{A}}{}_{\mathsf{i}}}f_{\mathsf{Ai}}+y_{\mathsf{Ai}}\right)\beta^{\mathsf{A}}=:-\tilde{P}^{\mathsf{A}}{}_{\mathsf{I}}\alpha_{\mathsf{A}}+O_{\mathsf{AI}}\beta^{\mathsf{A}},
}
and we used the collective notation (\ref{eq: Collective Notation for Even Cohomology})
to set 
\begin{equation}
\begin{array}{lcl}
O_{\mathsf{A}0}=r_{\mathsf{A}}, & \qquad & \tilde{P}^{\mathsf{A}}{}_{0}=\tilde{r}^{\mathsf{A}},\\
O_{\mathsf{A}}{}^{0}=h_{\mathsf{A}}, & \qquad & \tilde{P}^{\mathsf{A}0}=\tilde{h}^{\mathsf{A}}.\vphantom{^{X^{X^{X^{X^{X^{X}}}}}}}
\end{array}\label{eq: Fluxes Unified}
\end{equation}
Similarly to the previous sections, one can arrange the flux coefficients
in a collective notation that will greatly simplify calculations at
a later point. We define the matrices 
\begin{equation}
\mathcal{O}^{\mathbb{A}}{}_{\mathbb{I}}=\left(\begin{array}{cc}
-\tilde{P}^{\mathsf{A}}{}_{\mathsf{I}} & \tilde{P}^{\mathsf{AI}}\\
O_{\mathsf{AI}} & -O_{\mathsf{A}}{}^{\mathsf{I}}
\end{array}\right),\qquad\widetilde{\mathcal{O}}^{\mathbb{I}}{}_{\mathbb{A}}=\left(\begin{array}{cc}
(O^{T})^{\mathsf{I}}{}_{\mathsf{A}} & (\tilde{P}^{T}){}^{\mathsf{IA}}\\
(O^{T}){}_{\mathsf{IA}} & (\tilde{P}^{T}){}_{\mathsf{I}}{}^{\mathsf{A}}
\end{array}\right),\label{eq: Flux Matrix Definitions}
\end{equation}
such that the action of the twisted differential on the cohomology
bases can be expressed in the shorthand notation 
\begin{equation}
\mathcal{D}(\Sigma^{T})_{\mathbb{I}}=(\mathcal{O}^{T})_{\mathbb{I}}{}^{\mathbb{A}}(\Xi^{T})_{\mathbb{A}},\qquad\mathcal{D}(\Xi^{T})_{\mathbb{A}}=(\mathcal{\widetilde{\mathcal{O}}}^{T})_{\mathbb{A}}{}^{\mathbb{I}}(\Sigma^{T})_{\mathbb{I}}.\label{eq: Flux Matrix Axtions on Cohomology Bases - Components}
\end{equation}
They can be related by 
\begin{equation}
\widetilde{\mathcal{O}}=-S^{-1}\mathcal{O}^{T}S.\label{eq: Flux Matrix Symplectic Relation}
\end{equation}
Nilpotency of the twisted differential furthermore implies that the
relations 
\begin{equation}
\mathcal{D}^{2}(\Sigma^{T})_{\mathbb{I}}=0\qquad\textrm{and}\qquad\mathcal{D}^{2}(\Xi^{T})_{\mathbb{A}}=0
\end{equation}
have to be satisfied, giving rise to the constraints 
\begin{equation}
\widetilde{\mathcal{O}}^{\mathbb{I}}{}_{\mathbb{A}}\mathcal{O}^{\mathbb{A}}{}_{\mathbb{I}}=0,\qquad\mathcal{O}^{\mathbb{A}}{}_{\mathbb{I}}\widetilde{\mathcal{O}}^{\mathbb{I}}{}_{\mathbb{A}}=0,\label{eq: Nilpotency Constraint - Matrices}
\end{equation}
which take the role of a cohomology version of (\ref{eq:Bianchi-Identities-NSNS})
and will be important in (\ref{sec:5}).

\subsubsection{Integrating over the Internal Space - NS-NS Sector}

Proceeding in the same manner as for ordinary type II supergravity
theories, we now expand the fields of the scalar potential in the
cohomology bases (\ref{eq: Collective Notation for Even Cohomology})
and (\ref{eq: Cohomology Basis odd}) in order to filter out those
terms which become massive in four dimensions. For the NS-NS poly-forms,
we utilize the expansions (\ref{eq: DFT: chi expansion-1}) and (\ref{eq: Holomorphic Three-Form Expansion})
to arrange coefficients in vectors
\begin{equation}
\begin{array}{ccl}
V^{\mathbb{I}} & {\displaystyle \!\!\!\!=\!\!\!\!} & \left(\begin{array}{cccc}
{\displaystyle \frac{1}{3!}\mathcal{K}_{\mathsf{ijk}}t^{\mathsf{i}}t^{\mathsf{j}}t^{\mathsf{k}}},\: & t^{\mathsf{i}},\: & 1,\: & {\displaystyle \frac{1}{2!}\mathcal{K}_{\mathsf{ijk}}t^{\mathsf{i}}t^{\mathsf{j}}}\end{array}\right)^{T}\\
\vphantom{^{X^{X^{X^{X^{X^{X}}}}}}}W^{\mathbb{A}} & {\displaystyle \!\!\!\!=\!\!\!\!} & \left(\vphantom{\frac{1}{3!}}\begin{array}{cc}
X^{\mathsf{A}}, & -F_{\mathsf{A}}\end{array}\right)^{T}
\end{array}
\end{equation}
of dimension $\left(2h^{1,1}+2\right)$ and $\left(2h^{1,2}+2\right)$,
respectively, enabling us to use the shorthand notation
\begin{equation}
e^{B+iJ}=\Sigma{}_{\mathbb{I}}V^{\mathbb{I}},\qquad\Omega=\Xi{}_{\mathbb{A}}W^{\mathbb{A}}.
\end{equation}
Using the flux matrices (\ref{eq: Flux Matrix Definitions}) and the
relations (\ref{eq: Flux Matrix Axtions on Cohomology Bases - Components}),
the poly-forms $\chi$ and $\Psi$ can now be expressed as \label{sub:3.3.3}
\begin{equation}
\begin{array}{ccl}
\chi & = & {\displaystyle e^{-B}\Xi{}_{\mathbb{A}}\mathcal{O}^{\mathbb{A}}{}_{\mathbb{I}}}V^{\mathbb{I}},\\
\Psi & = & {\displaystyle e^{-B}\Sigma{}_{\mathbb{I}}\widetilde{\mathcal{O}}^{\mathbb{I}}{}_{\mathbb{A}}W^{\mathbb{A}}.\vphantom{^{X^{X^{X^{X^{X^{X}}}}}}}}
\end{array}\label{eq: Polyforms chi and Psi reformulated}
\end{equation}
When integrating the NS-NS action (\ref{eq:NS-NS-Scalar-Potential-Complete})
over $CY_{3}$, the first two terms of (\ref{eq: Polyforms chi and Psi reformulated})
combine to the matrices (\ref{eq:CollectiveMatrix-M}) and (\ref{eq:CollectiveMatrix-N}),
and one eventually obtains for the scalar potential 
\begin{equation}
\begin{array}{ccl}
V_{\textrm{scalar, NS-NS}} & {\displaystyle \!\!\!\!=\!\!\!\!} & {\displaystyle e^{-2\phi}\left[\vphantom{\frac{1}{2\mathcal{K}}}V^{\mathbb{I}}(\mathcal{O}^{T})_{\mathbb{I}}{}^{\mathbb{A}}\mathbb{M_{AB}}\mathcal{O}^{\mathbb{B}}{}_{\mathbb{J}}V^{\mathbb{J}}+W^{\mathbb{A}}(\widetilde{\mathcal{O}}^{T})_{\mathbb{A}}{}^{\mathbb{I}}\mathbb{N_{IJ}}\widetilde{\mathcal{O}}^{\mathbb{J}}{}_{\mathbb{B}}\overline{W}^{\mathbb{B}}\right.}\\
 & \vphantom{^{X^{X^{X^{X^{X^{X}}}}}}} & {\displaystyle \hphantom{e^{-2\phi}||}\left.-\frac{1}{4\mathcal{K}}\overline{W}^{\mathbb{A}}S_{\mathbb{AB}}\mathcal{O}^{\mathbb{B}}{}_{\mathbb{I}}\left(V^{\mathbb{I}}\overline{V}^{\mathbb{J}}+\overline{V}^{\mathbb{I}}V^{\mathbb{J}}\right)(\mathcal{O}^{T})_{\mathbb{J}}{}^{\mathbb{C}}(S^{T})_{\mathbb{CD}}\overline{W}^{\mathbb{D}}\right].}
\end{array}\label{eq:ScalarPotential-NSNS}
\end{equation}

\subsubsection{Integrating over the Internal Space - R-R Sector\label{sub:3.3.4}}

Following the same pattern for the R-R sector, we start by discarding
the cohomologically trivial (and thus massive) $C$-fields and expand
\begin{equation}
\begin{array}{ccl}
e^{B}\mathcal{C}^{{\scriptscriptstyle \left(\mathrm{IIA}\right)}} & {\displaystyle \!\!\!\!=\!\!\!\!} & C^{\left(3\right)}{}^{\mathsf{A}}\alpha_{\mathsf{A}}-C^{\left(3\right)}{}_{\mathsf{A}}\beta^{\mathsf{A}},\\
\vphantom{^{X^{X^{X^{X^{X^{X}}}}}}}e^{B}\mathcal{C}^{{\scriptscriptstyle \left(\mathrm{IIB}\right)}} & {\displaystyle \!\!\!\!=\!\!\!\!} & C^{\left(1\right)}{}_{0}\tilde{\omega}^{0}+C^{\left(2\right)}{}^{\mathsf{I}}\omega_{\mathsf{I}}+C^{\left(4\right)}{}_{\mathsf{I}}\tilde{\omega}^{\mathsf{I}}+C^{\left(6\right)0}\omega_{0}.
\end{array}
\end{equation}
The expansion coefficients are again arranged in vectors 
\begin{equation}
\begin{array}{cclc}
\mathsf{C}_{0}^{\mathbb{A}} & {\displaystyle \!\!\!\!=\!\!\!\!} & \left(\begin{array}{cc}
C^{\left(3\right)}{}^{\mathsf{A}}, & C^{\left(3\right)}{}^{\mathsf{A}}\end{array}\right) & \qquad\textrm{(type IIA theory)},\\
\vphantom{^{X^{X^{X^{X^{X^{X}}}}}}}\mathsf{C}_{0}^{\mathbb{I}} & {\displaystyle \!\!\!\!=\!\!\!\!} & \left(\begin{array}{cccc}
C^{\left(6\right)0}, & C^{\left(2\right)}{}^{\mathsf{I}}, & C^{\left(1\right)}{}_{0}, & C^{\left(4\right)}{}_{\mathsf{I}}\tilde{\omega}^{\mathsf{I}}\end{array}\right) & \qquad\textrm{(type IIB theory)},
\end{array}
\end{equation}
where the subscript index ``$0$'' denotes the number of external
components and is introduced for consistency with section~\ref{sec:5}. Similarly,
we write for the non-trivial R-R fluxes 
\begin{equation}
\begin{array}{ccl}
\mathcal{G}^{{\scriptscriptstyle \left(\mathrm{IIA}\right)}} & {\displaystyle \!\!\!\!=\!\!\!\!} & C^{\left(0\right)}{}_{0}\tilde{\omega}^{0}+C^{\left(2\right)}{}^{\mathsf{I}}\omega_{\mathsf{I}}+C^{\left(4\right)}{}_{\mathsf{I}}\tilde{\omega}^{\mathsf{I}}+C^{\left(6\right)0}\omega_{0},\\
\vphantom{^{X^{X^{X^{X^{X^{X}}}}}}}\mathcal{G}^{{\scriptscriptstyle \left(\mathrm{IIB}\right)}} & {\displaystyle \!\!\!\!=\!\!\!\!} & -G^{\left(3\right)}{}^{\mathsf{A}}\alpha_{\mathsf{A}}+G^{\left(3\right)}{}_{\mathsf{A}}\beta^{\mathsf{A}},
\end{array}
\end{equation}
and 
\begin{equation}
\begin{array}{cclc}
\mathsf{G}_{\textrm{flux}}^{\mathbb{I}} & {\displaystyle \!\!\!\!=\!\!\!\!} & \left(\begin{array}{cccc}
G^{\left(6\right)0}, & G^{\left(2\right)}{}^{\mathsf{I}}, & G^{\left(1\right)}{}_{0}, & G^{\left(4\right)}{}_{\mathsf{I}}\end{array}\right) & \qquad\textrm{(type IIA theory)},\\
\vphantom{^{X^{X^{X^{X^{X^{X}}}}}}}\mathsf{G}_{\textrm{flux}}^{\mathbb{A}} & {\displaystyle \!\!\!\!=\!\!\!\!} & \left(\begin{array}{cc}
G^{\left(3\right)}{}^{\mathsf{A}}, & G^{\left(3\right)}{}_{\mathsf{A}}\end{array}\right) & \qquad\textrm{(type IIB theory)},
\end{array}
\end{equation}
allowing us to reformulate the poly-forms (\ref{eq:RR-IIA-Poly-Form})
and (\ref{eq:RR-IIB-Poly-Form}) as 
\begin{equation}
\begin{array}{ccl}
\mathfrak{G}^{{\scriptscriptstyle \left(\mathrm{IIA}\right)}} & {\displaystyle \!\!\!\!=\!\!\!\!} & e^{-B}\left(\mathsf{G}_{\textrm{flux}}^{\mathbb{I}}+\widetilde{\mathcal{O}}^{\mathbb{I}}{}_{\mathbb{A}}\mathsf{C}_{0}^{\mathbb{A}}\right),\\
\vphantom{^{X^{X^{X^{X^{X^{X}}}}}}}\mathfrak{G}^{{\scriptscriptstyle \left(\mathrm{IIB}\right)}} & {\displaystyle \!\!\!\!=\!\!\!\!} & e^{-B}\left(\vphantom{\widetilde{\mathcal{O}}^{\mathbb{I}}{}_{\mathbb{A}}}\mathsf{G}_{\textrm{flux}}^{\mathbb{A}}+\mathcal{O}^{\mathbb{A}}{}_{\mathbb{I}}\mathsf{C}_{0}^{\mathbb{I}}\right).
\end{array}
\end{equation}
Integrating (\ref{eq:RR-IIA-ScalarPotential}) and (\ref{eq:RR-IIB-ScalarPotential})
over $CY_{3}$ and once more utilizing the relations (\ref{eq:CollectiveMatrix-M})
and (\ref{eq:CollectiveMatrix-N}), we eventually arrive at 
\begin{equation}
\begin{array}{ccl}
V_{\textrm{scalar, R-R}}^{{\scriptscriptstyle \left(\mathrm{IIA}\right)}} & {\displaystyle \!\!\!\!=\!\!\!\!} & {\displaystyle \frac{1}{2}\left(\mathsf{G}_{\textrm{flux}}^{\mathbb{I}}+\mathcal{\widetilde{O}}^{\mathbb{I}}{}_{\mathbb{A}}\mathsf{C}_{0}^{\mathbb{A}}\right)\mathbb{N}{}_{\mathbb{IJ}}\left(\mathsf{G}_{\textrm{flux}}^{\mathbb{J}}+\mathcal{\widetilde{O}}^{\mathbb{J}}{}_{\mathbb{B}}\mathsf{C}_{0}^{\mathbb{B}}\right)},\\
\vphantom{^{X^{X^{X^{X^{X^{X}}}}}}}V_{\textrm{scalar, R-R}}^{{\scriptscriptstyle \left(\mathrm{IIB}\right)}} & {\displaystyle \!\!\!\!=\!\!\!\!} & {\displaystyle \frac{1}{2}\left(\vphantom{\widetilde{\mathcal{O}}^{\mathbb{I}}{}_{\mathbb{A}}}\mathsf{G}_{\textrm{flux}}^{\mathbb{A}}+\mathcal{O}^{\mathbb{A}}{}_{\mathbb{I}}\mathsf{C}_{0}^{\mathbb{I}}\right)\mathbb{M}{}_{\mathbb{AB}}\left(\vphantom{\widetilde{\mathcal{O}}^{\mathbb{I}}{}_{\mathbb{A}}}\mathsf{G}_{\textrm{flux}}^{\mathbb{B}}+\mathcal{O}^{\mathbb{B}}{}_{\mathbb{J}}\mathsf{C}_{0}^{\mathbb{J}}\right)}.
\end{array}\label{eq:ScalarPotential-RR-IIAandIIB}
\end{equation}

\subsubsection{Mirror Symmetry}

Since DFT incorporates all fluxes of the T-duality chain presented
in \cite{Shelton:2005cf,Wecht:2007wu}, it is to be expected that
$\textrm{IIA}\leftrightarrow\textrm{IIB}$ Mirror Symmetry is restored
in this setting. Indeed, comparing the results (\ref{eq:ScalarPotential-RR-IIAandIIB})
for the type IIA and IIB cases, it is easy to verify that the theories
are related to each other as
\begin{equation}
\begin{array}{lclclcl}
\mathbb{M}{}_{\mathbb{AB}} & \leftrightarrow & \mathbb{N}{}_{\mathbb{IJ}},\qquad\qquad & \vphantom{^{X^{X^{X^{X^{X^{X}}}}}}} & h^{1,1} & \leftrightarrow & h^{1,2},\\
V^{\mathbb{I}} & \leftrightarrow & W^{\mathbb{A}}, & \vphantom{^{X^{X^{X^{X^{X^{X}}}}}}} & S_{\mathbb{IJ}} & \leftrightarrow & S_{\mathbb{AB}}\\
\mathsf{C}_{n}^{\mathbb{I}} & \leftrightarrow & \mathsf{C}_{n}^{\mathbb{A}}, & \vphantom{^{X^{X^{X^{X^{X^{X}}}}}}} & \mathsf{G}_{\textrm{flux}}^{\mathbb{I}} & \leftrightarrow & \mathsf{G}_{\textrm{flux}}^{\mathbb{A}},\\
\mathcal{O}^{\mathbb{A}}{}_{\mathbb{I}} & \leftrightarrow & \mathcal{\widetilde{\mathcal{O}}}^{\mathbb{I}}{}_{\mathbb{A}}. & \vphantom{^{X^{X^{X^{X^{X^{X}}}}}}}
\end{array}\label{eq:MirrorSymmetry-ScalarPotential}
\end{equation}
These transformations strongly resemble those
appearing in traditional Calabi-Yau compactifications of supergravity
theories \cite{Lerche:1989uy,Greene:1990ud}: The first two lines
resemble an exchange of roles between the K\"ahler class and complex
structure moduli spaces, while line three describes an obvious replacement
of the theory-specific R-R fields. The last line encodes mappings
between the fluxes, which in particular contain exchanges between
the geometric and non-geometric ones, once more illustrating how the
latter are required for preservation of $\textrm{IIA}\leftrightarrow\textrm{IIB}$
Mirror Symmetry. Taken as a whole, this implies that type IIA DFT
compactified on a Calabi-Yau three-fold $CY_{3}$ is physically equivalent
to its type IIB analogue compactified on a mirror Calabi-Yau three-fold
$\widetilde{CY}_{3}$, with the Hodge-diamonds of the
two manifolds being related by a reflection along their diagonal axes.

Note that the relations involving the expansion coefficients can be
lifted to ten dimensions, allowing for a more compact notation 
\begin{equation}
\chi\leftrightarrow\Psi,\qquad\mathfrak{\hat{G}}^{{\scriptscriptstyle \left(\mathrm{IIA}\right)}}\leftrightarrow\mathfrak{\hat{G}}^{{\scriptscriptstyle \left(\mathrm{IIB}\right)}}
\end{equation}
of the mirror mappings as an exchange of the poly-forms (\ref{eq:NS-NS-Scalar-Potential-Complete-Definitions}),
(\ref{eq:RR-IIA-Poly-Form}) and (\ref{eq:RR-IIB-Poly-Form}) we used
to reformulate the DFT action. Similarly to component notation, we
see that they precisely correspond to an exchange the terms encoding
the complexified K\"ahler-class ($\chi$) and complex structure ($\Psi$)
moduli, besides a mapping between the IIA and IIB R-R objects. In
particular, the structure of the theory remains invariant under Mirror

\section{The Scalar Potential on $K3\times T^{2}$\label{sec:4}}

We next repeat the process of dimensional reduction for DFT on $K3\times T^{2}$
and thereby show how the framework presented in the previous section
can straightforwardly be generalized to more complex cases of flux
compactifications. Much of the following discussion is completely
analogous to the Calabi-Yau setting, and we will therefore focus on
the specific features of $K3\times T^{2}$ instead. To simplify computations,
we will from now on set fluxes which cannot be supported on the internal
manifold to zero and ignore fields acquiring a mass in four dimensions.

In order to distinguish between $K3$ and $T^{2}$ indices, we split
the ``checked'' indices $\check{I,}\check{J},\ldots$ into $I,J,\ldots$
labeling $K3$ coordinates and $R,S\ldots$ labeling $T^{2}$ coordinates.
Their complex-geometric (undoubled) analogues are denoted by $a,\bar{a},b,\bar{b}$
and $g,\bar{g},h,\bar{h}$, respectively. For convenience, we accordingly
split the flux operators (\ref{eq: Flux Operators}) into their distinct
cohomologically nontrivial components,
\begin{equation}
\label{eq:FluxOperators-K3xT=0000B2-Split}
\begin{array}{crcl}
H\wedge: & \Omega^{p}\left(K3\times T^{2}\right) & \longrightarrow & \Omega^{p+3}\left(K3\times T^{2}\right)\\
 & \omega_{p} & \mapsto & {\displaystyle \frac{1}{2!}H_{ijr}\; dx^{i}\wedge dx^{j}\wedge dx^{r}\wedge\omega_{p}},\\
\\
F\circ: & \Omega^{p}\left(K3\times T^{2}\right) & \longrightarrow & \Omega^{p+1}\left(K3\times T^{2}\right)\\
 & \omega_{p} & \mapsto & {\displaystyle \left(\frac{1}{2!}F^{r}{}_{ij}\; dx^{i}\wedge dx^{j}\wedge\iota_{k}+F^{j}{}_{ir}\; dx^{i}\wedge dx^{r}\wedge\iota_{j}\right)\wedge\omega_{p}},
\\
Q\bullet: & \Omega^{p}\left(K3\times T^{2}\right) & \longrightarrow & \Omega^{p-1}\left(K3\times T^{2}\right)\\
 & \omega_{p} & \mapsto & {\displaystyle \left(\frac{1}{2!}Q_{r}{}^{ij}\; dx^{r}\wedge\iota_{i}\wedge\iota_{j}+Q_{i}{}^{jr}\; dx^{i}\wedge\iota_{j}\wedge\iota_{r}\right)\wedge\omega_{p}},\\
\\
R\llcorner: & \Omega^{p}\left(K3\times T^{2}\right) & \longrightarrow & \Omega^{p-3}\left(K3\times T^{2}\right)\\
 & \omega_{p} & \mapsto & {\displaystyle \frac{1}{3!}R^{ijr}\;\iota_{i}\wedge\iota_{j}\wedge\iota_{r}\wedge\omega_{p}}.,\\
\hphantom{F\circ:} &  &  & \hphantom{{\displaystyle \left(\frac{1}{2!}F^{r}{}_{ij}\; dx^{i}\wedge dx^{j}\wedge\iota_{k}+F^{j}{}_{ir}\; dx^{i}\wedge dx^{r}\wedge\iota_{j}\right)\wedge\omega_{p}},}\\
Y\wedge: & \Omega^{p}\left(K3\times T^{2}\right) & \longrightarrow & \Omega^{p+1}\left(K3\times T^{2}\right)\\
 & \omega_{p} & \mapsto & {\displaystyle \vphantom{\frac{1}{3!}}{\displaystyle Y_{r}\; dx^{r}\wedge\omega_{p}}},\\
\\
Z\blacktriangledown: & \Omega^{p}\left(K3\times T^{2}\right) & \longrightarrow & \Omega^{p-1}\left(K3\times T^{2}\right)\\
 & \omega_{p} & \mapsto & {\displaystyle \vphantom{\frac{1}{3!}}{\displaystyle Z^{r}\;\iota_{r}\wedge\omega_{p}}.}
\end{array}
\end{equation}
Finally, we again impose the strong constraint only for the background
and the field fluctuations, while applying the Bianchi identities
(\ref{eq:Bianchi-Identities-NSNS}) for the fluxes.

\subsection{Reformulating the Action}

The toolbox we used to reformulate the internal NS-NS action on $CY_{3}$
builds upon on the mathematical framework of generalized Calabi-Yau
structures \cite{Hitchin:2004ut} and can be straightforwardly extended
to arbitrary manifolds admitting such a one. For the case of $K3\times T^{2}$,
this can be done by utilizing the features of generalized $K3$ surfaces
\cite{Huybrechts:2003ak} and formally viewing $T^{2}$ as a complex
torus with a generalized Calabi-Yau structure. We therefore exploit
the product structure of $K3\times T^{2}$ and consider the K\"ahler
class and complex structure forms 
\begin{equation}
e^{b+iJ}=e^{b_{K3}+iJ_{K3}}\wedge e^{b_{T^{2}}+iJ_{T^{2}}},\qquad\qquad e^{b}\wedge\Omega=\left(\vphantom{e^{b_{T^{2}}}}e^{b_{K3}}\wedge\Omega_{K3}\right)\wedge\left(e^{b_{T^{2}}}\wedge\Omega_{T^{2}}\right),\label{eq:K3xT=0000B2-StructureForms}
\end{equation}
respectively. The reformulation of the scalar potential part of the
NS-NS sector (\ref{eq: Full Action NSNS-sector}) then follows a very
similar pattern as in the Calabi-Yau case. As an instructive example,
one can easily check that the only non-trivial contribution of the
pure $H$-flux setting is given by 
\begin{equation}
\star\mathcal{L}_{\textrm{NS-NS, scalar,}H}=\frac{e^{-2\phi}}{4}H_{ijr}H_{i'j'r'}g^{ii'}g^{jj'}g^{rr'}\star\mathbf{1}^{\left(6\right)},
\end{equation}
which can again be written as 
\begin{equation}
\star\mathcal{L}_{\textrm{NS-NS, scalar,}H}=-\frac{e^{-2\phi}}{2}H\wedge\star H,
\end{equation}
with $H$ now defined as in (\ref{eq:FluxOperators-K3xT=0000B2-Split}).
The $F$-flux allows for different nontrivial components and is therefore
slightly more involved. From the initial action (\ref{eq: Full Action NSNS-sector}),
we obtain 
\eq{
&\mathcal{L}_{\textrm{NS-NS, scalar,}F} = {\displaystyle -\frac{e^{-2\phi}}{4}\left(F^{r}{}_{ij}F^{r'}{}_{i'j'}g^{ii'}g^{jj'}g_{rr'}+2F^{i}{}_{jr}F^{i'}{}_{j'r'}g_{ii'}g^{jj'}g^{rr'}+2F^{m}{}_{nr}F^{n}{}_{mr'}g^{rr'}\right.}
\\
&\hspace{130pt}{\displaystyle \left.+4F^{m}{}_{mr}F^{m'}{}_{m'r'}g^{rr'}+4F^{r}{}_{mi}F^{m}{}_{ri'}g^{ii'}\right)},
}
Denoting the first and second component of $F\circ$ by $F_{1}\circ$
respectively $F_{2}\circ$ (based on the split employed in (\ref{eq:FluxOperators-K3xT=0000B2-Split})),
the first term can be rewritten similarly to the $H$-flux contribution
as 
\begin{equation}
-\frac{e^{-2\phi}}{4}F^{r}{}_{ij}F^{r'}{}_{i'j'}g^{ii'}g^{jj'}g_{rr'}\star\mathbf{1}^{\left(6\right)}=-\frac{e^{-2\phi}}{2}\left[\vphantom{Q_{r}{}^{ij}}F_{1}\circ\left(\star\mathbf{1}_{K3}\wedge\mathbf{1}_{T^{2}}\right)\right]\wedge\star\left[\vphantom{Q_{r}{}^{ij}}F_{1}\circ\left(\star\mathbf{1}_{K3}\wedge\mathbf{1}_{T^{2}}\right)\right],
\end{equation}
while a calculation analogous to the pure $F$-flux case in the Calabi-Yau
setting yields for the next three terms 
\begin{equation}
\begin{array}{cl}
 & {\displaystyle -\frac{e^{-2\phi}}{4}\left(2F^{i}{}_{jr}F^{i'}{}_{j'r'}g_{ii'}g^{jj'}g^{rr'}+2F^{m}{}_{nr}F^{n}{}_{mr'}g^{rr'}+4F^{m}{}_{mr}F^{m'}{}_{m'r'}g^{rr'}\right)}\\
\vphantom{^{X^{X^{X^{X^{X^{X}}}}}}}{\displaystyle \!\!\!\!=\!\!\!\!} & {\displaystyle -\frac{e^{-2\phi}}{2}\left\{ \vphantom{\frac{e^{-2\phi}}{2}}\left[\vphantom{Q_{r}{}^{ij}}F_{2}\circ\left(\vphantom{\overline{\Omega}_{K3}}iJ_{K3}\wedge\mathbf{1}_{T^{2}}\right)\right]\wedge\star\left[\vphantom{Q_{r}{}^{ij}}F_{2}\circ\left(\vphantom{\overline{\Omega}_{K3}}iJ_{K3}\wedge\mathbf{1}_{T^{2}}\right)\right]\right.}\\
\vphantom{^{X^{X^{X^{X^{X^{X}}}}}}} & {\displaystyle \hphantom{XXXX}\left.\vphantom{\frac{e^{-2\phi}}{2}}+\left[\vphantom{Q_{r}{}^{ij}}F_{2}\circ\left(\vphantom{\overline{\Omega}_{K3}}\star\mathbf{1}_{K3}\wedge\mathbf{1}_{T^{2}}\right)\right]\wedge\star\left[\vphantom{Q_{r}{}^{ij}}F_{2}\circ\left(\vphantom{\overline{\Omega}_{K3}}\star\mathbf{1}_{K3}\wedge\mathbf{1}_{T^{2}}\right)\right]\right.}\\
\vphantom{^{X^{X^{X^{X^{X^{X}}}}}}} & {\displaystyle \hphantom{XXXX}\left.\vphantom{\frac{e^{-2\phi}}{2}}+\left[\vphantom{Q_{r}{}^{ij}}F_{2}\circ\left(\vphantom{\overline{\Omega}_{K3}}\Omega_{K3}\wedge\Omega_{T^{2}}\right)\right]\wedge\star\left[\vphantom{Q_{r}{}^{ij}}F_{2}\circ\left(\overline{\Omega}_{K3}\wedge\overline{\Omega}_{T^{2}}\right)\right]\right.}\\
\vphantom{^{X^{X^{X^{X^{X^{X}}}}}}} & {\displaystyle \hphantom{XXXX}\left.\vphantom{\frac{e^{-2\phi}}{2}}-{\displaystyle \left[\vphantom{Q_{r}{}^{ij}}\left(\vphantom{\overline{\Omega}_{K3}}\Omega_{K3}\wedge\Omega_{T^{2}}\right)\wedge F_{2}\circ\left(iJ_{K3}\wedge\mathbf{1}_{T^{2}}\right)\right]\wedge\star\left[\vphantom{Q_{r}{}^{ij}}\left(\overline{\Omega}_{K3}\wedge\overline{\Omega}_{T^{2}}\right)F_{2}\circ\left(iJ_{K3}\wedge\mathbf{1}_{T^{2}}\right)\right]}\right\} }
\end{array}
\end{equation}
and the final one 
\begin{equation}
\begin{array}{cl}
 & {\displaystyle -e^{-2\phi}F^{r}{}_{mi}F^{m}{}_{ri'}g^{ii'}}\\
\vphantom{^{X^{X^{X^{X^{X^{X}}}}}}}{\displaystyle \!\!\!\!=\!\!\!\!} & {\displaystyle -e^{-2\phi}\left\{ \vphantom{\frac{e^{-2\phi}}{2}}\left[\vphantom{Q_{r}{}^{ij}}F_{1}\circ\left(\mathbf{1}_{K3}\wedge iJ_{T^{2}}\right)\right]\wedge\star\left[\vphantom{Q_{r}{}^{ij}}F_{2}\circ\left(iJ_{K3}\wedge\mathbf{1}_{T^{2}}\right)\right]\right.}\\
\vphantom{^{X^{X^{X^{X^{X^{X}}}}}}} & \hphantom{XXXX}-{\displaystyle \left.\vphantom{\frac{e^{-2\phi}}{2}}\left[\vphantom{Q_{r}{}^{ij}}\left(\Omega_{K3}\wedge\Omega_{T^{2}}\right)\wedge F_{1}\circ\left(\mathbf{1}_{K3}\wedge iJ_{T^{2}}\right)\right]\wedge\star\left[\vphantom{Q_{r}{}^{ij}}\left(\Omega_{K3}\wedge\Omega_{T^{2}}\right)F_{2}\circ\left(iJ_{K3}\wedge\mathbf{1}_{T^{2}}\right)\right]\right\} },
\end{array}
\end{equation}
showing that the $F$-contribution to the scalar potential takes the
form (\ref{eq:NS-NS-Lagrangian-PureF-Reformulated}) already known
from the Calabi-Yau setting. The discussion of the non-geometric and
generalized dilaton fluxes as well as the R-R sector is analogous.
For the most general setting, we eventually arrive at the familiar
expressions (\ref{eq:NS-NS-Scalar-Potential-Complete}), (\ref{eq:RR-IIA-ScalarPotential})
and (\ref{eq:RR-IIB-ScalarPotential}), with the fluxes adjusted according
to (\ref{eq:FluxOperators-K3xT=0000B2-Split}) and $e^{iJ}$ and $\Omega$
as in (\ref{eq:K3xT=0000B2-StructureForms}).

\subsection{\label{sub:4.2}Dimensional Reduction}

We next proceed as usual by expanding the fields and fluxes in terms
of the cohomology bases of $K3\times T^{2}$ before integrating over
the internal manifold.

\subsubsection{Special Geometry of $K3\times T^{2}$\label{sub:4.2.1}}

As in the Calabi-Yau case, it is convenient to treat the even and
odd cohomology groups of the compactification manifolds separately
in order to allow for a description of the K\"ahler class and complex
structure moduli spaces as well as Mirror Symmetry. Since all nontrivial
cohomology groups of $K3$ are of even degree, the property of a cohomologically
nontrivial differential form on $K3\times T^{2}$ being even or odd
depends purely on its $T^{2}$ component.

\subsubsection*{Even Cohomology}

The even cohomology bases of $T^{2}$ are precisely the identity $\mathbf{1}_{T^{2}}$
for the zero-forms and $\star\mathbf{1}_{T^{2}}$ for the two-forms
(the latter of which coincides with the normalized K\"ahler form), 
\eq{
\begin{array}{l}
\left\{ \vphantom{\vphantom{\frac{\sqrt{g_{CY_{3}}}}{\mathcal{K}_{t^{2}}}}}\mathbf{1}_{T^{2}}\right\} \in H^{0}\left(T^{2}\right),\\
\\
\left\{ \frac{\sqrt{g_{T^{2}}}}{\mathcal{K}_{T^{2}}}\star\mathbf{1}_{T^{2}}\right\} \in H^{2}\left(T^{2}\right).
\end{array}
}
and we denote them by $v_{0}$ respectively $v_{3}$ from now on.
The bases of the $K3$ de Rham cohomology groups are given by 
\begin{equation}
\begin{array}{lc}
\left\{ \vphantom{\frac{\sqrt{g_{K3}}}{\mathcal{K}_{K3}}}\mathbf{1}_{K3}\right\} \in H^{0}\left(T^{2}\right),\\
\\
\left\{ \vphantom{\frac{\sqrt{g_{K3}}}{\mathcal{K}_{K3}}}\sigma_{\mathsf{u}}\right\} \in H^{2}\left(T^{2}\right) & \qquad\textrm{with }\mathsf{u}=1,\ldots22\\
\\
\left\{ \frac{\sqrt{g_{K3}}}{\mathcal{K}_{K3}}\star\mathbf{1}_{K3}\right\} \in H^{4}\left(T^{2}\right),
\end{array},
\end{equation}
and we define $\sigma_{0}=\mathbf{1}^{\left(6\right)}$ and $\sigma_{23}=\star\mathbf{1}^{\left(6\right)}$
, enabling us to arrange the $K3$ bases in a collective notation
\begin{equation}
\sigma_{\mathsf{U}}=\left(\vphantom{\tilde{\omega}^{\mathsf{I}}}\begin{array}{ccc}
\sigma_{0} & \sigma_{\mathsf{u}} & \sigma_{\mathsf{23}}\end{array}\right).
\end{equation}
We furthermore define $\eta_{\mathsf{uv}}$ to be the intersection
metric 
\begin{equation}
\mathcal{\eta}_{\mathsf{uv}}=\int_{K3}\sigma_{\mathsf{u}}\wedge\sigma_{\mathsf{v}}.
\end{equation}
Its signature $\left(3,19\right)$ resembles the fact that there are
three antiselfdual two-forms (the K\"ahler form, the holomorphic two-form
and its antiholomorphic counterpart) and 19 selfdual ones. This metric
can serve as a building block of a matrix
\begin{equation}
L_{\mathsf{UV}}=\left(\begin{array}{ccc}
0 & 0 & -1\\
0 & \mathcal{\eta}_{\mathsf{uv}} & 0\\
-1 & 0 & 0
\end{array}\right),\qquad L^{\mathsf{UV}}=\left(\begin{array}{ccc}
0 & 0 & -1\\
0 & \eta^{\mathsf{uv}} & 0\\
-1 & 0 & 0
\end{array}\right),\label{eq:L-Matrix}
\end{equation}
which we use to lower and raise cohomological $K3$ indices, 
\begin{equation}
\sigma^{\mathsf{U}}=L^{\mathsf{UV}}\sigma_{\mathsf{V}}.
\end{equation}
Putting all of the above objects together, we can define a collective
basis for the even de Rham cohomology groups of $K3\times T^{2}$
by 
\begin{equation}
\begin{array}{rrl}
\omega_{\mathsf{I}} & ={\displaystyle \mbox{\ensuremath{\left(\vphantom{\tilde{\omega}^{\mathsf{I}}}\begin{array}{ccc}
\omega_{0} & \omega_{\mathsf{u}} & \omega_{23}\end{array}\right)}}} & ={\displaystyle \left(\vphantom{\tilde{\omega}^{\mathsf{I}}}\begin{array}{ccc}
v_{0}\wedge\sigma_{0} & v_{0}\wedge\sigma_{\mathsf{u}} & v_{0}\wedge\sigma_{23}\end{array}\right)},\\
\\
\widetilde{\omega}^{\mathsf{I}} & ={\displaystyle \left(\vphantom{\tilde{\omega}^{\mathsf{I}}}\begin{array}{ccc}
\widetilde{\omega}^{\mathsf{0}} & \widetilde{\omega}^{\mathsf{u}} & \widetilde{\omega}^{23}\end{array}\right)} & ={\displaystyle \left(\vphantom{\tilde{\omega}^{\mathsf{I}}}\begin{array}{ccc}
v_{3}\wedge\sigma^{0} & v_{3}\wedge\sigma^{\mathsf{u}} & v_{3}\wedge\sigma^{23}\end{array}\right),}
\end{array}\label{eq:Basis-Even-Cohomology_K3xT=0000B2}
\end{equation}
where the labeling $\mathsf{I,J,}\ldots$ was chosen to make it distinguishable
from its odd counterpart. The basis elements satisfy the normalization
condition 
\begin{equation}
\int_{K3\times T^{2}}\omega_{\mathsf{I}}\wedge\widetilde{\omega}^{\mathsf{J}}=\left(\begin{array}{ccc}
-1 & 0 & 0\\
0 & \delta_{\mathsf{u}}{}^{\mathsf{v}} & 0\\
0 & 0 & -1
\end{array}\right).
\end{equation}
We again use the collective notation 
\begin{equation}
\Sigma_{\mathbb{I}}=\left(\begin{array}{cc}
\omega_{\mathsf{I}} & \tilde{\omega}^{\mathsf{I}}\end{array}\right).
\end{equation}
Analogously to the Calabi-Yau case, this basis defines a symplectic
structure by 
\[
\int_{K3\times T^2}\left\langle \Sigma_{\mathbb{I}},\Sigma_{\mathbb{J}}\right\rangle =(S_{even})_{\mathbb{IJ}}=\left(\begin{array}{cc}
0 & \mathbbm{1}\\
-\mathbbm{1} & 0
\end{array}\right)\in Sp\left(48,\mathbb{R}\right).
\]
In order to describe the K\"ahler class moduli space of $K3\times T^{2}$,
we combine the K\"ahler form $J$ and the internal part $b$ of the
$\hat{B}$-field to the complexified K\"ahler form 
\begin{equation}
\mathfrak{J}=b+iJ=\left(b_{T^{2}}+iJ_{T^{2}}\right)+\left(\vphantom{b_{T^{2}}}b_{K3}+iJ_{K3}\right)=\rho\widetilde{\omega}^{\mathsf{0}}+t^{\mathsf{u}}\omega_{\mathsf{u}},
\end{equation}
where the latter splitting can be applied due to the vanishing first
Betti number of $K3$. The complex parameter $\rho=b^{0}+iw^{0}$
encodes the volume modulus $w^{0}$ of $T^{2}$ as well as the component
$b^{0}$ of $\hat{B}$ living purely in $T^{2}$. Analogously, the
$t^{\mathsf{u}}$ denote the moduli $w^{\mathsf{u}}$ of $J_{K3}$
and $b^{\mathsf{u}}$ spanning the complexified K\"ahler cone of $K3$.
In the upcoming discussion, we will mainly encounter the poly-form
$e^{\mathfrak{J}}$, which we will expand as $e^{\mathfrak{J}}=\Sigma{}_{\mathbb{I}}V^{\mathbb{I}}$
with 
\begin{equation}
V^{\mathbb{I}}=\left(\vphantom{\tilde{\omega}^{\mathsf{I}}}\begin{array}{cccccc}
1, & t^{\mathsf{u}}, & t^{\mathsf{u}}t^{\mathsf{v}}\eta_{\mathsf{uv}}, & \rho t_{\mathsf{u}}t_{\mathsf{v}}\eta^{\mathsf{uv}}, & \rho t_{\mathsf{u}}, & \rho\end{array}\right)^{T}.
\end{equation}

\subsubsection*{Odd Cohomology}

A basis for the odd cohomology groups can be constructed in a similar
manner by replacing the even basis elements of $T^{2}$ by two one-form
basis elements 
\begin{equation}
\left\{ \vphantom{\frac{\sqrt{g_{K3}}}{\mathcal{K}_{K3}}}v_{1},v_{2}\right\} \in H^{1}\left(T^{2}\right)\qquad\textrm{with }\int_{T^2}v_{1}\wedge v_{2}=1
\end{equation}
and defining 
\begin{equation}
\begin{array}{lll}
\alpha_{\mathsf{A}} & =\ensuremath{\left(\vphantom{\frac{\sqrt{g_{K3}}}{\mathcal{K}_{K3}}}\begin{array}{ccc}
\alpha_{0} & \alpha_{\mathsf{u}} & \alpha_{23}\end{array}\right)} & =\left(\vphantom{\frac{\sqrt{g_{K3}}}{\mathcal{K}_{K3}}}\begin{array}{ccc}
v_{1}\wedge\sigma_{0} & v_{1}\wedge\sigma_{\mathsf{u}} & v_{1}\wedge\sigma_{23}\end{array}\right),\\
\\
\beta^{\mathsf{A}} & =\left(\vphantom{\frac{\sqrt{g_{K3}}}{\mathcal{K}_{K3}}}\begin{array}{ccc}
\beta^{0} & \beta^{\mathsf{u}} & \beta^{23}\end{array}\right) & =\left(\vphantom{\frac{\sqrt{g_{K3}}}{\mathcal{K}_{K3}}}\begin{array}{ccc}
v_{2}\wedge\sigma^{0} & v_{2}\wedge\sigma^{\mathsf{u}} & v_{2}\wedge\sigma^{23}\end{array}\right).
\end{array}\label{eq:Basis-Odd-Cohomology}
\end{equation}
They satisfy the normalization condition 
\begin{equation}
\int_{K3\times T^{2}}\alpha_{\mathsf{A}}\wedge\beta^{\mathsf{A}}=\left(\begin{array}{ccc}
-1 & 0 & 0\\
0 & \delta_{\mathsf{u}}{}^{\mathsf{v}} & 0\\
0 & 0 & -1
\end{array}\right)
\end{equation}
and can be arranged in a collective basis 
\begin{equation}
\Xi_{\mathbb{A}}=\left(\begin{array}{cc}
\alpha_{\mathsf{A}} & \beta^{\mathsf{A}}\end{array}\right)
\end{equation}
to define a symplectic structure by 
\begin{equation}
\int_{K3\times T^2}\left\langle \Xi_{\mathbb{A}},\Xi_{\mathbb{B}}\right\rangle =(S_{odd})_{\mathbb{IJ}}=\left(\begin{array}{cc}
0 & \mathbbm{1}\\
-\mathbbm{1} & 0
\end{array}\right)\in Sp\left(48,\mathbb{R}\right).
\end{equation}
Notice that we again incorporated a relative minus sign into the expansions
in terms of the even and odd cohomology bases for later convenience.
More specifically, we expand an arbitrary poly-form field $A$ as
\begin{equation}
A=A^{\mathbb{I}}\Sigma_{\mathbb{I}}+A^{\mathbb{A}}\Xi_{\mathbb{A}}=A^{\mathsf{I}}\omega_{\mathsf{I}}+A_{\mathsf{I}}\tilde{\omega}^{\mathsf{I}}+A^{\mathsf{A}}\alpha_{\mathsf{A}}-A_{\mathsf{A}}\beta^{\mathsf{A}}.
\end{equation}
Similarly to the K\"ahler class case, the complex structure moduli space
of $K3\times T^{2}$ can be described by its holomorphic three-form
$\Omega$, which on its part can be split into a holomorphic one-form
$\Omega_{T^{2}}$ living in $T^{2}$ and a holomorphic two-form $\Omega_{K3}$
living in $K3$. Viewing $T^{2}$ as a one-dimensional complex torus,
the former encodes the modular (complex structure) parameter $\tau$
by 
\begin{equation}
\Omega_{T^{2}}=v_{1}-\tau v_{2},
\end{equation}
where 
\begin{equation}
\tau=\int_{T^{2}}\Omega_{T^{2}}\wedge v_{1}.
\end{equation}
Similarly, the latter can be expanded as 
\begin{equation}
\Omega_{K3}=T^{\mathsf{u}}\sigma_{\mathsf{u}},
\end{equation}
allowing us to expand the complete holomorphic three-form $\Omega$
in the basis (\ref{eq:Basis-Odd-Cohomology}). In the following, we
will be mainly concerned with the expression $e^{b}\Omega$, which
can be expanded as $e^{b}\Omega=\Xi{}_{\mathbb{A}}W^{\mathbb{A}}$
with
\begin{equation}
W^{\mathbb{A}}=\left(\vphantom{\tilde{\omega}^{\mathsf{I}}}\begin{array}{cccccc}
0, & T^{\mathsf{u}}, & T^{\mathsf{u}}b^{\mathsf{v}}\eta_{\mathsf{uv}}, & \tau T_{\mathsf{u}}b_{\mathsf{v}}\eta^{\mathsf{uv}}, & \tau T_{\mathsf{u}}, & 0\end{array}\right)^{T}.
\end{equation}

\subsubsection*{Gauge Coupling Matrices}

As in the Calabi-Yau setting, we again define a gauge coupling matrix
\begin{equation}
\mathbb{M_{AB}}={\displaystyle \int_{K3\times T^2}\left(\begin{array}{cc}
\,\,-\left\langle \vphantom{\beta^{\mathsf{B}}}\alpha_{\mathsf{A}},\star_{b}\alpha_{\mathsf{B}}\right\rangle \, & \,\left\langle \alpha_{\mathsf{A}},\star_{b}\beta^{\mathsf{B}}\right\rangle \,\,\\
\vphantom{^{X^{X^{X^{X^{X^{X}}}}}}}\,\,\left\langle \beta^{\mathsf{A}},\star_{b}\alpha_{\mathsf{B}}\right\rangle \, & \,-\left\langle \beta^{\mathsf{A}},\star_{b}\beta^{\mathsf{B}}\right\rangle \,\,
\end{array}\right)},
\end{equation}
which can be written as 
\begin{equation}
\mathbb{M_{AB}}={\displaystyle \frac{1}{\textrm{Im}\tau}\left(\begin{array}{cc}
\,\,\left|\tau\right|^{2}\widetilde{\mathbb{N}}_{\mathsf{AB}}\, & \,\textrm{Re}\tau\widetilde{\mathbb{N}}_{\mathsf{A}}{}^{\mathsf{B}}\,\,\,\\
\vphantom{^{X^{X^{X^{X^{X^{X}}}}}}}\,\,\textrm{Re}\tau\widetilde{\mathbb{N}}^{\mathsf{A}}{}_{\mathsf{B}}\, & \,\widetilde{\mathbb{N}}^{\mathbf{\mathsf{AB}}}\,\,\,
\end{array}\right),}\label{eq:eq:Gauge-Coupling-Matrix-M-K2xT=0000B2}
\end{equation}
where 
\begin{equation}
\widetilde{\mathbb{N}}_{\mathbb{AB}}= \int_{K3}\left(\begin{array}{cc}
\,\,\left\langle \vphantom{\beta^{\mathsf{B}}}\sigma_{\mathsf{U}},\star_{b_{K3}}\sigma_{\mathsf{V}}\right\rangle \, & \,\left\langle \sigma_{\mathsf{U}},\star_{b_{K3}}\sigma^{\mathsf{V}}\right\rangle \,\,\\
\vphantom{^{X^{X^{X^{X^{X^{X}}}}}}}\,\,\left\langle \sigma^{\mathsf{U}},\star_{b_{K3}}\sigma_{\mathsf{V}}\right\rangle \, & \,\left\langle \sigma^{\mathsf{U}},\star_{b_{K3}}\sigma^{\mathsf{V}}\right\rangle \,\,
\end{array}\right)\label{eq:Gauge-Coupling-Matrix-K3}
\end{equation}
is the $K3$ analogue of  (\ref{eq:CollectiveMatrix-N}) (recall that
the indices $\mathsf{A},\mathsf{B},\ldots$, $\mathsf{I},\mathsf{J},\ldots$
and $\mathsf{U},\mathsf{V},\ldots$ run over the same values). Similarly,
we define for the even cohomology groups

\begin{equation}
\mathbb{N_{IJ}}=\int_{K3\times T^2}\left(\begin{array}{cc}
\,\,\left\langle \omega_{\mathsf{I}},\star_{b}\omega_{\mathsf{J}}\vphantom{\tilde{\omega}^{\mathsf{J}}}\right\rangle \, & \,\left\langle \omega_{\mathsf{I}},\star_{b}\tilde{\omega}^{\mathsf{J}}\right\rangle \,\,\\
\vphantom{^{X^{X^{X^{X^{X^{X}}}}}}}\,\,\left\langle \tilde{\omega}^{\mathsf{I}},\star_{b}\omega_{\mathsf{J}}\right\rangle \, & \,\left\langle \tilde{\omega}^{\mathsf{I}},\star_{b}\tilde{\omega}^{\mathsf{J}}\right\rangle \,\,
\end{array}\right),
\end{equation}
which can be reformulated as 
\begin{equation}
\mathbb{N_{IJ}}=\frac{1}{\textrm{Im}\rho}\left(\begin{array}{cc}
\,\,\left|\rho\right|^{2}\widetilde{\mathbb{N}}_{\mathsf{IJ}}\, & \,\textrm{Re}\rho\widetilde{\mathbb{N}}_{\mathsf{I}}{}^{\mathsf{J}}\,\,\,\\
\vphantom{^{X^{X^{X^{X^{X^{X}}}}}}}\,\,\textrm{Re}\rho\widetilde{\mathbb{N}}^{\mathsf{I}}{}_{\mathsf{J}}\, & \,\mathbb{\widetilde{\mathbb{N}}}^{\mathbf{\mathsf{IJ}}}\,\,\,
\end{array}\right),\label{eq:eq:Gauge-Coupling-Matrix-N-K2xT=0000B2}
\end{equation}
with $\widetilde{\mathbb{N}}_{\mathbb{IJ}}$ taking the same form
as (\ref{eq:Gauge-Coupling-Matrix-K3}).

\subsubsection{Fluxes and Cohomology Bases}

To relate the flux operators (\ref{eq:FluxOperators-K3xT=0000B2-Split})
to the gaugings of four-dimensional supergravity, we once more proceed
analogously to the Calabi-Yau setting. The action of the twisted differential
(\ref{eq: Twisted Differential}) on the cohomology bases be summarized
by the relations 
\begin{equation}
\mathcal{D}(\Sigma^{T})_{\mathbb{I}}=(\mathcal{O}^{T})_{\mathbb{I}}{}^{\mathbb{A}}(\Xi^{T})_{\mathbb{A}},\qquad\mathcal{D}(\Xi^{T})_{\mathbb{A}}=(\mathcal{\widetilde{\mathcal{O}}}^{T})_{\mathbb{A}}{}^{\mathbb{I}}(\Sigma^{T})_{\mathbb{I}},
\end{equation}
where the charge matrices
\begin{equation}
\mathcal{O}^{\mathbb{A}}{}_{\mathbb{I}}=\left(\begin{array}{cc}
-\tilde{P}^{\mathsf{A}}{}_{\mathsf{I}} & \tilde{P}^{\mathsf{AI}}\\
O_{\mathsf{AI}} & -O_{\mathsf{A}}{}^{\mathsf{I}}
\end{array}\right),\qquad\widetilde{\mathcal{O}}^{\mathbb{I}}{}_{\mathbb{A}}=\left(\begin{array}{cc}
(O^{T})^{\mathsf{I}}{}_{\mathsf{A}} & (\tilde{P}^{T}){}^{\mathsf{IA}}\\
(O^{T}){}_{\mathsf{IA}} & (\tilde{P}^{T}){}_{\mathsf{I}}{}^{\mathsf{A}}
\end{array}\right)
\end{equation}
comprise the flux expansion coefficients. Their components read
\begin{equation}
\begin{array}{l}
\tilde{P}^{\mathsf{A}}{}_{\mathsf{I}}=\left(\begin{array}{ccc}
(f+y)^{0}{}_{0} & q^{0}{}_{\mathsf{u}} & 0\\
h^{\mathsf{u}}{}_{0} & (f+y)^{\mathsf{u}}{}_{\mathsf{u}} & q^{\mathsf{u}}{}_{\mathsf{23}}\\
0 & h^{\mathsf{23}}{}_{\mathsf{u}} & (f+y)_{23\,23}
\end{array}\right),\\
\\
\tilde{P}^{\mathsf{AI}}=\left(\begin{array}{ccc}
0 & r{}^{0\mathsf{u}} & \left(q+z\right){}^{0\,23}\\
r{}^{\mathsf{u}0} & \left(q+z\right){}^{\mathsf{uu}} & f{}^{\mathsf{u}\,23}\\
\left(q+z\right){}^{23\,0} & f{}^{23\,\mathsf{u}} & 0
\end{array}\right),\\
\\
O_{\mathsf{AI}}=\left(\begin{array}{ccc}
0 & h_{0\mathsf{u}} & \left(f+y\right)_{0\,23}\\
h_{\mathsf{u}0} & \left(f+y\right)_{\mathsf{uu}} & q_{\mathsf{u}\,23}\\
\left(f+y\right)_{23\,0} & q_{23\,\mathsf{u}} & 0
\end{array}\right),\\
\\
O_{\mathsf{A}}{}^{\mathsf{I}}=\left(\begin{array}{ccc}
(q+z)_{0}{}^{0} & f_{0}{}^{\mathsf{u}} & 0\\
r_{\mathsf{u}}{}^{0} & (q+z)_{\mathsf{u}}{}^{\mathsf{u}} & f_{\mathsf{u}}{}^{23}\\
0 & r_{\mathsf{23}}{}^{\mathsf{u}} & (q+z)_{23}{}^{23}
\end{array}\right),
\end{array}
\end{equation}
 once more satisfying the relation 
\begin{equation}
\widetilde{\mathcal{O}}=-S^{-1}\mathcal{O}^{T}S.\label{eq: Flux Matrix Symplectic Relation-K3xT=0000B2}
\end{equation}
The notation was chosen such that the small letters in the charge
matrices indicate the fluxes they descend from. While their origin
should be clear for most cases, there are some caveats for the $F$-
and $Q$-fluxes: Here, the coefficients with unequal indices arise
from the flux components with two sub- respectively superscript $K3$
indices, while the coefficients with matching indices originate from
the components with one sub- and one superscript index in $K3$.

\subsubsection{Integrating over the Internal Space}

With everything formulated in the same framework as the Calabi-Yau
setting, it is now an easy exercise to integrate over the internal
manifold. Similar considerations as in subsection~\ref{sub:3.3.3} and \ref{sub:3.3.4}
eventually lead to the results 
\eq{
\begin{array}{ccl}
V_{\textrm{scalar, NS-NS}}^{{\scriptscriptstyle \left(\mathrm{IIA}\right)}} & = & {\displaystyle e^{-2\phi}\left[\vphantom{\frac{1}{2\mathcal{K}}}V^{\mathbb{I}}(\mathcal{O}^{T})_{\mathbb{I}}{}^{\mathbb{A}}\mathbb{M_{AB}}\mathcal{O}^{\mathbb{B}}{}_{\mathbb{J}}V^{\mathbb{J}}+W^{\mathbb{A}}(\widetilde{\mathcal{O}}^{T})_{\mathbb{A}}{}^{\mathbb{I}}\mathbb{N_{IJ}}\widetilde{\mathcal{O}}^{\mathbb{J}}{}_{\mathbb{B}}\overline{W}^{\mathbb{B}}\right.}\\
 & \vphantom{^{X^{X^{X^{X^{X^{X}}}}}}} & {\displaystyle \hphantom{e^{-2\phi}||}\left.-\frac{1}{4\mathcal{K}}\overline{W}^{\mathbb{A}}S_{\mathbb{AB}}\mathcal{O}^{\mathbb{B}}{}_{\mathbb{I}}\left(V^{\mathbb{I}}\overline{V}^{\mathbb{J}}+\overline{V}^{\mathbb{I}}V^{\mathbb{J}}\right)(\mathcal{O}^{T})_{\mathbb{J}}{}^{\mathbb{C}}(S^{T})_{\mathbb{CD}}\overline{W}^{\mathbb{D}}\right]}\\
 & \vphantom{^{X^{X^{X^{X^{X^{X}}}}}}} & +{\displaystyle \frac{1}{2}\left(\mathsf{G}_{\textrm{flux}}^{\mathbb{I}}+\mathcal{\widetilde{O}}^{\mathbb{I}}{}_{\mathbb{A}}\mathsf{C}_{0}^{\mathbb{A}}\right)\mathbb{N}{}_{\mathbb{IJ}}\left(\mathsf{G}_{\textrm{flux}}^{\mathbb{J}}+\mathcal{\widetilde{O}}^{\mathbb{J}}{}_{\mathbb{B}}\mathsf{C}_{0}^{\mathbb{B}}\right)},
\end{array}
}
for the type IIA case and 
\eq{
\begin{array}{ccl}
V_{\textrm{scalar, NS-NS}}^{{\scriptscriptstyle \left(\mathrm{IIA}\right)}} & = & {\displaystyle e^{-2\phi}\left[\vphantom{\frac{1}{2\mathcal{K}}}V^{\mathbb{I}}(\mathcal{O}^{T})_{\mathbb{I}}{}^{\mathbb{A}}\mathbb{M_{AB}}\mathcal{O}^{\mathbb{B}}{}_{\mathbb{J}}V^{\mathbb{J}}+W^{\mathbb{A}}(\widetilde{\mathcal{O}}^{T})_{\mathbb{A}}{}^{\mathbb{I}}\mathbb{N_{IJ}}\widetilde{\mathcal{O}}^{\mathbb{J}}{}_{\mathbb{B}}\overline{W}^{\mathbb{B}}\right.}\\
 & \vphantom{^{X^{X^{X^{X^{X^{X}}}}}}} & {\displaystyle \hphantom{e^{-2\phi}||}\left.-\frac{1}{4\mathcal{K}}\overline{W}^{\mathbb{A}}S_{\mathbb{AB}}\mathcal{O}^{\mathbb{B}}{}_{\mathbb{I}}\left(V^{\mathbb{I}}\overline{V}^{\mathbb{J}}+\overline{V}^{\mathbb{I}}V^{\mathbb{J}}\right)(\mathcal{O}^{T})_{\mathbb{J}}{}^{\mathbb{C}}(S^{T})_{\mathbb{CD}}\overline{W}^{\mathbb{D}}\right]}\\
 & \vphantom{^{X^{X^{X^{X^{X^{X}}}}}}} & {\displaystyle +\frac{1}{2}\left(\vphantom{\widetilde{\mathcal{O}}^{\mathbb{I}}{}_{\mathbb{A}}}\mathsf{G}_{\textrm{flux}}^{\mathbb{A}}+\mathcal{O}^{\mathbb{A}}{}_{\mathbb{I}}\mathsf{C}_{0}^{\mathbb{I}}\right)\mathbb{M}{}_{\mathbb{AB}}\left(\vphantom{\widetilde{\mathcal{O}}^{\mathbb{I}}{}_{\mathbb{A}}}\mathsf{G}_{\textrm{flux}}^{\mathbb{B}}+\mathcal{O}^{\mathbb{B}}{}_{\mathbb{J}}\mathsf{C}_{0}^{\mathbb{J}}\right)}
\end{array}
}
for the type IIB case. Comparing the results reveals the same set
of Mirror Transformations (\ref{eq:MirrorSymmetry-ScalarPotential})
already known from the Calabi-Yau setting (including a self-reflection
of the Hodge diamond. One can
furthermore see from the structure of the $K3\times T^{2}$ gauge
coupling matrices (\ref{eq:eq:Gauge-Coupling-Matrix-M-K2xT=0000B2})
and (\ref{eq:eq:Gauge-Coupling-Matrix-N-K2xT=0000B2}) that the mappings
$\mathbb{M}{}_{\mathbb{AB}}\leftrightarrow\mathbb{N}{}_{\mathbb{IJ}}$
can  be realized by
\begin{equation}
\label{mirror_001}
\tau\leftrightarrow\rho.
\end{equation}
In the bases employed above, the explicit mirror mapping between the moduli fields 
is not obvious. However, for $T^2$ mirror symmetry acts as \eqref{mirror_001} -- whereas
for the $K3$-part there are $19$ complex-structure moduli plus a complex scalar 
consisting of the $(2,0)$- and $(0,2)$-components of the $B$-field, which are 
interchanged with the $20$ complexified K\"ahler moduli.

\section{Obtaining the Full Action of $\mathcal{N}=2$ Gauged Supergravity\label{sec:5}}

We next show how the framework can be extended to the kinetic terms
by deriving the full four-dimensional action of $\mathcal{N}=2$ gauged
supergravity from the Calabi-Yau setting. The analysis for $K3\times T^{2}$
is more involved due to the appearance of additional Kaluza-Klein
vectors and will be saved for future work.

\subsection{NS-NS Sector}

Due to the vanishing first and fifth Betti numbers of Calabi-Yau three-folds,
there do not exist any non-trivial one- or five-cycles on $CY_{3}$.
It follows that all fields with effectively one or five free internal
indices acquire mass in four dimensions and can be ignored in the
low-energy limit. One immediate effect is that all components of the
metric and the Kalb-Ramond field with mixed indices can be discarded,
which drastically simplifies the expressions (\ref{eq: NSNS Gauge Field Strengths})
and (\ref{eq: NSNS Covariant Derivative}) building up the NS-NS contribution
(\ref{eq: Full Action NSNS-sector}) to the action,
\begin{equation}
\mathcal{B}_{\mu\nu\rho}\widetilde{\mathcal{F}}^{I}{}_{\mu\nu}\rightarrow0,\qquad\mathcal{B}_{\mu\nu\rho}\rightarrow\partial_{[\underline{\mu}}B_{\underline{\nu\rho}]},\qquad D_{\mu}\mathcal{H}_{IJ}\rightarrow\partial_{\mu}\mathcal{H}_{IJ},
\end{equation}
leaving us with

\begin{equation}
\begin{array}{@{}cll@{}}
S_{\textrm{NS-NS}} & {\displaystyle \!\!=\!\!\!\!\!} & {\displaystyle \frac{1}{2}\int_{M^{10}}\textrm{d}^{4}x\textrm{d}^{12}Y\,\sqrt{g^{\left(4\right)}}e^{-2\phi}\left[\vphantom{\frac{1}{4}}\right.}\\
 & \vphantom{^{X^{X^{X^{X^{X^{X}}}}}}} & {\displaystyle R^{\left(4\right)}+4g^{\mu\nu}\partial_{\mu}\hat{\phi}\partial_{\nu}\hat{\phi}-\frac{1}{12}g^{\mu\nu}g^{\rho\sigma}g^{\tau\lambda}\partial_{[\underline{\mu}}B_{\underline{\rho\tau}]}\partial_{[\underline{\nu}}B_{\underline{\sigma\lambda}]}+\frac{1}{8}g^{\mu\nu}\partial_{\mu}\mathcal{H}_{IJ}\partial_{\nu}\mathcal{H}^{IJ}}\\
 & \vphantom{^{X^{X^{X^{X^{X^{X}}}}}}} & +\sqrt{g^{\left(6\right)}}{\displaystyle \mathcal{F}_{IJK}\mathcal{F}_{I'J'K'}\left(-\frac{1}{12}\mathcal{H}^{II'}\mathcal{H}^{JJ'}\mathcal{H}^{KK'}+\frac{1}{4}\mathcal{H}^{II'}\eta^{JJ'}\eta^{KK'}-\frac{1}{6}\eta^{II'}\eta^{JJ'}\eta^{KK'}\right)}\\
 & \vphantom{^{X^{X^{X^{X^{X^{X}}}}}}} & \left.{\displaystyle +\sqrt{g^{\left(6\right)}}\mathcal{F}_{I}\mathcal{F}_{I'}\left(\vphantom{\frac{1}{4}S^{AA'}}\eta^{II'}-\mathcal{H}^{II'}\right)}\right].
\end{array}\label{eq: NS-NS Action on Calabi-Yau}
\end{equation}
The first three terms are known from normal type II supergravities,
while the last two lines were shown to correctly give rise to the
scalar potential of $\mathcal{N}=2$ gauged supergravity in section~\ref{sec:3}.
It is therefore to be expected that the remaining term $\frac{1}{8}g^{\mu\nu}\partial_{\mu}\mathcal{H}_{IJ}\partial_{\nu}\mathcal{H}^{IJ}$
gives rise to the kinetic terms of the K\"ahler class and complex structure
moduli. Indeed, inserting (\ref{eq: Generalized Metric}) and using
antisymmetry of the Kalb-Ramond field, one obtains 
\begin{equation}
\frac{1}{8}\partial_{\mu}\mathcal{H}_{IJ}\partial^{\mu}\mathcal{H}^{IJ}=\frac{1}{4}g^{\mu\nu}\left(\partial_{\mu}g_{ij}\partial_{\nu}g^{ij}-g^{ik}g^{jl}\partial_{\mu}b_{ij}\partial_{\nu}b_{kl}\right).
\end{equation}
The first term encodes the dynamics of the internal metric, which
is fully described by its fluctuations. Similarly to Calabi-Yau compactifications
of supergravity theories, these can be expanded in terms of the K\"ahler
class and complex structure moduli. For the Kalb-Ramond field, one
can proceed analogously by using the expansion (\ref{eq: Expansion K=0000E4hler Form and B-Field}),
which combines with the K\"ahler class moduli to form the complexified
K\"ahler moduli.

Using this as a starting point, the rest of the dimensional reduction
follows the same principles as in Calabi-Yau compactifications of
type II supergravities. A review of the topic in general can be found
in chapter two of \cite{Gurrieri:2003st}, a similar discussion concerning
manifolds with $SU(3)\!\times\! SU(3)$ structure in \cite{Cassani:2007pq,Cassani:2008rb}.
After switching to Einstein frame via Weyl-rescaling 
\begin{equation}
g_{\mu\nu}\rightarrow e^{-2\phi}g_{\mu\nu}\label{eq: Weyl-Rescaling}
\end{equation}
of the external metric, one eventually arrives at  
\begin{equation}
S_{\textrm{NS-NS, kin}}=\int_{M^{1,4}}\frac{1}{2}R^{\left(4\right)}\star\mathbf{1}^{\left(4\right)}-\textrm{d}\phi\wedge\star\textrm{d}\phi-\frac{1}{2}e^{-4\phi}\textrm{d}B\wedge\star\textrm{d}B-g_{\mathsf{ij}}\textrm{d}t^{\mathsf{i}}\wedge\star\textrm{d}t^{\mathsf{j}}-g_{\mathsf{a\bar{b}}}\textrm{d}z^{\mathsf{a}}\wedge\star\textrm{d}\bar{z}^{\mathsf{\bar{b}}},
\end{equation}
where we switched to differential form notation for the sake of clarity.
The expansion coefficients $t^{\mathsf{i}}$ (cf. (\ref{eq: Expansion Complexified K=0000E4hler Form}))
parametrize the K\"ahler class moduli space $M_{\textrm{KC }}$with
metric $g_{\mathsf{ij}}$, and $z^{\mathsf{a}}$ the complex structure
moduli space $M_{\textrm{CS}}$ with metric $g_{\mathsf{a\bar{b}}}$.

\subsection{R-R Sector}

The most obvious way to proceed for the R-R sector would be to evaluate
the corresponding action of (\ref{eq: DFT: RR Action}) in four dimensions
and then implement the duality relations (\ref{eq: Duality Relations General})
in order to recover the action of $\mathcal{N}=2$ gauged supergravity.
Since handling these duality relations in four dimensions turns out
rather demanding, we will, however, pursue a different approach and
consider the reduced equations of motion instead. Notice that this
has been done for compactifications on $SU(3)\!\times\! SU(3)$ structure
manifolds in \cite{Cassani:2008rb}, and many of the following technical
steps are close to the ones employed in this work.

\subsubsection{Type IIA Setting}

\subsubsection*{Relation to Democratic Type IIA Supergravity}

Starting from (\ref{eq: DFT: RR Action}), a first step is to write
down the pseudo-action explicitly in terms of poly-form fields and
obtain a form similar to (\ref{eq:RR-IIA-ScalarPotential}). In doing
so, we again neglect all cohomologically trivial expressions and,
thus, take into account only those components with zero, two, three,
four or six internal indices. Applying the methods presented in section~4 of \cite{Blumenhagen:2013hva} to evaluate the expressions found
in (\ref{eq: DFT: RR Action}) and arranging the (now ten-dimensional)
$\hat{C}$-fields and R-R fluxes in poly-forms 
\begin{equation}
\begin{array}{lcl}
\mathcal{\hat{C}}^{{\scriptscriptstyle \left(\mathrm{IIA}\right)}} & {\displaystyle \!\!\!\!=\!\!\!\!} & \hat{C}_{1}+\hat{C}_{3}+\hat{C}_{5}+\hat{C}_{7}+\hat{C}_{9},\\
\mathcal{G}^{{\scriptscriptstyle \left(\mathrm{IIA}\right)}} & {\displaystyle \!\!\!\!=\!\!\!\!} & G_{0}+G_{2}+G_{4}+G_{6},\vphantom{^{X^{X^{X^{X^{X^{X}}}}}}}
\end{array}
\end{equation}
we can define 
\begin{equation}
\mathfrak{\hat{G}}^{{\scriptscriptstyle \left(\mathrm{IIA}\right)}}=e^{-\hat{B}}\mathcal{G}^{{\scriptscriptstyle \left(\mathrm{IIA}\right)}}+\hat{\mathfrak{D}}\mathcal{\hat{C}}^{{\scriptscriptstyle \left(\mathrm{IIA}\right)}}=e^{-\hat{B}}\mathcal{G}^{{\scriptscriptstyle \left(\mathrm{IIA}\right)}}+e^{-\hat{B}}\hat{\mathcal{D}}\left(e^{\hat{B}}\mathcal{\hat{C}}^{{\scriptscriptstyle \left(\mathrm{IIA}\right)}}\right),\label{eq: R-R Poly-Form IIA}
\end{equation}
with the ten-dimensional twisted differential 
\begin{equation}
\hat{\mathcal{D}}=\hat{\textrm{d}}-H\wedge-F\circ-Q\bullet-R\llcorner-Y\wedge-Z\blacktriangledown,
\end{equation}
to write the complete type IIA R-R pseudo-Lagrangian (\ref{eq: DFT: RR Action})
as 
\begin{equation}
\star\mathcal{L}_{\textrm{R-R}}=-\frac{1}{2}\hat{\mathfrak{G}}^{{\scriptscriptstyle \left(\mathrm{IIA}\right)}}\wedge\star\hat{\mathfrak{G}}^{{\scriptscriptstyle \left(\mathrm{IIA}\right)}}.\label{eq: IIA R-R Lagrangian reformulated}
\end{equation}
Notice that this resembles the R-R sector of democratic type IIA supergravity
\cite{Bergshoeff:2001pv}, up to an exchange of signs in the exponential
factors and the inclusion of additional background fluxes. Since the
action depends on all R-R potentials explicitly, their duality relations
(\ref{eq: Duality Relations General}) have to be imposed by hand.
For the type IIA case, these are equivalent to 
\begin{equation}
\mathfrak{\hat{G}}^{{\scriptscriptstyle \left(\mathrm{IIA}\right)}}=\lambda\left(\star\mathfrak{\hat{G}}^{{\scriptscriptstyle \left(\mathrm{IIA}\right)}}\right),\label{eq: 10D Duality Constraints}
\end{equation}
where $\lambda$ denotes the involution operator defined in (\ref{eq:MukaiPairing-Involution}).
Varying the corresponding action of (\ref{eq: IIA R-R Lagrangian reformulated})
with respect to the R-R fields, one obtains the poly-form equation
\begin{equation}
\left(\hat{\textrm{d}}-\textrm{d}\hat{B}\wedge+\mathfrak{H}\wedge+\mathfrak{F}\circ+\mathfrak{Q}\bullet+\mathfrak{R}\llcorner\right)\star\mathfrak{\hat{G}}^{{\scriptscriptstyle \left(\mathrm{IIA}\right)}}=0.\label{eq: IIA R-R Equations of Motion old}
\end{equation}
Employing the duality relations (\ref{eq: 10D Duality Constraints}),
these can be recast to take the form of the Bianchi identities 
\begin{equation}
{\displaystyle e^{-\hat{B}}\hat{\mathcal{D}}\left(e^{\hat{B}}\mathfrak{\hat{G}}^{{\scriptscriptstyle \left(\mathrm{IIA}\right)}}\right)=0},\label{eq: IIA R-R Equations of Motion}
\end{equation}
where the prefactor of $e^{-\hat{B}}$ was included for later convenience.
They are automatically satisfied when imposing nilpotency of the twisted
differential by hand, and the nontrivial equations of motion in four
dimensions can be obtained by implementation of the duality constraints
(\ref{eq: 10D Duality Constraints}).

\subsubsection*{Reduced Equations of Motion }

In order to evaluate the equations of motion in four dimensions, we
next express the appearing objects in a way that the framework of
special geometry presented in subsection~\ref{sub:3.3.1} can be applied. This
can be achieved by switching to the so-called ``$A$-basis''%
\footnote{The naming was chosen based on the notation used in the original work
\cite{Bergshoeff:2001pv} and will not play any role in the upcoming
discussion.%
} introduced in \cite{Bergshoeff:2001pv}, for which we define
\begin{equation}
e^{\hat{B}}\mathcal{C}^{{\scriptscriptstyle \left(\mathrm{IIA}\right)}}=\left(\mathsf{C}_{1}^{\mathsf{I}}+\mathsf{C}_{3}^{\mathsf{I}}\right)\omega_{\mathsf{I}}+\left(\mathsf{C}_{0}^{\mathsf{A}}+\mathsf{C}_{2}^{\mathsf{A}}+\mathsf{C}_{4}^{\mathsf{A}}\right)\alpha_{\mathsf{A}}-\left(\vphantom{\mathsf{C}_{4}^{\mathsf{A}}}\mathsf{C}_{0\,\mathsf{A}}+\mathsf{C}_{2\,\mathsf{A}}+\mathsf{C}_{4\,\mathsf{A}}\right)\beta^{\mathsf{A}}+\left(\vphantom{\mathsf{C}_{4}^{\mathsf{A}}}\mathsf{C}_{1\,\mathsf{I}}+\mathsf{C}_{3\,\mathsf{I}}\right)\widetilde{\omega}^{\mathsf{I}}\label{eq: IIA R-R 4D C-Poly-Form}
\end{equation}
and 
\begin{equation}
G_{0}=\mathsf{G}_{\textrm{flux\,0}}\widetilde{\omega}^{0},\quad G_{2}=\mathsf{G}_{\textrm{flux}}^{\mathsf{i}}\omega_{\mathsf{i}},\quad G_{4}=\mathsf{G}{}_{\textrm{flux}\,\mathsf{i}}\widetilde{\omega}^{\mathsf{i}},\quad G_{6}=\mathsf{G}_{\textrm{flux}\,}^{0}\omega_{\text{0}},\label{eq: IIA R-R 4D G-Forms}
\end{equation}
where the objects $\mathsf{C}_{n}$ now denote differential $n$-forms
living in four dimensional spacetime. The R-R poly-form (\ref{eq: R-R Poly-Form IIA})
can then be expressed as 
\begin{equation}
\mathfrak{\hat{G}}^{{\scriptscriptstyle \left(\mathrm{IIA}\right)}}=e^{-\hat{B}}\hat{\mathsf{G}}^{{\scriptscriptstyle \left(\mathrm{IIA}\right)}}=e^{-\hat{B}}\left(\hat{\mathsf{G}}_{0}^{{\scriptscriptstyle \left(\mathrm{IIA}\right)}}+\hat{\mathsf{G}}_{2}^{{\scriptscriptstyle \left(\mathrm{IIA}\right)}}+\hat{\mathsf{G}}_{4}^{{\scriptscriptstyle \left(\mathrm{IIA}\right)}}+\hat{\mathsf{G}}_{6}^{{\scriptscriptstyle \left(\mathrm{IIA}\right)}}+\hat{\mathsf{G}}_{8}^{{\scriptscriptstyle \left(\mathrm{IIA}\right)}}+\hat{\mathsf{G}}_{10}^{{\scriptscriptstyle \left(\mathrm{IIA}\right)}}\right).\label{eq: IIA R-R Poly-Form A-Basis}
\end{equation}
Using the flux matrices (\ref{eq: Flux Matrix Definitions}) and the
relations (\ref{eq: Flux Matrix Axtions on Cohomology Bases - Components}),
the appearing poly-forms can be expanded in terms four-dimensional
differential form fields, 
\begin{equation}
\begin{array}{ccl}
\hat{\mathsf{G}}_{0}^{{\scriptscriptstyle \left(\mathrm{IIA}\right)}} & {\displaystyle \!\!\!\!=\!\!\!\!} & \mathsf{G}_{0\,0}\widetilde{\omega}^{0},\\
\hat{\mathsf{G}}_{2}^{{\scriptscriptstyle \left(\mathrm{IIA}\right)}} & {\displaystyle \!\!\!\!=\!\!\!\!} & {\displaystyle \mathsf{G}_{2\,0}\widetilde{\omega}^{0}}+{\displaystyle \mathsf{G}_{0}^{\mathsf{i}}\omega_{\mathsf{i}}},\vphantom{^{X^{X^{X^{X^{X^{X}}}}}}}\\
\hat{\mathsf{G}}_{4}^{{\scriptscriptstyle \left(\mathrm{IIA}\right)}} & {\displaystyle \!\!\!\!=\!\!\!\!} & \mathsf{G}_{4\,0}\widetilde{\omega}^{0}+{\displaystyle \mathsf{G}_{2}^{\mathsf{i}}\wedge\omega_{\mathsf{i}}}-{\displaystyle \mathsf{G}_{1}^{\mathsf{A}}\wedge\alpha_{\mathsf{A}}}+{\displaystyle \mathsf{G}_{1\,\mathsf{A}}\wedge\beta^{\mathsf{A}}}+{\displaystyle \mathsf{G}_{0\,\mathsf{i}}\widetilde{\omega}^{\mathsf{i}}},\vphantom{^{X^{X^{X^{X^{X^{X}}}}}}}\\
\hat{\mathsf{G}}_{6}^{{\scriptscriptstyle \left(\mathrm{IIA}\right)}} & {\displaystyle \!\!\!\!=\!\!\!\!} & {\displaystyle \mathsf{G}_{4}^{\mathsf{i}}\wedge\omega_{\mathsf{i}}}-{\displaystyle \mathsf{G}_{3}^{\mathsf{A}}\wedge\alpha_{\mathsf{A}}}+{\displaystyle \mathsf{G}_{3\,\mathsf{A}}\wedge\beta^{\mathsf{A}}}+{\displaystyle {\displaystyle \mathsf{G}_{2\,\mathsf{i}}}\wedge\widetilde{\omega}^{\mathsf{i}}}+{\displaystyle \mathsf{G}_{0}^{0}}\wedge\omega_{0},\vphantom{^{X^{X^{X^{X^{X^{X}}}}}}}\\
\hat{\mathsf{G}}_{8}^{{\scriptscriptstyle \left(\mathrm{IIA}\right)}} & {\displaystyle \!\!\!\!=\!\!\!\!} & {\displaystyle \mathsf{G}_{4\,\mathsf{i}}\wedge\widetilde{\omega}^{\mathsf{i}}}+{\displaystyle {\displaystyle \mathsf{G}_{2}^{0}}\wedge\omega_{0}},\vphantom{^{X^{X^{X^{X^{X^{X}}}}}}}\\
\hat{\mathsf{G}}_{10}^{{\scriptscriptstyle \left(\mathrm{IIA}\right)}} & {\displaystyle \!\!\!\!=\!\!\!\!} & {\displaystyle {\displaystyle \mathsf{G}_{4}^{0}}\wedge\omega_{0}},\vphantom{^{X^{X^{X^{X^{X^{X}}}}}}}
\end{array}\label{eq: IIA R-R Forms in A-Basis}
\end{equation}
with the expansion coefficients given by 
\begin{equation}
\begin{array}{lcl}
\mathsf{G}_{0}^{\mathbb{I}} & {\displaystyle \!\!\!\!=\!\!\!\!} & \mathsf{G}_{\textrm{flux}}^{\mathbb{I}}+\mathcal{\widetilde{\mathcal{O}}}^{\mathbb{I}}{}_{\mathbb{A}}\mathsf{C}_{0}^{\mathbb{A}},\\
\mathsf{G}_{1}^{\mathbb{A}} & {\displaystyle \!\!\!\!=\!\!\!\!} & \textrm{d}\mathsf{C}_{0}^{\mathbb{A}}+\mathcal{O}^{\mathbb{A}}{}_{\mathbb{I}}\mathsf{C}_{1}^{\mathbb{I}},\vphantom{^{X^{X^{X^{X^{X^{X}}}}}}}\\
\mathsf{G}_{2}^{\mathbb{I}} & {\displaystyle \!\!\!\!=\!\!\!\!} & {\displaystyle \textrm{d}\mathsf{C}_{1}^{\mathbb{I}}}+\mathcal{\widetilde{\mathcal{O}}}^{\mathbb{I}}{}_{\mathbb{A}}\mathsf{\mathsf{C}_{2}^{\mathbb{A}}},\vphantom{^{X^{X^{X^{X^{X^{X}}}}}}}\\
\mathsf{G}_{3}^{\mathbb{A}} & {\displaystyle \!\!\!\!=\!\!\!\!} & \textrm{d}\mathsf{C}_{2}^{\mathbb{A}}+\mathcal{O}^{\mathbb{A}}{}_{\mathbb{I}}\mathsf{C}_{3}^{\mathbb{I}},\vphantom{^{X^{X^{X^{X^{X^{X}}}}}}}\\
\mathsf{G}_{4}^{\mathbb{I}} & {\displaystyle \!\!\!\!=\!\!\!\!} & {\displaystyle \textrm{d}\mathsf{C}_{3}^{\mathbb{I}}}+\mathcal{\widetilde{\mathcal{O}}}^{\mathbb{I}}{}_{\mathbb{A}}\mathsf{\mathsf{C}_{4}^{\mathbb{A}}},.\vphantom{^{X^{X^{X^{X^{X^{X}}}}}}}
\end{array}\label{eq: IIA R-R Field Expansion Coefficients}
\end{equation}
 This expansion can be used as a starting point to compute the reduced
equations of motion descending from (\ref{eq: IIA R-R Equations of Motion}).
Substituting the definition (\ref{eq: IIA R-R Poly-Form A-Basis})
into (\ref{eq: IIA R-R Equations of Motion}), one obtains in $A$-basis
notation 
\begin{equation}
\hat{\mathcal{D}}\hat{\mathsf{G}}^{{\scriptscriptstyle \left(\mathrm{IIA}\right)}}=0.
\end{equation}
After separating different components and integrating over $CY_{3}$,
this gives rise to the four-dimensional equations of motion 
\begin{equation}
\begin{array}{rrl}
\mathcal{O}^{\mathbb{A}}{}_{\mathbb{I}}{\displaystyle \mathsf{G}_{0}^{\mathbb{I}}} & {\displaystyle \!\!\!\!=\!\!\!\!} & 0,\\
\textrm{d}\mathsf{G}_{0}^{\mathbb{I}}-\mathcal{\widetilde{\mathcal{O}}}^{\mathbb{I}}{}_{\mathbb{A}}\mathsf{G}_{1}^{\mathbb{A}} & {\displaystyle \!\!\!\!=\!\!\!\!} & 0,\vphantom{^{X^{X^{X^{X^{X^{X}}}}}}}\\
\textrm{d}\mathsf{G}_{1}^{\mathbb{A}}-\mathcal{O}^{\mathbb{A}}{}_{\mathbb{I}}\mathsf{G}_{2}^{\mathbb{I}} & {\displaystyle \!\!\!\!=\!\!\!\!} & 0,\vphantom{^{X^{X^{X^{X^{X^{X}}}}}}}\\
\textrm{d}\mathsf{G}_{2}^{\mathbb{I}}-\mathcal{\widetilde{\mathcal{O}}}^{\mathbb{I}}{}_{\mathbb{A}}\mathsf{G}_{3}^{\mathbb{A}} & {\displaystyle \!\!\!\!=\!\!\!\!} & 0,\vphantom{^{X^{X^{X^{X^{X^{X}}}}}}}\\
\textrm{d}\mathsf{G}_{3}^{\mathbb{A}}-\mathcal{O}^{\mathbb{A}}{}_{\mathbb{I}}\mathsf{G}_{4}^{\mathbb{I}} & {\displaystyle \!\!\!\!=\!\!\!\!} & 0.\vphantom{^{X^{X^{X^{X^{X^{X}}}}}}}
\end{array}\label{eq: IIA R-R 4D Equations of Motion}
\end{equation}
Since the Kalb-Ramond field couples with the $C$-fields, one furthermore
has to take into account the (non-trivial) equation of motion obtained
by varying the complete ten-dimensional action with respect to $\hat{B}$,
which yields an eight-form equation
\begin{equation}
\textrm{d}\left(e^{-4\phi}\star\textrm{d}B\right)+\frac{1}{2}\left[\mathfrak{\hat{G}}^{{\scriptscriptstyle \left(\mathrm{IIA}\right)}}\wedge\star\mathfrak{\hat{G}}^{{\scriptscriptstyle \left(\mathrm{IIA}\right)}}\right]_{8}=0.\label{eq: IIA B-Field Equation of Motion}
\end{equation}

\subsubsection*{Reduced Duality Constraints}

Our aim is now to implement the duality constraints (\ref{eq: 10D Duality Constraints})
into the equations of motion (\ref{eq: IIA R-R 4D Equations of Motion})
and (\ref{eq: IIA B-Field Equation of Motion}) in an appropriate
way in order to recover the $D=4$ $\mathcal{N}=2$ gauged supergravity
action found in formula (35) of \cite{DAuria:2007axr}. In particular,
we want the fundamental (but not necessarily propagating) degrees
of freedom to be given by%
\footnote{We preliminarily adopt the notation of \cite{DAuria:2007axr} and
identify the correct definitions in the course of the following discussion.%
} $2h^{1,2}+2$ scalars $\hat{Z}^{\mathbb{A}}$, $h^{1,1}+1$ one-forms
$A_{1}^{\mathsf{I}}$, $2h^{1,2}+2$ two-forms $B^{\mathbb{A}}$ and
the external Kalb-Ramond field $B$.

Up to conventions, the reduced duality constraints can be obtained
completely analogous to \cite{Cassani:2008rb}. Inserting the expansion
\begin{equation}
e^{-\hat{B}}\hat{\mathsf{G}}^{{\scriptscriptstyle \left(\mathrm{IIA}\right)}}=e^{-b^{\mathsf{i}}\omega_{\mathsf{i}}}\left(K^{\mathsf{I}}\omega_{\mathsf{I}}+K_{\mathsf{I}}\widetilde{\omega}^{\mathsf{I}}+L^{\mathsf{A}}\alpha_{\mathsf{A}}-L_{\mathsf{A}}\beta^{\mathsf{A}}\right)\label{eq: IIA Duality Relations - Expansion}
\end{equation}
into (\ref{eq: 10D Duality Constraints}), one obtains 
\begin{equation}
K^{\mathsf{I}}\omega_{\mathsf{I}}+K_{\mathsf{I}}\widetilde{\omega}^{\mathsf{I}}+L^{\mathsf{A}}\alpha_{\mathsf{A}}-L_{\mathsf{A}}\beta^{\mathsf{A}}=-\star\lambda\left(K^{\mathsf{I}}\right)\star_{b}\omega_{\mathsf{I}}-\star\lambda\left(K_{\mathsf{I}}\right)\star_{b}\widetilde{\omega}^{\mathsf{I}}-\star\lambda\left(L^{\mathsf{A}}\right)\star_{b}\alpha_{\mathsf{A}}+\star\lambda\left(L_{\mathsf{A}}\right)\star_{b}\beta^{\mathsf{A}}.
\end{equation}
Applying the operators $\int_{CY_{3}}\left\langle \widetilde{\omega}^{\mathsf{I}},\star_{b}\cdot\right\rangle $
and $\int_{CY_{3}}\left\langle \beta^{\mathsf{A}},\star_{b}\cdot\right\rangle $
to both sides of the equation and using (\ref{eq: Gauge Coupling Matrices: M})-(\ref{eq: Gauge Coupling Matrices: N}),
one can separate different internal components and obtain the reduced
duality constraints 
\begin{equation}
\begin{array}{ccl}
K_{\mathsf{I}} & {\displaystyle \!\!\!\!=\!\!\!\!} & -\textrm{Im}\mathcal{N}_{\mathsf{IJ}}\star\lambda\left(K^{\mathsf{I}}\right)+\textrm{Re}\mathcal{N}_{\mathsf{IJ}}K^{\mathsf{I}},\\
L_{\mathsf{A}} & {\displaystyle \!\!\!\!=\!\!\!\!} & -\textrm{Im}\mathcal{M}_{\mathsf{AB}}\star\lambda\left(L^{\mathsf{A}}\right)+\textrm{Re}\mathcal{M}_{\mathsf{AB}}L^{\mathsf{A}}.\vphantom{^{X^{X^{X^{X^{X^{X}}}}}}}
\end{array}\label{eq: IIA Duality Constraints original}
\end{equation}
The $K$- and $L$-poly-forms still contain four-dimension differential
forms of different degrees. Separating components by hand and performing
a Weyl-rescaling (\ref{eq: Weyl-Rescaling}) according to (\ref{eq: Weyl-Rescaling}),
we eventually arrive at 
\begin{equation}
\begin{array}{rcl}
{\displaystyle \mathsf{G}_{2\,\mathsf{I}}-B\mathsf{G}_{0\,\mathsf{I}}} & {\displaystyle \!\!\!\!=\!\!\!\!} & {\displaystyle \textrm{Im}\mathcal{N}_{\mathsf{IJ}}\star\left(\mathsf{G}_{2}^{\mathsf{J}}-B\wedge\mathsf{G}_{0}^{\mathsf{J}}\right)+\textrm{Re}\mathcal{N}_{\mathsf{IJ}}\left(\mathsf{G}_{2}^{\mathsf{J}}-B\wedge\mathsf{G}_{0}^{\mathsf{J}}\right),}\\
{\displaystyle \mathsf{G}_{4}^{\mathbb{I}}-B\wedge\mathsf{G}_{2}^{\mathbb{I}}+\frac{1}{2}B^{2}\mathsf{G}_{0}^{\mathbb{I}}} & {\displaystyle \!\!\!\!=\!\!\!\!} & -{\displaystyle e^{4\phi}\left(S^{-1}\right){}^{\mathbb{IJ}}\mathbb{N}{}_{\mathbb{JK}}\mathsf{G}_{0}^{\mathbb{K}}}\star\mathbf{1}^{\left(4\right)},\vphantom{^{X^{X^{X^{X^{X^{X}}}}}}}\\
\mathsf{G}_{3}^{\mathbb{A}}-B\wedge\mathsf{G}_{1}^{\mathbb{A}} & {\displaystyle \!\!\!\!=\!\!\!\!} & e^{2\phi}\left(S^{-1}\right)^{\mathbb{AB}}\mathbb{M}{}_{\mathbb{BC}}\star\mathsf{G}_{1}^{\mathbb{C}}.\vphantom{^{X^{X^{X^{X^{X^{X}}}}}}}
\end{array}\label{eq: 4D Duality Constraints}
\end{equation}

\subsubsection*{Evaluating the Equations of Motion - Constraints on Fluxes}

Before implementing the duality constraints, it makes sense to take
a closer look at the first line of (\ref{eq: IIA R-R 4D Equations of Motion}).
Unlike the remaining equations of motion, the left hand side does
not vanish trivially when imposing the nilpotency conditions (\ref{eq: Nilpotency Constraint - Matrices}).
Instead, we are left with an additional constraint, which after integration
over $CY_{3}$ via $\int_{CY_{3}}\left\langle \Sigma_{\mathbb{I}},\cdot\right\rangle $
reads 
\begin{equation}
\mathcal{O}^{\mathbb{A}}{}_{\mathbb{I}}\mathsf{G}_{\textrm{flux}}^{\mathbb{I}}=0\label{eq: IIA Flux Constraints}
\end{equation}
and resembles the conditions found in (37) of \cite{DAuria:2007axr}.
Notice that these arise automatically from the DFT framework and do
not have to be imposed by hand.

\subsubsection*{Evaluating the Equations of Motion - $\mathsf{C}_{1}^{\mathsf{I}}$}

The simplest equations of motion to derive are those of the one-forms
$A_{1}^{\mathsf{I}}$, which we will be able to identify with the
fields $\mathsf{C}_{1}^{\mathsf{I}}$ at the end of this subsection.
In order to get some intuition for the way of proceeding, we will
treat this example in more detail. The underlying idea can then easily
be transferred to the remaining degrees of freedom.

Many of the technical steps in the following discussion are again
very close to the ones employed in \cite{Cassani:2008rb}. The essential
difference is that in the present setting, the expressions (\ref{eq: IIA R-R Field Expansion Coefficients})
are completely determined by the DFT action, whereas in the case of
$SU(3)\!\times\! SU(3)$ manifolds, their structure is governed only
by the equations of motion (\ref{eq: IIA R-R 4D Equations of Motion}).
This leads to slight redefinitions of the encountered objects, and
we will in particular go without additional assumptions regarding
the flux matrices (\ref{eq: Flux Matrix Definitions}) and the existence
of corresponding operators.

Before presenting explicit calculations, it is helpful to motivate
our ansatz to derive the desired equations of motion for $\mathsf{C}_{1}^{\mathsf{I}}$.
For this purpose, we take a look at the corresponding expression obtained
by varying the action found in \cite{DAuria:2007axr} with respect
to the $A_{1}^{\mathsf{I}}$, 
\begin{equation}
\textrm{d}\left(\textrm{Im}\mathcal{N}_{\mathsf{IJ}}\star\mathsf{F}_{2}^{\mathsf{J}}+\textrm{Re}\mathcal{N}_{\mathsf{IJ}}\mathsf{F}_{2}^{\mathsf{J}}-e_{\mathsf{I}\,\mathbb{A}}B^{\mathbb{A}}-c_{\mathsf{I}}B\right)=0.\label{eq: IIA R-R C1 EoM Expected Structure}
\end{equation}
The first two terms strongly resemble the first line of (\ref{eq: 4D Duality Constraints}),
and since $\mathsf{G}_{0\,\mathsf{I}}$ contains only expressions
which we expect to appear in the four-dimensional action, a viable
ansatz might be to replace $\mathsf{G}_{2\,\mathsf{I}}$ in one of
the equations of motion (\ref{eq: IIA R-R 4D Equations of Motion}).
Reverting to the expected structure (\ref{eq: IIA R-R C1 EoM Expected Structure})
of the final equation of motion once more, we see that the most obvious
way to do this is by considering the lower-index components of
the fourth equation of motion of (\ref{eq: IIA R-R 4D Equations of Motion}).
Applying the nilpotency constraint (\ref{eq: Nilpotency Constraint - Matrices})
of $\mathcal{D}$ and integrating over $CY_{3}$ similarly to the
previous case, this can be written as 
\begin{equation}
\textrm{d}\mathsf{G}_{2\,\mathsf{I}}-\mathcal{\widetilde{\mathcal{O}}}{}_{\mathsf{I}\,\mathbb{A}}\textrm{d}\mathsf{C}_{2}^{\mathbb{A}}=0.
\end{equation}
Using the first line of (\ref{eq: 4D Duality Constraints}) to substitute
$\mathsf{G}_{2\,\mathsf{I}}$ yields 
\begin{equation}
\textrm{d}\left(\textrm{Im}\mathcal{N}_{\mathsf{IJ}}\star\mathsf{F}_{2}^{\mathsf{J}}+\textrm{Re}\mathcal{N}_{\mathsf{IJ}}\mathsf{F}_{2}^{\mathsf{J}}-\mathcal{\widetilde{\mathcal{O}}}{}_{\mathsf{I}\,\mathbb{A}}\mathsf{C}_{2}^{\mathbb{A}}+B\wedge\mathsf{G}_{0\,\mathsf{I}}\right)=0,\label{eq: Equation of Motion C1 - Proto}
\end{equation}
where
\begin{equation}
\mathsf{F}_{2}^{\mathsf{I}}:=\mathsf{G}_{2}^{\mathsf{I}}-B\wedge\mathsf{G}_{0}^{\mathsf{I}}.
\end{equation}
This can be further simplified by pulling out a factor of $B\wedge$
from the definition (\ref{eq: IIA R-R 4D C-Poly-Form}) of $\mathsf{C}_{2}^{\mathbb{A}}$.
We do this by employing the alternative expansion 
\begin{equation}
\begin{array}{ccl}
e^{b^{\mathsf{i}}\omega_{\mathsf{i}}}\mathcal{\mathcal{\hat{C}}^{{\scriptscriptstyle \left(\mathrm{IIA}\right)}}} & {\displaystyle \!\!\!\!=\!\!\!\!} & \left(\mathsf{\widetilde{C}}_{1}^{\mathsf{I}}+\mathsf{\widetilde{C}}_{3}^{\mathsf{I}}\right)\omega_{\mathsf{I}}\\
 & \vphantom{^{X^{X^{X^{X^{X^{X}}}}}}} & +\left(\mathsf{\widetilde{C}}_{0}^{\mathsf{A}}+\mathsf{\mathsf{\widetilde{C}}_{2}^{\mathsf{A}}}+\mathsf{\widetilde{C}}_{4}^{\mathsf{A}}\right)\alpha_{\mathsf{A}}-\left(\mathsf{\widetilde{C}}_{0\,\mathsf{A}}+\mathsf{\widetilde{C}}_{2\,\mathsf{A}}+\mathsf{\widetilde{C}}_{4\,\mathsf{A}}\right)\beta^{\mathsf{A}}\\
 & \vphantom{^{X^{X^{X^{X^{X^{X}}}}}}} & +\left(\mathsf{\widetilde{C}}_{1\,\mathsf{I}}+\mathsf{\widetilde{C}}_{3\,\mathsf{I}}\right)\widetilde{\omega}^{\mathsf{I}},
\end{array}\label{eq: IIA R-R 4D C-Poly-Form alternative}
\end{equation}
from which we infer the relation 
\begin{equation}
\mathsf{C}_{2}^{\mathbb{A}}=\mathsf{\widetilde{C}}_{2}^{\mathbb{A}}+B\wedge\mathsf{C}_{0}^{\mathbb{A}},
\end{equation}
while the other fields appearing in (\ref{eq: Equation of Motion C1 - Proto})
remain unaffected. Inserting the definitions (\ref{eq: IIA R-R Field Expansion Coefficients})
for the $\mathsf{G}_{0\,\mathsf{I}}$, we are left with 
\begin{equation}
\mathsf{F}_{2}^{\mathsf{I}}={\displaystyle \textrm{d}\mathsf{C}_{1}^{\mathbb{\mathsf{I}}}}+\mathcal{\widetilde{\mathcal{O}}}^{\mathbb{\mathsf{I}}}{}_{\mathbb{A}}\mathsf{\widetilde{C}}_{2}^{\mathbb{A}}-B\wedge\mathsf{G}_{\textrm{flux}}^{\mathbb{\mathsf{I}}}
\end{equation}
and the equations of motion 
\begin{equation}
\textrm{d}\left(\textrm{Im}\mathcal{N}_{\mathsf{IJ}}\star\mathsf{F}_{2}^{\mathsf{J}}+\textrm{Re}\mathcal{N}_{\mathsf{IJ}}\mathsf{F}_{2}^{\mathsf{J}}-\mathcal{\widetilde{\mathcal{O}}}{}_{\mathsf{I}\,\mathbb{A}}\mathsf{\widetilde{C}}_{2}^{\mathbb{A}}+B\wedge\mathsf{G}_{\mathsf{I}\,\textrm{flux}}\right)=0,
\end{equation}
which, up to sign convention for $B$, take precisely the form of
the corresponding ones obtained from the action of \cite{DAuria:2007axr}
when identifying $A_{1}^{\mathsf{I}}=\mathsf{C}_{1}^{\mathsf{I}}$,
$B^{\mathbb{A}}=\mathsf{\widetilde{C}}_{2}^{\mathbb{A}}$, $e_{\mathsf{I}\,\mathbb{A}}=\mathcal{\widetilde{\mathcal{O}}}{}_{\mathsf{I}\,\mathbb{A}}$
and $c_{\mathsf{I}}=\mathsf{G}_{\mathsf{I}\,\textrm{flux}}$.

\subsubsection*{Evaluating the Equations of Motion - $\mathsf{\widetilde{C}}_{2}^{\mathbb{A}}$}

A similar analysis for the fields $B^{\mathbb{A}}$ in \cite{DAuria:2007axr}
implies that a viable strategy is to use lines one and three of the
duality constraints (\ref{eq: 4D Duality Constraints}) in order to
eliminate the expressions $\mathcal{O}^{\mathbb{A}}{}_{\mathbb{I}}\mathsf{C}_{1}^{\mathbb{I}}$
and $\mathsf{G}_{2\,\mathsf{I}}$ from the third equation of motion
of (\ref{eq: IIA R-R 4D Equations of Motion}). This can be done by
first left-multiplying line three of (\ref{eq: 4D Duality Constraints})
with $\mathcal{\widetilde{O}}{}_{\mathsf{I}\,\mathbb{A}}$, yielding
\begin{equation}
\mathcal{\widetilde{\mathcal{O}}}{}_{\mathsf{I}\,\mathbb{A}}\textrm{d}\mathsf{C}_{2}^{\mathbb{A}}-B\wedge\textrm{d}(\mathcal{\widetilde{\mathcal{O}}}{}_{\mathsf{I}\,\mathbb{A}}\mathsf{C}_{0}^{\mathbb{A}})=e^{2\phi}\mathcal{\widetilde{\mathcal{O}}}{}_{\mathsf{I}\,\mathbb{A}}\left(S^{-1}\right)^{\mathbb{AB}}\mathbb{M}{}_{\mathbb{BC}}\star\mathsf{G}_{1}^{\mathbb{C}}.
\end{equation}
Employing the expansion (\ref{eq: IIA R-R 4D C-Poly-Form alternative})
and solving for $\mathcal{O}^{\mathbb{A}}{}_{\mathbb{I}}\mathsf{C}_{1}^{\mathbb{I}}$,
we obtain 
\begin{equation}
\mathcal{O}^{\mathbb{A}}{}_{\mathbb{I}}\mathsf{C}_{1}^{\mathbb{I}}=-\mathcal{O}^{\mathbb{A}}{}_{\mathsf{I}}(\Delta^{-1})^{\mathsf{IJ}}\left(\star\textrm{d}(\mathcal{\widetilde{\mathcal{O}}}{}_{\mathsf{J}\,\mathbb{B}}\mathsf{\widetilde{C}}_{2}^{\mathbb{B}})+\mathcal{\widetilde{\mathcal{O}}}{}_{\mathsf{J}\,\mathbb{B}}\mathsf{C}_{0}^{\mathbb{B}}\star\textrm{d}B+e^{2\phi}(\mathcal{O}^{T})_{\mathsf{J}}\vphantom{(\mathcal{O}^{T})}^{\mathbb{B}}\mathbb{M_{BC}}\textrm{d}\mathsf{C}_{0}^{\mathbb{C}}\right),\label{eq: IIA R-R OC1-Relation}
\end{equation}
with 
\begin{equation}
\Delta_{\mathsf{IJ}}=e^{2\phi}(\mathcal{O}^{T})_{\mathsf{I}}\vphantom{(\mathcal{O}^{T})}^{\mathbb{A}}\mathbb{M_{AB}\mathcal{O}^{\mathbb{B}}{}_{\mathsf{J}}}.
\end{equation}
Starting from line three of (\ref{eq: IIA R-R 4D Equations of Motion}),
we separate desired and undesired components to get 
\begin{equation}
\textrm{d}(\mathcal{O}^{\mathbb{A}}{}_{\mathbb{I}}\mathsf{C}_{1}^{\mathbb{I}})-\textrm{d}(\mathcal{O}^{\mathbb{A}}{}_{\mathsf{I}}\mathsf{C}_{1}^{\mathsf{I}})-\mathcal{O}^{\mathbb{A}}{}_{\mathsf{I}}\mathcal{\widetilde{\mathcal{O}}}^{\mathsf{I}}{}_{\mathbb{B}}\mathsf{\mathsf{C}_{2}^{\mathbb{B}}}-\mathcal{O}^{\mathbb{A}\,\mathsf{I}}\mathsf{G}_{2\,\mathsf{I}}=0.
\end{equation}
The first term can be substituted by (\ref{eq: IIA R-R OC1-Relation}),
the third term by the relation 
\begin{equation}
-\Xi_{\mathbb{A}}\mathcal{O}^{\mathbb{A}}{}_{\mathsf{I}}\mathcal{\widetilde{\mathcal{O}}}^{\mathsf{I}}{}_{\mathbb{B}}\mathsf{\mathsf{C}_{2}^{\mathbb{B}}}=\left(\Xi_{\mathbb{A}}\mathcal{O}^{\mathbb{A\,\mathsf{I}}}\mathsf{\mathcal{\widetilde{\mathcal{O}}}{}_{\mathsf{I}\,\mathbb{B}}}+\Sigma_{\mathbb{I}}\wedge\textrm{d}_{\textrm{int}}\mathcal{\widetilde{\mathcal{O}}}^{\mathbb{I}}{}_{\mathbb{B}}\right)\mathsf{\mathsf{C}_{2}^{\mathbb{B}}}
\end{equation}
derived from (\ref{eq: Nilpotency Constraint - Matrices}), and the
fourth term by the line two of (\ref{eq: 4D Duality Constraints}).
Integration over $CY_{3}$ then yields after left-multiplication with
$S_{\mathbb{AB}}$, 
\begin{equation}
\begin{array}{rcl}
0 & {\displaystyle \!\!\!\!=\!\!\!\!} & {\displaystyle -\textrm{d}\left[(\mathcal{\widetilde{O}}^{T}){}_{\mathbb{A}\,\mathsf{I}}(\Delta^{-1})^{\mathsf{IJ}}\left(\star\textrm{d}(\mathcal{\widetilde{\mathcal{O}}}{}_{\mathsf{J}\,\mathbb{B}}\mathsf{\widetilde{C}}_{2}^{\mathbb{B}})+\mathcal{\widetilde{\mathcal{O}}}{}_{\mathsf{J}\,\mathbb{B}}\mathsf{C}_{0}^{\mathbb{B}}\star\textrm{d}B+e^{2\phi}(\mathcal{O}^{T})_{\mathsf{J}}\vphantom{(\mathcal{O}^{T})}^{\mathbb{B}}\mathbb{M_{BC}}\textrm{d}\mathsf{C}_{0}^{\mathbb{C}}\right)\right]}\\
 & \vphantom{^{X^{X^{X^{X^{X^{X}}}}}}} & {\displaystyle -\textrm{d}(\mathcal{\widetilde{O}}^{T}){}_{\mathbb{A}\,\mathsf{I}}\mathsf{C}_{1}^{\mathsf{I}}{\displaystyle +(\mathcal{\widetilde{O}}^{T})_{\mathbb{A}}{}^{\mathsf{I}}\left(\textrm{Im}\mathcal{N}_{\mathsf{IJ}}\star\mathsf{F}_{2}^{\mathsf{J}}+\textrm{Re}\mathcal{N}_{\mathsf{IJ}}\mathsf{F}_{2}^{\mathsf{J}}+B\wedge\mathsf{G}_{\mathsf{I}\,\textrm{flux}}-\mathcal{\widetilde{\mathcal{O}}}{}_{\mathsf{I}\,\mathbb{B}}\mathsf{\widetilde{C}}_{2}^{\mathbb{B}}\right)}},
\end{array}
\end{equation}
revealing that we can identify $\hat{Z}^{\mathbb{A}}=\mathsf{C}_{0}^{\mathbb{A}}$.

\subsubsection*{Evaluating the Equations of Motion - $\mathsf{C}_{0}^{\mathbb{A}}$}

Following the same procedure once more, we implement lines two and
three of (\ref{eq: 4D Duality Constraints}) into the fifth equation
of motion of (\ref{eq: IIA R-R 4D Equations of Motion}). Simplifying
via equations of motion one and three, we obtain after integrating
over $CY_{3}$ 
\begin{equation}
\textrm{d}\left[e^{2\phi}(S^{-1})^{\mathbb{AB}}\mathbb{M}{}_{\mathbb{BC}}\star\mathsf{G}_{1}^{\mathbb{C}}\right]+\textrm{d}B\wedge\mathsf{G}_{1}^{\mathbb{A}}+e^{4\phi}\mathcal{\mathcal{O}^{\mathbb{A}}{}_{\mathbb{I}}}\left(S^{-1}\right){}^{\mathbb{IJ}}\mathbb{N}{}_{\mathbb{JK}}\mathsf{G}_{0}^{\mathbb{K}}\star\mathbf{1}_{4}=0.\label{eq: IIA R-R EoM Line 5}
\end{equation}
Substituting (\ref{eq: IIA R-R OC1-Relation}) and lowering symplectic
indices with $S\mathbb{_{AB}}$, we arrive at 
\begin{equation}
\begin{array}{ccl}
0 & {\displaystyle \!\!\!\!=\!\!\!\!} & -{\displaystyle \textrm{d}\left[\widetilde{\Delta}_{\mathbb{AB}}\star\textrm{d}\mathsf{C}_{0}^{\mathbb{B}}-e^{2\phi}\mathbb{M_{AB}}\mathcal{O}^{\mathbb{B}}{}_{\mathsf{I}}(\Delta^{-1})^{\mathsf{IJ}}\left(\textrm{d}(\mathcal{\widetilde{\mathcal{O}}}{}_{\mathsf{J}\,\mathbb{C}}\mathsf{\widetilde{C}}_{2}^{\mathbb{C}})+\mathcal{\widetilde{\mathcal{O}}}{}_{\mathsf{J}\,\mathbb{C}}\mathsf{C}_{0}^{\mathbb{C}}\textrm{d}B\right)\right]}\\
 & \vphantom{^{X^{X^{X^{X^{X^{X}}}}}}} & {\displaystyle -\textrm{d}B\wedge\left[S\mathbb{_{AB}}\textrm{d}\mathsf{C}_{0}^{\mathbb{B}}-(\mathcal{\widetilde{\mathcal{O}}}^{T}){}_{\mathbb{A}\,\mathsf{I}}(\Delta^{-1})^{\mathsf{IJ}}\right.}\\
 & \vphantom{^{X^{X^{X^{X^{X^{X}}}}}}} & {\displaystyle \phantom{+\textrm{d}B\wedge|}\cdot\left.\left(\star\textrm{d}(\mathcal{\widetilde{\mathcal{O}}}{}_{\mathsf{J}\,\mathbb{C}}\mathsf{\widetilde{C}}_{2}^{\mathbb{C}})+\mathcal{\widetilde{\mathcal{O}}}{}_{\mathsf{J}\,\mathbb{C}}\mathsf{C}_{0}^{\mathbb{C}}\star\textrm{d}B+e^{2\phi}(\mathcal{O}^{T})_{\mathsf{J}}\vphantom{(\mathcal{O}^{T})}^{\mathbb{C}}\mathbb{M_{CD}}\textrm{d}\mathsf{C}_{0}^{\mathbb{D}}\right)\right]}\\
 & \vphantom{^{X^{X^{X^{X^{X^{X}}}}}}} & {\displaystyle +e^{4\phi}(\widetilde{\mathcal{O}}^{T})_{\mathbb{A}}{}^{\mathbb{I}}\mathbb{N}{}_{\mathbb{IJ}}\left(\mathsf{G}_{\textrm{flux}}^{\mathbb{J}}+\mathcal{\widetilde{\mathcal{O}}}^{\mathbb{J}}{}_{\mathbb{B}}\mathsf{C}_{0}^{\mathbb{B}}\right)\star\mathbf{1}_{4}},
\end{array}
\end{equation}
where 
\begin{equation}
\widetilde{\Delta}_{\mathbb{AB}}=e^{2\phi}\left(\mathbb{M_{AB}}-e^{2\phi}\mathbb{M_{AC}}\mathcal{O}^{\mathbb{C}}{}_{\mathsf{I}}(\Delta^{-1})^{\mathsf{IJ}}(\mathcal{O}^{T})_{\mathsf{J}}\vphantom{(\mathcal{O}^{T})}^{\mathbb{D}}\mathbb{M_{DB}}\right).
\end{equation}

\subsubsection*{Evaluating the Equations of Motion - $B$}

The equations of motion (\ref{eq: IIA B-Field Equation of Motion})
of $\hat{B}$ are already non-trivial and only need to be reformulated
in a way that the undesired degrees of freedom disappear. We here
consider only the relevant part with two external and six internal
indices. Using the expansion (\ref{eq: IIA Duality Relations - Expansion})
and manually inserting involution operators (\ref{eq:MukaiPairing-Involution}),
we can use (\ref{eq: Gauge Coupling Matrices: M}) and (\ref{eq: Gauge Coupling Matrices: N})
to integrate over $CY_{3}$, and after another Weyl-rescaling according
to (\ref{eq: Weyl-Rescaling}), we arrive at
\begin{equation}
\frac{1}{2}\textrm{d}\left(e^{-4\phi}\star\textrm{d}B\right)-\mathsf{G}_{0}^{\mathsf{I}}\mathsf{G}_{2\,\mathsf{I}}+\mathsf{G}_{0\,\mathsf{I}}\mathsf{G}_{2}^{\mathsf{I}}+\mathsf{G}_{1\,\mathsf{A}}\wedge\mathsf{G}_{1}^{\mathsf{A}}=0.\label{eq: IIA R-R EoM B simple}
\end{equation}
Substituting the corresponding expressions from (\ref{eq: IIA R-R Field Expansion Coefficients}),
we eventually find 
\begin{equation}
\begin{array}{ccl}
0 & {\displaystyle \!\!\!\!=\!\!\!\!} & {\displaystyle \frac{1}{2}\textrm{d}\left(e^{-4\phi}\star\textrm{d}B\right)-\mathsf{G}_{\textrm{flux}}^{\mathsf{I}}\left(\textrm{Im}\mathcal{N}_{\mathsf{IJ}}\star\mathsf{F}_{2}^{\mathsf{J}}+\textrm{Re}\mathcal{N}_{\mathsf{IJ}}\mathsf{F}_{2}^{\mathsf{J}}\right)+\mathsf{G}_{\mathsf{I}\,\textrm{flux}}\mathsf{F}_{2}^{\mathsf{I}}+\frac{1}{2}\textrm{d}\mathsf{C}_{0}^{\mathbb{A}}S_{\mathbb{AB}}\textrm{d}\mathsf{C}_{0}^{\mathbb{B}}}\\
 & \vphantom{^{X^{X^{X^{X^{X^{X}}}}}}} & {\displaystyle -\textrm{d}\left[\mathsf{C}_{0}^{\mathbb{A}}(\mathcal{\widetilde{O}}^{T}){}_{\mathbb{A}\,\mathsf{I}}(\Delta^{-1})^{\mathsf{IJ}}\left(\star\textrm{d}(\mathcal{\widetilde{\mathcal{O}}}{}_{\mathsf{J}\,\mathbb{B}}\mathsf{\widetilde{C}}_{2}^{\mathbb{B}})-\mathcal{\widetilde{\mathcal{O}}}{}_{\mathsf{J}\,\mathbb{B}}\mathsf{C}_{0}^{\mathbb{B}}\star\textrm{d}B+e^{2\phi}(\mathcal{O}^{T})_{\mathsf{J}}\vphantom{(\mathcal{O}^{T})}^{\mathbb{B}}\mathbb{M_{BC}}\textrm{d}\mathsf{C}_{0}^{\mathbb{C}}\right)\right].}
\end{array}
\end{equation}

\subsubsection*{Reconstructing the Action of $D=4$ $\mathcal{N}=2$ Gauged Supergravity}

Taking into account conventions and field identifications, we expect
the complete four-dimensional action to take the form 
\begin{equation}
\begin{array}{ccl}
S_{\textrm{IIA}} & {\displaystyle \!\!\!\!=\!\!\!\!} & {\displaystyle \int_{M^{1,4}}\frac{1}{2}R^{(4)}\star\mathbf{1}^{(4)}-\textrm{d\ensuremath{\phi}}\wedge\star\textrm{d\ensuremath{\phi}}-\frac{e^{-4\phi}}{4}\textrm{d}B\wedge\star\textrm{d}B-g_{\mathsf{ij}}\textrm{d}t^{\mathsf{i}}\wedge\star\textrm{d}t^{\mathsf{j}}-g_{\mathsf{ab}}\textrm{d}z^{\mathsf{a}}\wedge\star\textrm{d}\bar{z}^{\mathsf{\bar{b}}}}\\
 & \vphantom{^{X^{X^{X^{X^{X^{X}}}}}}} & {\displaystyle +\frac{1}{2}\textrm{Im}\mathcal{N}_{\mathsf{IJ}}\mathsf{F}_{2}^{\mathsf{J}}\wedge\star\mathsf{F}_{2}^{\mathsf{J}}+\frac{1}{2}\textrm{Re}\mathcal{N}_{\mathsf{IJ}}\mathsf{F}_{2}^{\mathsf{J}}\wedge\mathsf{F}_{2}^{\mathsf{J}}+\frac{1}{2}\widetilde{\Delta}_{\mathbb{AB}}\textrm{d}\mathsf{C}_{0}^{\mathbb{A}}\wedge\star\textrm{d}\mathsf{C}_{0}^{\mathbb{B}}}\\
 & \vphantom{^{X^{X^{X^{X^{X^{X}}}}}}} & {\displaystyle +\frac{1}{2}(\Delta^{-1})^{\mathsf{IJ}}\left(\textrm{d}(\mathcal{\widetilde{\mathcal{O}}}{}_{\mathsf{I}\,\mathbb{A}}\mathsf{\widetilde{C}}_{2}^{\mathbb{A}})+\mathcal{\widetilde{\mathcal{O}}}{}_{\mathsf{I}\,\mathbb{A}}\mathsf{C}_{0}^{\mathbb{A}}\textrm{d}B\right)\wedge\star\left(\textrm{d}(\mathcal{\widetilde{\mathcal{O}}}{}_{\mathsf{J}\,\mathbb{B}}\mathsf{\widetilde{C}}_{2}^{\mathbb{B}})+\mathcal{\widetilde{\mathcal{O}}}{}_{\mathsf{J}\,\mathbb{B}}\mathsf{C}_{0}^{\mathbb{B}}\textrm{d}B\right)}\\
 & \vphantom{^{X^{X^{X^{X^{X^{X}}}}}}} & {\displaystyle +\left(\textrm{d}(\mathcal{\widetilde{\mathcal{O}}}{}_{\mathsf{I}\,\mathbb{A}}\mathsf{\widetilde{C}}_{2}^{\mathbb{A}})+\mathcal{\widetilde{\mathcal{O}}}{}_{\mathsf{I}\,\mathbb{A}}\mathsf{C}_{0}^{\mathbb{A}}\textrm{d}B\right)\wedge\left(e^{2\phi}(\Delta^{-1})^{\mathsf{IJ}}(\mathcal{O}^{T})_{\mathsf{J}}\vphantom{(\mathcal{O}^{T})}^{\mathbb{B}}\mathbb{M_{BC}}\textrm{d}\mathsf{C}_{0}^{\mathbb{C}}\right)-\frac{1}{2}\textrm{d}B\wedge\mathsf{C}_{0}^{\mathbb{A}}S_{\mathbb{AB}}\textrm{d}\mathsf{C}_{0}^{\mathbb{B}}}\\
 & \vphantom{^{X^{X^{X^{X^{X^{X}}}}}}} & -\left(\mathcal{\widetilde{\mathcal{O}}}{}_{\mathsf{I}\,\mathbb{A}}\mathsf{\widetilde{C}}_{2}^{\mathbb{A}}-\mathsf{G}_{\mathsf{I}\,\textrm{flux}}B\right)\wedge\left(\textrm{d}\mathsf{C}_{1}^{\mathbb{\mathsf{I}}}+\frac{1}{2}\mathcal{\widetilde{\mathcal{O}}}^{\mathbb{\mathsf{I}}}{}_{\mathbb{B}}\mathsf{\widetilde{C}}_{2}^{\mathbb{B}}-\frac{1}{2}\mathsf{G}_{\textrm{flux}}^{\mathsf{I}}B\right)+V_{\textrm{scalar}}\star\mathbf{1}^{\left(4\right)},
\end{array}\label{eq: IIA Full Action}
\end{equation}
with
\begin{equation}
\begin{array}{ccl}
V_{\textrm{scalar}} & = & V_{\textrm{NSNS}}+V_{\textrm{RR}}\\
\vphantom{^{X^{X^{X^{X^{X^{X}}}}}}} & = & {\displaystyle +\frac{e^{-2\phi}}{2}V^{\mathbb{I}}(\mathcal{O}^{T}){}_{\mathbb{I}}\,\vphantom{(\widetilde{\mathcal{O}}^{T})}^{\mathbb{A}}\mathbb{M_{AB}}\mathcal{\mathcal{O}^{\mathbb{B}}{}_{\mathbb{J}}}V^{\mathbb{J}}}{\displaystyle +\frac{e^{-2\phi}}{2}W^{\mathbb{A}}(\mathcal{\widetilde{O}}^{T}){}_{\mathbb{A}}\,\vphantom{(\mathcal{O}^{T})}^{\mathbb{I}}\mathbb{N}{}_{\mathbb{IJ}}\mathcal{\widetilde{O}}^{\mathbb{J}}{}_{\mathbb{B}}\overline{W}^{\mathbb{B}}}\\
\vphantom{^{X^{X^{X^{X^{X^{X}}}}}}} &  & {\displaystyle -\frac{e^{-2\phi}}{4\mathcal{K}}W^{\mathbb{A}}S_{\mathbb{AC}}\mathcal{\mathcal{O}^{\mathbb{C}}{}_{\mathbb{I}}}\left(V^{\mathbb{I}}\overline{V}^{\mathbb{J}}+\overline{V}^{\mathbb{I}}V^{\mathbb{J}}\right)(\mathcal{O}^{T}){}_{\mathbb{J}}\,\vphantom{(\widetilde{\mathcal{O}}^{T})}^{\mathbb{D}}S_{\mathbb{DB}}\overline{W}^{\mathbb{B}}}\\
\vphantom{^{X^{X^{X^{X^{X^{X}}}}}}} &  & {\displaystyle +\frac{e^{4\phi}}{2}\left(\mathsf{G}_{\textrm{flux}}^{\mathbb{I}}+\mathcal{\widetilde{O}}^{\mathbb{I}}{}_{\mathbb{A}}\mathsf{C}_{0}^{\mathbb{A}}\right)\mathbb{N}{}_{\mathbb{IJ}}\left(\mathsf{G}_{\textrm{flux}}^{\mathbb{J}}+\mathcal{\widetilde{O}}^{\mathbb{J}}{}_{\mathbb{B}}\mathsf{C}_{0}^{\mathbb{B}}\right)}.
\end{array}\label{eq: IIA Full Action Scalar Potential}
\end{equation}
One can now verify by direct calculation and use of the relations
(\ref{eq: Flux Matrix Symplectic Relation}) and (\ref{eq: IIA Flux Constraints})
that one indeed obtains the previously derived equations of motion
when varying with respect to the corresponding fields. Up to different
conventions and additional terms from the remaining sectors, this
replicates the structure of (35) from \cite{DAuria:2007axr}.

A similar result was derived for $SU(3)\!\times\! SU(3)$ structure
manifolds in \cite{Cassani:2008rb}, where the main difference is
that the authors used projectors to render the fields $\mathcal{\widetilde{\mathcal{O}}}{}_{\mathsf{I}\,\mathbb{A}}\mathsf{\widetilde{C}}_{2}^{\mathbb{A}}$
rather than $\mathsf{\widetilde{C}}_{2}^{\mathbb{A}}$ the fundamental
degrees of freedom. This was done in accordance with the fact that
$\mathsf{\widetilde{C}}_{2}^{\mathbb{A}}$ appears as propagating
degree of freedom only in conjunction with the fluxes (or charges).
Although this is certainly a desirable feature, we intentionally abstain
from making any further assumptions regarding $CY_{3}$ and the flux
matrices (\ref{eq: Flux Matrix Definitions}). While this comes with
the drawback that $\mathsf{\widetilde{C}}_{2}^{\mathbb{A}}$ appears
explicitly as a fundamental degree of freedom of the action (\ref{eq: IIA Full Action}),
an obvious advantage is that one can directly read off the ten-dimensional
origin of the four-dimensional fields.

To conclude the discussion of the type IIA setting, let us briefly
illustrate how this result relates to the standard formulation of
$D=4$ $\mathcal{N}=2$ gauged supergravity. As we have remarked at
the beginning of this paper, the action constructed in \cite{DAuria:2007axr}
poses an alternative formulation of gauged supergravity in which a
subset of the axions is dualized to two-forms. More precisely, the
four-dimensional component $B$ of the Kalb-Ramond field appears explicitly,
in addition to different combinations of the NS-NS fluxes with the
two-form fields $\mathsf{\widetilde{C}}_{2}^{\mathbb{A}}$. It was
shown in \cite{DAuria:2007axr} that under the assumption that $h^{1,1}\leq h^{1,2}$,
the expressions $\mathcal{\widetilde{\mathcal{O}}}{}_{\mathsf{I}\,\mathbb{A}}\mathsf{\widetilde{C}}_{2}^{\mathbb{A}}$
arise as duals of a subset of axions containing $h^{1,1}+1$ of the
corresponding $h^{1,2}+1$ scalars of the original formulation. It
is precisely the presence of the flux coefficients $q_{\mathsf{A}}{}^{\mathsf{I}},\,\tilde{q}{}^{\mathsf{AI}}$
that prevents this dualization procedure from being reversible. Similarly,
in the context of \cite{Louis:2002ny,Gurrieri:2002iw,Gurrieri2003}
it was found that the dualization of $B$ to an axion $a$ using Lagrange
multipliers does not work out as straightforward when non-vanishing
R-R fluxes are considered.

Before attempting to reconstruct the standard formulation of gauged
supergravity, it is important to bear in mind that we did not perform
any a posteriori dualizations of four-dimensional fields to obtain
(\ref{eq: IIA Full Action}). Instead, the two-forms $\mathsf{\widetilde{C}}_{2}^{\mathbb{A}}$
descended naturally from the ten-dimensional field $\hat{C}_{5}$
dual to the ``parent'' $\hat{C}_{3}$ of the $\mathsf{C}_{0}^{\mathbb{A}}$
as well as $B\wedge\hat{C}_{3}$. In order to obtain a dual formulation,
it therefore makes sense to again consider the ten-dimensional equations
of motion and assume vanishing coefficients $q_{\mathsf{A}}{}^{\mathsf{I}},\,\tilde{q}{}^{\mathsf{AI}}$.
This is equivalent to setting 
\begin{equation}
\mathcal{O}^{\mathbb{A}\,\mathsf{I}}=0,\qquad\widetilde{\mathcal{O}}^{\mathsf{I}}{}_{\mathbb{A}}=0,
\end{equation}
and most of the undesired degrees of freedom found in (\ref{eq: IIA R-R Field Expansion Coefficients})
to vanish immediately. One can then proceed differently from the general
case by substituting lines one and three of (\ref{eq: 4D Duality Constraints})
into the lower-index components of the fourth equation of motion of
(\ref{eq: IIA R-R 4D Equations of Motion}). After integrating over
$CY_{3}$, this yields the non-trivial equation of motion 
\begin{equation}
\textrm{d}\left(\textrm{Im}\mathcal{N}_{\mathsf{IJ}}\star\mathsf{F}_{2}^{\mathsf{J}}+\textrm{Re}\mathcal{N}_{\mathsf{IJ}}\mathsf{F}_{2}^{\mathsf{J}}\right)+\left(\mathsf{G}_{\mathsf{I}\,\textrm{flux}}+\widetilde{\mathcal{O}}{}_{\mathsf{I}\,\mathbb{A}}\mathsf{C}_{0}^{\mathbb{A}}\right)\textrm{d}B+e^{2\phi}(\mathcal{O}^{T}){}_{\mathsf{I}}{}^{\mathbb{A}}\mathbb{M}{}_{\mathbb{AB}}\star\left(\textrm{d}\mathsf{C}_{0}^{\mathbb{A}}+\mathcal{O}^{\mathbb{A}}{}_{\mathsf{I}}\mathsf{C}_{1}^{\mathit{\mathsf{I}}}\right)=0
\end{equation}
with 
\begin{equation}
\mathsf{F}_{2}^{\mathsf{I}}={\displaystyle \textrm{d}\mathsf{C}_{1}^{\mathsf{I}}}-B\wedge\mathsf{G}_{\textrm{flux}}^{\mathsf{I}}.
\end{equation}
The first steps for line five of (\ref{eq: IIA R-R 4D Equations of Motion})
and the equation of motion (\ref{eq: IIA B-Field Equation of Motion})
of $\hat{B}$ are analogous to the general case. There is no need
for a reformulation of the duality constraints in this simplified
setting, and they can evaluated in the forms found in (\ref{eq: IIA R-R EoM Line 5})
and (\ref{eq: IIA R-R EoM B simple}), respectively. After inserting
the duality relations (\ref{eq: 4D Duality Constraints}) once more,
it is easy to check that these equations of motion descend from the
action 
\begin{equation}
\begin{array}{ccl}
S_{\textrm{IIA}} & {\displaystyle \!\!\!\!=\!\!\!\!} & {\displaystyle \int_{M^{1,4}}\frac{1}{2}R^{(4)}\star\mathbf{1}^{(4)}-\textrm{d\ensuremath{\phi}}\wedge\star\textrm{d\ensuremath{\phi}}-\frac{e^{-4\phi}}{4}\textrm{d}B\wedge\star\textrm{d}B-g_{\mathsf{ij}}\textrm{d}t^{\mathsf{i}}\wedge\star\textrm{d}t^{\mathsf{j}}-g_{\mathsf{a\bar{b}}}\textrm{d}z^{\mathsf{a}}\wedge\star\textrm{d}\bar{z}^{\mathsf{\bar{b}}}}\\
 & \vphantom{^{X^{X^{X^{X^{X^{X}}}}}}} & {\displaystyle +\frac{1}{2}\textrm{Im}\mathcal{N}_{\mathsf{IJ}}\mathsf{F}_{2}^{\mathsf{J}}\wedge\star\mathsf{F}_{2}^{\mathsf{J}}+\frac{1}{2}\textrm{Re}\mathcal{N}_{\mathsf{IJ}}\mathsf{F}_{2}^{\mathsf{J}}\wedge\mathsf{F}_{2}^{\mathsf{J}}+\frac{e^{2\phi}}{2}\mathbb{M}{}_{\mathbb{AB}}D\mathsf{C}_{0}^{\mathbb{A}}\wedge\star D\mathsf{C}_{0}^{\mathbb{B}}}\\
 & \vphantom{^{X^{X^{X^{X^{X^{X}}}}}}} & {\displaystyle -\frac{1}{2}\textrm{d}B\wedge\left[\mathsf{C}_{0}^{\mathbb{A}}S_{\mathbb{AB}}D\mathsf{C}_{0}^{\mathbb{B}}+\left(2\mathsf{G}_{\mathsf{I}\,\textrm{flux}}+\widetilde{\mathcal{O}}{}_{\mathsf{I}\,\mathbb{A}}\mathsf{C}_{0}^{\mathbb{A}}\right)\mathsf{C}_{1}^{\mathit{\mathsf{I}}}\right]-\frac{1}{2}\mathsf{G}_{\mathsf{I}\,\textrm{flux}}\mathsf{G}_{\textrm{flux}}^{\mathsf{I}}B\wedge B}\\
 & \vphantom{^{X^{X^{X^{X^{X^{X}}}}}}} & {\displaystyle +V_{\textrm{scalar}}\star\mathbf{1}^{\left(4\right)}},
\end{array}\label{eq: IIA Full Action Half-Fluxes}
\end{equation}
where $V_{\textrm{scalar}}$ takes the same form as in (\ref{eq: IIA Full Action Scalar Potential})
and we defined the covariant derivative $D$ by 
\begin{equation}
D\mathsf{C}_{0}^{\mathbb{A}}=\textrm{d}\mathsf{C}_{0}^{\mathbb{A}}+\mathcal{O}^{\mathbb{A}}{}_{\mathsf{I}}\mathsf{C}_{1}^{\mathit{\mathsf{I}}},
\end{equation}
such that the corresponding expression $D\mathsf{C}_{0}^{\mathbb{A}}$
matches with the field strength $\mathsf{G}_{1}^{\mathbb{A}}$. Notice
that the second term does not appear in (\ref{eq: IIA Full Action}).
This is closely related to the dualization procedure described in
\cite{DAuria:2007axr}, where the original action contained additional
scalars $e_{\mathsf{I}}{}^{\mathbb{A}}Z^{\mathsf{I}}$ orthogonal
to the $\hat{Z}^{\mathbb{A}}$, the former of which were then dualized
in order to obtain the two-form fields needed to account for the case
of non-vanishing geometric and non-geometric fluxes.

From (\ref{eq: Fluxes - Cohomology Bases}) and (\ref{eq: Fluxes Unified}),
we can infer that this setting corresponds to dimensional reduction
of type IIA supergravity on $CY_{3}$ with non-vanishing $F$- and
$R$-flux as well as R-R fluxes. The appearance of the non-geometric
$R$-flux is due to the conventions we used for the collective notation
(\ref{eq: Collective Notation for Even Cohomology}), and one can
obtain an analogous expression for non-vanishing $F$- and $H$-fluxes
by exchanging the roles of the identity $\mathbf{1}^{\left(6\right)}$
and the volume form $\star\mathbf{1}^{\left(6\right)}$. Again, a
similar result was found in \cite{Cassani:2008rb} and identified
as the effective action of compactifications on $SU(3)$ structure
manifolds.

Parts of the action (\ref{eq: IIA Full Action Half-Fluxes}) already
resemble the standard formulation of $D=4$ $\mathcal{N}=2$ gauged
supergravity. In a final step, we would like to dualize the four-dimensional
Kalb-Ramond field $B$ to an axion $a$. However, since the presence
of non-vanishing R-R fluxes gives rise to a mass term for $B$, the
simple recipe for dualization via Lagrange multipliers does not apply.
This was already discussed in the context of \cite{Louis:2002ny,Gurrieri2003,Gurrieri:2002iw}
for simpler settings, and we will spare the details here. For the
purpose of this paper, it is sufficient to just consider the case
\begin{equation}
\mathsf{G}_{\textrm{flux}}^{\mathsf{I}}=0.
\end{equation}
Implementing the axion $a$ as Lagrange multiplier, the standard procedure
for dualization (see, e.g. \cite{Louis:2002ny} for explicit calculations)
then brings us to 
\begin{equation}
\begin{array}{ccl}
S_{\textrm{IIA}} & {\displaystyle \!\!\!\!=\!\!\!\!} & {\displaystyle \int_{M^{1,4}}\frac{1}{2}R^{(4)}\star\mathbf{1}^{(4)}-\textrm{d\ensuremath{\phi}}\wedge\star\textrm{d\ensuremath{\phi}}-g_{\mathsf{ij}}\textrm{d}t^{\mathsf{i}}\wedge\star\textrm{d}t^{\mathsf{j}}-g_{\mathsf{ab}}\textrm{d}z^{\mathsf{a}}\wedge\star\textrm{d}\bar{z}^{\mathsf{\bar{b}}}}\\
 & \vphantom{^{X^{X^{X^{X^{X^{X}}}}}}} & {\displaystyle +\frac{1}{2}\textrm{Im}\mathcal{N}_{\mathsf{IJ}}\mathsf{F}_{2}^{\mathsf{J}}\wedge\star\mathsf{F}_{2}^{\mathsf{J}}+\frac{1}{2}\textrm{Re}\mathcal{N}_{\mathsf{IJ}}\mathsf{F}_{2}^{\mathsf{J}}\wedge\mathsf{F}_{2}^{\mathsf{J}}+\frac{e^{2\phi}}{2}\mathbb{M}{}_{\mathbb{AB}}D\mathsf{C}_{0}^{\mathbb{A}}\wedge\star D\mathsf{C}_{0}^{\mathbb{B}}}\\
 & \vphantom{^{X^{X^{X^{X^{X^{X}}}}}}} & {\displaystyle -\frac{e^{4\phi}}{4}\left(Da+\mathsf{C}_{0}^{\mathbb{A}}S_{\mathbb{AB}}D\mathsf{C}_{0}^{\mathbb{B}}\right)\wedge\star\left(Da+\mathsf{C}_{0}^{\mathbb{A}}S_{\mathbb{AB}}D\mathsf{C}_{0}^{\mathbb{B}}\right)}\\
 & \vphantom{^{X^{X^{X^{X^{X^{X}}}}}}} & {\displaystyle +V_{\textrm{scalar}}\star\mathbf{1}^{\left(4\right)}},
\end{array}
\end{equation}
where the covariant derivative of the axion reads 
\begin{equation}
Da=\textrm{d}a-\left(2\mathsf{G}_{\mathsf{I}\,\textrm{flux}}+\widetilde{\mathcal{O}}{}_{\mathsf{I}\,\mathbb{A}}\mathsf{C}_{0}^{\mathbb{A}}\right)\mathsf{C}_{1}^{\mathit{\mathsf{I}}}.
\end{equation}
This strongly resembles the well-known form of $D=4$ $\mathcal{N}=2$
supergravity, with additional gaugings descending from the non-vanishing
NS-NS fluxes. When setting the remaining fluxes to zero, the contributions
of $\mathsf{G}_{\mathsf{I}\,\textrm{flux}}$ as well as the matrices
$\mathcal{O}$ and $\widetilde{\mathcal{O}}$ vanish, and one obtains
ungauged $D=4$ $\mathcal{N}=2$ supergravity as expected.

\subsubsection{Type IIB Setting}

The discussion for the type IIB case follows a very similar pattern,
and we will only sketch the most important steps here.

\subsubsection*{Relation to Democratic Type IIB Supergravity}

Our ansatz is again to reformulate the type IIB R-R pseudo-action
(\ref{eq: DFT: RR Action}) in poly-form notation. The computations
are mostly analogous to the type IIA case, and we obtain 
\begin{equation}
\star\mathcal{L}_{R\, R}^{{\scriptscriptstyle \left(\mathrm{IIB}\right)}}=-\frac{1}{2}\mathfrak{\hat{G}}^{{\scriptscriptstyle \left(\mathrm{IIB}\right)}}\wedge\star\mathfrak{\hat{G}}^{{\scriptscriptstyle \left(\mathrm{IIB}\right)}}
\end{equation}
with 
\begin{equation}
\mathfrak{\hat{G}}^{{\scriptscriptstyle \left(\mathrm{IIB}\right)}}=e^{-\hat{B}}\mathcal{G}^{{\scriptscriptstyle \left(\mathrm{IIB}\right)}}+\mathfrak{D}\mathcal{\hat{C}}^{{\scriptscriptstyle \left(\mathrm{IIB}\right)}}=e^{-\hat{B}}\mathcal{G}^{{\scriptscriptstyle \left(\mathrm{IIB}\right)}}+e^{-\hat{B}}\mathcal{D}\left(e^{\hat{B}}\mathcal{\hat{C}}^{{\scriptscriptstyle \left(\mathrm{IIB}\right)}}\right),\label{eq: IIB R-R Poly-Form}
\end{equation}
and
\begin{equation}
\begin{array}{lcl}
\mathcal{G}^{{\scriptscriptstyle \left(\mathrm{IIB}\right)}} & {\displaystyle \!\!\!\!=\!\!\!\!} & G_{3},\\
\mathcal{\hat{C}}^{{\scriptscriptstyle \left(\mathrm{IIB}\right)}} & {\displaystyle \!\!\!\!=\!\!\!\!} & \hat{C}_{0}+\hat{C}_{2}+\hat{C}_{4}+\hat{C}_{6}+\hat{C}_{8},\vphantom{^{X^{X^{X^{X^{X^{X}}}}}}}
\end{array}
\end{equation}
Notice that we consider only the three-form R-R flux since the one-
and five-forms are trivial in cohomology on $CY_{3}$. The factor
$e^{-\hat{B}}$ in front of $\mathcal{\hat{G}}^{{\scriptscriptstyle \left(\mathrm{IIB}\right)}}$
thus has no effect and is included only for later convenience. The
duality constraints (\ref{eq: Duality Relations General}) for the
type IIB case can be written as 
\begin{equation}
\mathfrak{\hat{G}}^{{\scriptscriptstyle \left(\mathrm{IIB}\right)}}=-\lambda\left(\star\mathfrak{\hat{G}}^{{\scriptscriptstyle \left(\mathrm{IIB}\right)}}\right),\label{eq: IIB R-R 10D Duality Constraints}
\end{equation}
and varying the action with respect to the $C$-field components yields
the equations of motion 
\begin{equation}
\left(\textrm{d}-\textrm{d}\hat{B}\wedge+\mathfrak{H}\wedge+\mathfrak{F}\circ+\mathfrak{Q}\bullet+\mathfrak{R}\llcorner\right)\star\mathfrak{\hat{G}}^{{\scriptscriptstyle \left(\mathrm{IIB}\right)}}=0,
\end{equation}
which are equivalent to the Bianchi identities 
\begin{equation}
{\displaystyle e^{-\hat{B}}\mathcal{D}\left(e^{\hat{B}}\mathfrak{\hat{G}}^{{\scriptscriptstyle \left(\mathrm{IIB}\right)}}\right)=0}.\label{eq: IIB R-R 10D Equations of Motion}
\end{equation}

\subsubsection*{Reduced Equations of Motion and Duality Constraints}

In order to employ the framework of special geometry, we again rewrite
the above expressions in $A$-basis notation. We define 
\begin{equation}
e^{\hat{B}}\mathcal{C}^{{\scriptscriptstyle \left(\mathrm{IIB}\right)}}=\left(\mathsf{C}_{0}^{\mathsf{I}}+\mathsf{C}_{2}^{\mathsf{I}}+\mathsf{C}_{4}^{\mathsf{I}}\right)\omega_{\mathsf{I}}+\left(\mathsf{C}_{1}^{\mathsf{A}}+\mathsf{C}_{3}^{\mathsf{A}}\right)\alpha_{\mathsf{A}}-\left(\vphantom{\mathsf{C}_{1}^{\mathsf{A}}}\mathsf{C}_{1\,\mathsf{A}}+\mathsf{C}_{3\,\mathsf{A}}\right)\beta^{\mathsf{A}}+\left(\vphantom{\mathsf{C}_{1}^{\mathsf{A}}}\mathsf{C}_{0\,\mathsf{I}}+\mathsf{C}_{2\,\mathsf{I}}+\mathsf{C}_{4\,\mathsf{I}}\right)\widetilde{\omega}^{\mathsf{I}}
\end{equation}
and 
\begin{equation}
G_{3}=-\mathsf{G}_{\textrm{flux}}^{\mathsf{A}}\alpha_{\mathsf{A}}+\mathsf{G}_{\textrm{flux}\,\mathsf{A}}\beta^{\mathsf{A}},
\end{equation}
which can be utilized to reformulate the type IIB R-R poly-form (\ref{eq: IIB R-R Poly-Form})
as
\begin{equation}
\mathfrak{\hat{G}}^{{\scriptscriptstyle \left(\mathrm{IIB}\right)}}=e^{-\hat{B}}\hat{\mathsf{G}}^{{\scriptscriptstyle \left(\mathrm{IIB}\right)}}=e^{-\hat{B}}\left(\hat{\mathsf{G}}_{1}^{{\scriptscriptstyle \left(\mathrm{IIB}\right)}}+\hat{\mathsf{G}}_{3}^{{\scriptscriptstyle \left(\mathrm{IIB}\right)}}+\hat{\mathsf{G}}_{5}^{{\scriptscriptstyle \left(\mathrm{IIB}\right)}}+\hat{\mathsf{G}}_{7}^{{\scriptscriptstyle \left(\mathrm{IIB}\right)}}+\hat{\mathsf{G}}_{9}^{{\scriptscriptstyle \left(\mathrm{IIB}\right)}}\right).\label{eq: IIB R-R Poly-Form in A-Basis}
\end{equation}
Notice that this strongly resembles the corresponding expressions
of the type IIA case (cf. (\ref{eq: IIA R-R 4D C-Poly-Form}), (\ref{eq: IIA R-R 4D G-Forms})
and (\ref{eq: IIA R-R Poly-Form A-Basis}))  with exchanged roles
of the even and odd cohomology components. We once more employ a shorthand
notation 
\begin{equation}
\begin{array}{ccl}
\hat{\mathsf{G}}_{1}^{{\scriptscriptstyle \left(\mathrm{IIB}\right)}} & {\displaystyle \!\!\!\!=\!\!\!\!} & \mathsf{G}_{1\,0}\widetilde{\omega}^{0},\\
\hat{\mathsf{G}}_{3}^{{\scriptscriptstyle \left(\mathrm{IIB}\right)}} & {\displaystyle \!\!\!\!=\!\!\!\!} & {\displaystyle \mathsf{G}_{3\,0}\widetilde{\omega}^{0}}+{\displaystyle \mathsf{G}_{1}^{\mathsf{i}}\omega_{\mathsf{i}}}-{\displaystyle \mathsf{G}_{0}^{\mathsf{A}}\wedge\alpha_{\mathsf{A}}}+{\displaystyle \mathsf{G}_{0\,\mathsf{A}}\wedge\beta^{\mathsf{A}}},\vphantom{^{X^{X^{X^{X^{X^{X}}}}}}}\\
\hat{\mathsf{G}}_{5}^{{\scriptscriptstyle \left(\mathrm{IIB}\right)}} & {\displaystyle \!\!\!\!=\!\!\!\!} & {\displaystyle \mathsf{G}_{3}^{\mathsf{i}}\wedge\omega_{\mathsf{i}}}-{\displaystyle \mathsf{G}_{2}^{\mathsf{A}}\wedge\alpha_{\mathsf{A}}}+{\displaystyle \mathsf{G}_{2\,\mathsf{A}}\wedge\beta^{\mathsf{A}}}+{\displaystyle \mathsf{G}_{1\,\mathsf{i}}\widetilde{\omega}^{\mathsf{i}}},\vphantom{^{X^{X^{X^{X^{X^{X}}}}}}}\\
\hat{\mathsf{G}}_{7}^{{\scriptscriptstyle \left(\mathrm{IIB}\right)}} & {\displaystyle \!\!\!\!=\!\!\!\!} & -{\displaystyle \mathsf{G}_{4}^{\mathsf{A}}\wedge\alpha_{\mathsf{A}}}+{\displaystyle \mathsf{G}_{4\,\mathsf{A}}\wedge\beta^{\mathsf{A}}}+{\displaystyle {\displaystyle \mathsf{G}_{3\,\mathsf{i}}}\wedge\widetilde{\omega}^{\mathsf{i}}}+{\displaystyle \mathsf{G}_{1}^{0}}\wedge\omega_{0},\vphantom{^{X^{X^{X^{X^{X^{X}}}}}}}\\
\hat{\mathsf{G}}_{9}^{{\scriptscriptstyle \left(\mathrm{IIB}\right)}} & {\displaystyle \!\!\!\!=\!\!\!\!} & {\displaystyle {\displaystyle \mathsf{G}_{3}^{0}}\wedge\omega_{0}},\vphantom{^{X^{X^{X^{X^{X^{X}}}}}}}
\end{array}\label{eq: IIB R-R Forms in A-Basis}
\end{equation}
where the expansion coefficients 
\begin{equation}
\begin{array}{lcl}
\mathsf{G}_{0}^{\mathbb{A}} & {\displaystyle \!\!\!\!=\!\!\!\!} & \mathsf{G}_{\textrm{flux}}^{\mathbb{A}}+\mathcal{O}^{\mathbb{A}}{}_{\mathbb{I}}\mathsf{\mathsf{C}_{0}^{\mathbb{I}}},\\
\mathsf{G}_{1}^{\mathbb{I}} & {\displaystyle \!\!\!\!=\!\!\!\!} & \textrm{d}\mathsf{C}_{0}^{\mathbb{I}}+\mathcal{\widetilde{\mathcal{O}}}^{\mathbb{I}}{}_{\mathbb{A}}\mathsf{C}_{1}^{\mathbb{A}},\vphantom{^{X^{X^{X^{X^{X^{X}}}}}}}\\
\mathsf{G}_{2}^{\mathbb{A}} & {\displaystyle \!\!\!\!=\!\!\!\!} & {\displaystyle \textrm{d}\mathsf{C}_{1}^{\mathbb{A}}}+\mathcal{O}^{\mathbb{A}}{}_{\mathbb{I}}\mathsf{\mathsf{C}_{2}^{\mathbb{I}}},\vphantom{^{X^{X^{X^{X^{X^{X}}}}}}}\\
\mathsf{G}_{3}^{\mathbb{I}} & {\displaystyle \!\!\!\!=\!\!\!\!} & \textrm{d}\mathsf{C}_{2}^{\mathbb{I}}+\mathcal{\widetilde{\mathcal{O}}}^{\mathbb{I}}{}_{\mathbb{A}}\mathsf{C}_{3}^{\mathbb{A}},\vphantom{^{X^{X^{X^{X^{X^{X}}}}}}}\\
\mathsf{G}_{4}^{\mathbb{A}} & {\displaystyle \!\!\!\!=\!\!\!\!} & {\displaystyle \textrm{d}\mathsf{C}_{3}^{\mathbb{A}}}+\mathcal{O}^{\mathbb{A}}{}_{\mathbb{I}}\mathsf{\mathsf{C}_{4}^{\mathbb{I}}}\vphantom{^{X^{X^{X^{X^{X^{X}}}}}}}
\end{array}\label{eq: IIB R-R Field Expansion Coefficients}
\end{equation}
can be derived by using the flux matrix relations (\ref{eq: Flux Matrix Definitions})-(\ref{eq: Flux Matrix Axtions on Cohomology Bases - Components}).
The equations of motion (\ref{eq: IIB R-R 10D Equations of Motion})
reduce to 
\begin{equation}
\mathcal{D}\hat{\mathsf{G}}^{{\scriptscriptstyle \left(\mathrm{IIB}\right)}}=0
\end{equation}
in $A$-basis notation, giving rise to the set of four-dimensional
equations\\
 
\begin{equation}
\begin{array}{rrl}
\Sigma_{\mathbb{I}}\mathcal{\widetilde{\mathcal{O}}}^{\mathbb{I}}{}_{\mathbb{A}}{\displaystyle \mathsf{G}_{0}^{\mathbb{A}}} & {\displaystyle \!\!\!\!=\!\!\!\!} & 0,\\
\textrm{d}\mathsf{G}_{0}^{\mathbb{A}}-\mathcal{O}^{\mathbb{A}}{}_{\mathbb{I}}\mathsf{G}_{1}^{\mathbb{I}} & {\displaystyle \!\!\!\!=\!\!\!\!} & 0,\vphantom{^{X^{X^{X^{X^{X^{X}}}}}}}\\
\textrm{d}\mathsf{G}_{1}^{\mathbb{I}}-\mathcal{\widetilde{\mathcal{O}}}^{\mathbb{I}}{}_{\mathbb{A}}\mathsf{G}_{2}^{\mathbb{A}} & {\displaystyle \!\!\!\!=\!\!\!\!} & 0,\vphantom{^{X^{X^{X^{X^{X^{X}}}}}}}\\
\textrm{d}\mathsf{G}_{2}^{\mathbb{A}}-\mathcal{O}^{\mathbb{A}}{}_{\mathbb{I}}\mathsf{G}_{3}^{\mathbb{I}} & {\displaystyle \!\!\!\!=\!\!\!\!} & 0,\vphantom{^{X^{X^{X^{X^{X^{X}}}}}}}\\
\textrm{d}\mathsf{G}_{3}^{\mathbb{I}}-\mathcal{\widetilde{\mathcal{O}}}^{\mathbb{I}}{}_{\mathbb{A}}\mathsf{G}_{4}^{\mathbb{A}} & {\displaystyle \!\!\!\!=\!\!\!\!} & 0\vphantom{^{X^{X^{X^{X^{X^{X}}}}}}}
\end{array}\label{eq: IIB R-R Reduced Equations of Motion}
\end{equation}
after applying the same methods we already used to derive (\ref{eq: IIA R-R 4D Equations of Motion}).
 The equation of motion for $\hat{B}$ reads after Weyl-rescaling
according to (\ref{eq: Weyl-Rescaling}),
\begin{equation}
\textrm{d}\left(e^{-4\phi}\star\textrm{d}B\right)+\frac{1}{2}\left[\mathfrak{\hat{G}}^{{\scriptscriptstyle \left(\mathrm{IIB}\right)}}\wedge\star\mathfrak{\hat{G}}^{{\scriptscriptstyle \left(\mathrm{IIB}\right)}}\right]_{8}=0.\label{eq: IIB B-Field Equations of MoTion}
\end{equation}
For the duality constraints (\ref{eq: IIB R-R 10D Duality Constraints}),
we follow the same pattern as for (\ref{eq: 10D Duality Constraints})
and obtain 
\begin{equation}
\begin{array}{rcl}
{\displaystyle \mathsf{G}_{2\,\mathsf{A}}-B\mathsf{G}_{0\,\mathsf{A}}} & {\displaystyle \!\!\!\!=\!\!\!\!} & {\displaystyle \textrm{Im}\mathcal{M}_{\mathsf{AB}}\star\left(\mathsf{G}_{2}^{\mathsf{B}}-B\wedge\mathsf{G}_{0}^{\mathsf{B}}\right)+\textrm{Re}\mathcal{M}_{\mathsf{AB}}\left(\mathsf{G}_{2}^{\mathsf{B}}-B\wedge\mathsf{G}_{0}^{\mathsf{B}}\right)},\\
{\displaystyle \mathsf{G}_{4}^{\mathbb{A}}-B\wedge\mathsf{G}_{2}^{\mathbb{A}}}+\frac{1}{2}B^{2}\mathsf{G}_{0}^{\mathbb{A}} & {\displaystyle \!\!\!\!=\!\!\!\!} & {\displaystyle -e^{4\phi}\left(S^{-1}\right){}^{\mathbb{AB}}\mathbb{M}{}_{\mathbb{BC}}\mathsf{G}_{0}^{\mathbb{C}}}\star\mathbf{1}^{\left(4\right)},\vphantom{^{X^{X^{X^{X^{X^{X}}}}}}}\\
\mathsf{G}_{3}^{\mathbb{I}}-B\wedge\mathsf{G}_{1}^{\mathbb{I}} & {\displaystyle \!\!\!\!=\!\!\!\!} & e^{2\phi}\left(S^{-1}\right)^{\mathbb{IJ}}\mathbb{N}{}_{\mathbb{JK}}\star\mathsf{G}_{1}^{\mathbb{K}}.\vphantom{^{X^{X^{X^{X^{X^{X}}}}}}}
\end{array}\label{eq: IIB Reduced Duality Constraints}
\end{equation}
\\
~

\subsubsection*{Reconstructing the Action}

As the structural analogies between the two settings suggest, the
equations of motion can be evaluated by following the same pattern
as in the type IIA case, eventually leading to the effective four-dimensional
action 
\begin{equation}
\begin{array}{ccl}
S_{\textrm{IIB}} & {\displaystyle \!\!\!\!=\!\!\!\!} & {\displaystyle \int_{M^{1,4}}\frac{1}{2}R^{(4)}\star\mathbf{1}^{(4)}-\textrm{d\ensuremath{\phi}}\wedge\star\textrm{d}\phi-\frac{e^{-4\phi}}{4}\textrm{d}B\wedge\star\textrm{d}B-g_{\mathsf{ij}}\textrm{d}t^{\mathsf{i}}\wedge\star\textrm{d}t^{\mathsf{j}}-g_{\mathsf{ab}}\textrm{d}z^{\mathsf{a}}\wedge\star\textrm{d}\bar{z}^{\mathsf{\bar{b}}}}\\
 & \vphantom{^{X^{X^{X^{X^{X^{X}}}}}}} & {\displaystyle +\frac{1}{2}\textrm{Im}\mathcal{M}_{\mathsf{AB}}\mathsf{F}_{2}^{\mathsf{B}}\wedge\star\mathsf{F}_{2}^{\mathsf{B}}+\frac{1}{2}\textrm{Re}\mathcal{M}_{\mathsf{AB}}\mathsf{F}_{2}^{\mathsf{B}}\wedge\mathsf{F}_{2}^{\mathsf{B}}+\frac{1}{2}\widetilde{\Delta}_{\mathbb{IJ}}\textrm{d}\mathsf{C}_{0}^{\mathbb{I}}\wedge\star\textrm{d}\mathsf{C}_{0}^{\mathbb{J}}}\\
 & \vphantom{^{X^{X^{X^{X^{X^{X}}}}}}} & {\displaystyle +\frac{1}{2}(\Delta^{-1})^{\mathsf{AB}}\left(\textrm{d}(\mathcal{\mathcal{O}}{}_{\mathsf{A}\,\mathbb{I}}\mathsf{\widetilde{C}}_{2}^{\mathbb{I}})+\mathcal{\mathcal{O}}{}_{\mathsf{A}\,\mathbb{I}}\mathsf{C}_{0}^{\mathbb{I}}\textrm{d}B\right)\wedge\star\left(\textrm{d}(\mathcal{\mathcal{O}}{}_{\mathsf{B}\,\mathbb{J}}\mathsf{\widetilde{C}}_{2}^{\mathbb{J}})+\mathcal{\mathcal{O}}{}_{\mathsf{B}\,\mathbb{J}}\mathsf{C}_{0}^{\mathbb{J}}\textrm{d}B\right)}\\
 & \vphantom{^{X^{X^{X^{X^{X^{X}}}}}}} & {\displaystyle +\left(\textrm{d}(\mathcal{O}{}_{\mathsf{A}\,\mathbb{I}}\mathsf{\widetilde{C}}_{2}^{\mathbb{I}})+\mathcal{O}{}_{\mathsf{A}\,\mathbb{I}}\mathsf{C}_{0}^{\mathbb{I}}\textrm{d}B\right)\wedge\left(e^{2\phi}(\Delta^{-1})^{\mathsf{AB}}(\mathcal{\widetilde{O}}^{T})_{\mathsf{B}}\vphantom{(\mathcal{\widetilde{O}}^{T})}^{\mathbb{J}}\mathbb{N_{JK}}\textrm{d}\mathsf{C}_{0}^{\mathbb{K}}\right)+\frac{1}{2}\textrm{d}B\wedge\mathsf{C}_{0}^{\mathbb{I}}S_{\mathbb{IJ}}\textrm{d}\mathsf{C}_{0}^{\mathbb{J}}}\\
 & \vphantom{^{X^{X^{X^{X^{X^{X}}}}}}} & -\left(\mathcal{O}{}_{\mathsf{A}\,\mathbb{I}}\mathsf{\widetilde{C}}_{2}^{\mathbb{I}}-\mathsf{G}_{\mathsf{A}\,\textrm{flux}}B\right)\wedge\left(\textrm{d}\mathsf{C}_{1}^{\mathbb{\mathsf{A}}}+\frac{1}{2}\mathcal{\mathcal{O}}^{\mathbb{\mathsf{A}}}{}_{\mathbb{J}}\mathsf{\widetilde{C}}_{2}^{\mathbb{J}}-\frac{1}{2}\mathsf{G}_{\textrm{flux}}^{\mathsf{A}}B\right)+V_{\textrm{scalar}}\star\mathbf{1}^{\left(4\right)}
\end{array}\label{eq: IIB Full Action}
\end{equation}
with
\begin{equation}
\begin{array}{ccl}
V_{\textrm{scalar}} & {\displaystyle \!\!\!\!=\!\!\!\!} & V_{\textrm{NSNS}}+V_{\textrm{RR}}\\
\vphantom{^{X^{X^{X^{X^{X^{X}}}}}}} & {\displaystyle \!\!\!\!=\!\!\!\!} & {\displaystyle +\frac{e^{-2\phi}}{2}V^{\mathbb{I}}(\mathcal{O}^{T}){}_{\mathbb{I}}\,\vphantom{(\widetilde{\mathcal{O}}^{T})}^{\mathbb{A}}\mathbb{M_{AB}}\mathcal{\mathcal{O}^{\mathbb{B}}{}_{\mathbb{J}}}V^{\mathbb{J}}}{\displaystyle +\frac{e^{-2\phi}}{2}W^{\mathbb{A}}(\mathcal{\widetilde{O}}^{T}){}_{\mathbb{A}}\,\vphantom{(\mathcal{O}^{T})}^{\mathbb{I}}\mathbb{N}{}_{\mathbb{IJ}}\mathcal{\widetilde{O}}^{\mathbb{J}}{}_{\mathbb{B}}\overline{W}^{\mathbb{B}}}\\
\vphantom{^{X^{X^{X^{X^{X^{X}}}}}}} &  & {\displaystyle -\frac{e^{-2\phi}}{4\mathcal{K}}W^{\mathbb{A}}S_{\mathbb{AC}}\mathcal{\mathcal{O}^{\mathbb{C}}{}_{\mathbb{I}}}\left(V^{\mathbb{I}}\overline{V}^{\mathbb{J}}+\overline{V}^{\mathbb{I}}V^{\mathbb{J}}\right)(\mathcal{O}^{T}){}_{\mathbb{J}}\,\vphantom{(\widetilde{\mathcal{O}}^{T})}^{\mathbb{D}}S_{\mathbb{DB}}\overline{W}^{\mathbb{B}}}\\
\vphantom{^{X^{X^{X^{X^{X^{X}}}}}}} &  & {\displaystyle +\frac{e^{4\phi}}{2}\left(\mathsf{G}_{\textrm{flux}}^{\mathbb{A}}+\mathcal{O}^{\mathbb{A}}{}_{\mathbb{I}}\mathsf{C}_{0}^{\mathbb{I}}\right)\mathbb{M}{}_{\mathbb{AB}}\left(\mathsf{G}_{\textrm{flux}}^{\mathbb{B}}+\mathcal{O}^{\mathbb{B}}{}_{\mathbb{J}}\mathsf{C}_{0}^{\mathbb{J}}\right)}.
\end{array}
\end{equation}
Comparing this to (\ref{eq: IIA Full Action}), one can again construct
a set of mirror mappings by extending (\ref{eq:MirrorSymmetry-ScalarPotential})
to 
\begin{equation}
\begin{array}{lclclcl}
t^{\mathsf{i}} & \leftrightarrow & z^{\mathsf{a}}, & \qquad\qquad & g_{\mathsf{ij}} & \leftrightarrow & g_{\mathsf{a\bar{b}}},\\
\mathbb{M}{}_{\mathbb{AB}} & \leftrightarrow & \mathbb{N}{}_{\mathbb{IJ}}, & \vphantom{^{X^{X^{X^{X^{X^{X}}}}}}} & h^{1,1} & \leftrightarrow & h^{1,2},\\
V^{\mathbb{I}} & \leftrightarrow & W^{\mathbb{A}}, & \vphantom{^{X^{X^{X^{X^{X^{X}}}}}}} & S_{\mathbb{IJ}} & \leftrightarrow & S_{\mathbb{AB}}\\
\mathsf{C}_{n}^{\mathbb{I}} & \leftrightarrow & \mathsf{C}_{n}^{\mathbb{A}}, & \vphantom{^{X^{X^{X^{X^{X^{X}}}}}}} & \mathsf{G}_{\textrm{flux}}^{\mathbb{I}} & \leftrightarrow & \mathsf{G}_{\textrm{flux}}^{\mathbb{A}},\\
\mathcal{O}^{\mathbb{A}}{}_{\mathbb{I}} & \leftrightarrow & \mathcal{\widetilde{\mathcal{O}}}^{\mathbb{I}}{}_{\mathbb{A}}, & \vphantom{^{X^{X^{X^{X^{X^{X}}}}}}}
\end{array}
\end{equation}
once more confirming preservation of $\textrm{IIA}\leftrightarrow\textrm{IIB}$
Mirror Symmetry in the presence of both geometric and non-geometric
fluxes.

\section{Conclusion\label{sec:6}}

Let us summarize the results obtain in this work. 
In section~\ref{sec:2} we derived the scalar potential of four-dimensional
$\mathcal{N}=2$ gauged supergravity from dimensional reduction of
the purely internal type IIA and IIB DFT action on a Calabi-Yau three-fold
$CY_{3}$. Building upon the elaborations of \cite{Blumenhagen:2015lta},
we extended the discussed setting by including cohomologically trivial
terms and relaxing the primitivity constraints, revealing a more general
structure of the reformulated DFT action which strongly resembles
that of type II supergravities on $SU(3)\!\times\! SU(3)$ structure
manifolds (cf. \cite{Cassani:2008rb}).

It was then exemplified through  $K3\times T^{2}$ (cf. section~\ref{sec:3})
how the framework can be generalized beyond the Calabi-Yau setting.
This was done by utilizing the features of generalized Calabi-Yau
and $K3$ structures \cite{Hitchin:2004ut,Huybrechts:2003ak} to
allow for a special geometric description of the $K3\times T^{2}$
moduli space, eventually leading to a scalar potential term resembling
that of $\mathcal{N}=4$ gauged supergravity formulated in the $\mathcal{N}=2$
formalism first discussed in \cite{DAuria:2007axr}. The essential
idea here was to exploit the Calabi-Yau property of $K3$ and $T^{2}$
to formally construct $K3\times T^{2}$ analogues of the structure
forms of $CY_{3}$, 
\eq{
e^{b_{CY_{3}}+iJ_{CY_{3}}}&\longleftrightarrow e^{b_{K3}+iJ_{K3}}\wedge e^{b_{T^{2}}+iJ_{T^{2}}},
\\
e^{b_{CY_{3}}}\wedge\Omega_{CY_{3}}&\longleftrightarrow\left(\vphantom{e^{b_{T^{2}}}}e^{b_{K3}}\wedge\Omega_{K3}\right)\wedge\left(e^{b_{T^{2}}}\wedge\Omega_{T^{2}}\right),
}
where $J$ denotes the K\"ahler form of the respective manifold and
$\Omega$ its holomorphic one-, two- or three-form. While the constructed
scalar potential shows characteristic features of $\mathcal{N}=4$
gauged supergravity, relating the result to its standard formulation
explicitly turned out to be a nontrivial task and will therefore be
saved for future work. We expect that the discussion for arbitrary
manifolds allowing for a generalized Calabi-Yau structure in the sense
of \cite{Hitchin:2004ut,Huybrechts:2003ak} follows the same pattern.

Another novel feature of the setting discussed in this paper is its
capability of describing generalized dilaton fluxes and  non-vanishing
trace-terms of the geometric and non-geometric fluxes. While the role
of these additional fluxes remains unclear for the Calabi-Yau setting
(cf. page \pageref{sec:3} for more details on the issue of cohomology
and fluxes of DFT), it is to be expected that they serve as a ten-dimensional
origin of the non-unimodular gaugings of $\mathcal{N}=4$ gauged supergravity
\cite{Derendinger:2007xp,DallAgata:2009wsi} in
the $K3\times T^{2}$ setting (cf. also section 4.2.3 of \cite{Aldazabal:2011nj}
for a brief discussion in the DFT context). 
Integrating the dilaton flux operators into the twisted differential
of DFT did not require including a rescaling charge operator as done
in \cite{DallAgata:2009wsi}, which  is in accordance with the result of \cite{Cassani:2008rb}
for $SU(3)\!\times\! SU(3)$ structure manifolds.

Finally, in both the $CY_{3}$ and the $K3\times T^{2}$ setting,
a set of mirror mappings relating the results for type IIA and IIB
DFT could be read off and featured the characteristic exchange of
roles between the K\"ahler class and complex structure moduli spaces
in the former and between the two modular parameters of $T^{2}$ in
the latter.

In section~\ref{sec:5} we reconstructed the full bosonic part of the four-dimensional
$\mathcal{N}=2$ gauged supergravity action by including the kinetic
terms into the Calabi-Yau setting. Our results replicate the findings
of \cite{DAuria:2007axr} and once more illustrate how simultaneous
treatment of all NS-NS and R-R fluxes not only gives rise to gaugings
in the effective four-dimensional theory, but also requires a dualization
of a subset of the axions in order to account for all fluxes. Turning
off half of the fluxes correctly led to the standard formulation of
$\mathcal{N}=2$ gauged supergravity, which could be further reduced
to its ungauged version when setting the remaining fluxes to zero.
The $\textrm{IIA}\leftrightarrow\textrm{IIB}$ mirror mappings constructed
in the context of the scalar potential discussion could be straightforwardly
generalized to the full action.

Our analysis of the R-R sector strongly resembles that of \cite{Cassani:2008rb}
for $SU(3)\!\times\! SU(3)$ manifolds, where the essential difference
is that in the discussion of the present paper the field strengths
are determined by the DFT action. This leads to a slightly altered
formulation of the action in which the ten-dimensional origin of the
four-dimensional fields becomes evident. In particular, rather than
only the actual propagating fields, the reduced action contains fundamental
degrees of freedom which appear in the equations of motion only in
conjunction with the flux charges.

It would be interesting to use the procedure elaborated here
to derive the remaining four-dimensional gauged supergravities. 
The next step is to see how the framework can be applied
to the full action compactified on $K3\times T^{2}$. Since dimensional
reduction on Calabi-Yau three-folds leads to a partially dualized
formulation of gauged $\mathcal{N}=2$ supergravity, an important
question in this context is whether the action of half-maximal supersymmetric
gauged supergravity obtained via $K3\times T^{2}$ shows similar properties
in the case of non-vanishing non-geometric fluxes. We plan to address
these questions in future work by extending the discussion
to manifolds with $SU(2)$ structure \cite{Spanjaard:2008zz,Triendl:2009ap,Danckaert:2011ju}.

Other possible directions include extensions of the orientifold
setting discussed in \cite{Blumenhagen:2015lta} or dimensional reduction
of heterotic DFT. Finally, a particularly interesting question is
whether the framework can be generalized to the U-duality covariant
exceptional field theory (EFT) and, if so, in which way the
additional fluxes that are not part of the T-duality chain will manifest themselves
in four dimensions.

\vspace{0.5cm}
\subsubsection*{Acknowledgments} 

We thank Ralph Blumenhagen and Dieter L\"ust for very helpful and stimulating discussions. The work of P.B. is 
supported by the ``Excellence Cluster Universe''.

\newpage{}

\appendix

\section{\label{sec:A}Notation and Conventions}

\subsection{Spacetime Geometry and Indices}

Throughout this paper we make use of various kinds of
indices, which are structured as follows:
\begin{itemize}
\item We distinguish between serif letters $A,a,\ldots$ denoting spacetime
indices and sanserif letters $\mathsf{A,a},\ldots$ labeling the coordinates
of moduli spaces. We furthermore introduce blackboard typeface capital
letters $\mathbb{A},\mathbb{B},\ldots$, $\mathbb{I},\mathbb{J},\ldots$
for collective notation summarizing several de Rham cohomology bases,
which are specified in subsection~\ref{sub:3.3.1} and \ref{sub:4.2.1}.
\item For spacetime indices, capital letters denote doubled coordinates,
and small letters denote normal coordinates. 
\item For spacetime indices, ten-dimensional indices (including doubled
ones) are labeled with a hat symbol, external indices are denoted
by small Greek letters and internal indices by checked or normal Latin
letters as specified below.
\end{itemize}
Using this as a guideline, we define the following indices:
\begin{itemize}
\item Hatted Latin capital letters $\hat{M},\hat{N},\ldots$ and $\hat{A},\hat{B},\ldots$
label the curved respectively tangent coordinates of twenty-dimensional
doubled spacetime.
\item Small hatted letters $\hat{m},\hat{n},\ldots$ and $\hat{a},\hat{b},\ldots$
label the curved respectively tangent coordinates of ten-dimensional
spacetime.
\item Small Greek letters $\mu,\nu,\ldots$ and small Latin letters $e,f,\ldots$
label the curved respectively tangent coordinates of four-dimensional
external spacetime.
\item Checked capital Latin letters $\check{I},\check{J},\ldots$ and $\check{A},\check{B},\ldots$
label the curved respectively tangent coordinates of a general twelve-dimensional
doubled internal space. 
\item Checked small Latin letters $\check{i},\check{j},\ldots$ and $\check{a},\check{b},\ldots$
label the curved respectively tangent coordinates of a general six-dimensional
internal space.
\item Coordinates of specific internal manifolds or their components (e.g.
$CY_{3}$, $K3$ and $T^{2}$) are denoted by normal Latin letters
specified in the corresponding sections of this paper.
\item On $CY_{3}$, small Latin letters $a,\bar{a},b,\bar{b}\ldots$ denote
complex curved coordinates of six-dimensional internal spacetime.
It will be clear from the context whether the letters $a,b,\ldots$
without bars denote holomorphic curved coordinates or normal tangent
coordinates. On $K3\times T^{2}$, $a,\bar{a},b,\bar{b}\ldots$ denote
complex curved coordinates of $K3$ and $g,\bar{g},h,\bar{h}\ldots$
those of $T^{2}$.
\item Moduli space or cohomological indices are specified in the sections
the bases are defined.
\end{itemize}

\subsection{Tensor Formalism and Differential Forms}

For general tensors, differential forms and related operators, we
apply the following conventions:
\begin{itemize}
\item The antisymmetrization of a tensor $A$ is is defined by 
\begin{equation}
A_{\left[\underline{\hat{m}_{1}\hat{m}_{2}}\ldots\underline{\hat{m}_{n}}\right]}:=\frac{1}{n!}\sum_{\pi\in S_{n}}\left(-1\right)^{\textrm{sign}\left(\pi\right)}A_{\pi\left(\hat{m}_{1}\right)\pi\left(\hat{m}{}_{2}\right)\ldots\pi\left(\hat{m}_{n}\right)},
\end{equation}
where $S_{n}$ denotes the set of permutations of $\left\{ 1,2,\ldots n\right\} $.
\item The Levi-Civita tensor\textit{ }$\varepsilon^{\hat{m}_{1}\ldots\hat{m}_{D}}$
in $D$ dimensions is defined as the totally antisymmetric tensor
with $\varepsilon^{012\ldots\left(D-1\right)}=1$ (Lorentzian signature)
or $\varepsilon^{123\ldots D}$ (Euclidean signature). It satisfies
the relations
\begin{equation}
\begin{array}{lclcl}
\varepsilon^{\hat{m}_{1}\ldots\hat{m}_{D}}\varepsilon_{\hat{n}_{1}\ldots\hat{n}_{D}} & = & D!\delta_{\hat{n}_{1}}^{[\underline{\hat{m}_{1}}}\ldots\delta_{\hat{n}_{D}}^{\underline{\hat{m}_{D}}]} & = & \delta_{\hat{n}_{1}\ldots\hat{n}_{D}}^{\hat{m}_{1}\ldots\hat{m}_{D}}\\
\varepsilon^{\hat{m}_{1}\ldots\hat{m}_{p}\hat{m}_{p+1}\ldots\hat{m}_{D}}\varepsilon_{\hat{m}_{1}\ldots\hat{m}_{p}\hat{n}_{p+1}\ldots\hat{n}_{D}} & = & p!\left(D-p\right)!\delta_{\hat{n}_{p+1}}^{[\underline{\hat{m}_{p+1}}}\ldots\delta_{\hat{n}_{D}}^{\underline{\hat{m}_{D}}]} & = & p!\delta_{\hat{n}_{p+1}\ldots\hat{n}_{D}}^{\hat{m}_{p+1}\ldots\hat{m}_{D}}\vphantom{^{X^{X^{X^{X^{X^{X}}}}}}}\\
\varepsilon^{\hat{m}_{1}\ldots\hat{m}_{D}}\varepsilon_{\hat{m}_{1}\ldots\hat{m}_{D}} & = & D!.\vphantom{^{X^{X^{X^{X^{X^{X}}}}}}}
\end{array}\label{eq:Levi-Civita-Delta-Relation}
\end{equation}

\item The components of a differential $p$-form are defined as 
\begin{equation}
\hat{\omega}_{p}=\frac{1}{p\text{!}}\omega_{\hat{m}_{1}\ldots\hat{m}_{p}}\textrm{d}x^{\hat{m}_{1}}\wedge\ldots\wedge dx^{\hat{m}_{p}}.
\end{equation}

\item The exterior product of a $p$-form $\hat{\omega}_{p}$ and a $q$-form
$\hat{\chi}_{q}$ is given by 
\begin{equation}
\begin{array}{ccccl}
\wedge: & \Omega^{p}\left(\mathcal{M}\right)\times\Omega^{q}\left(\mathcal{M}\right) & \rightarrow & \Omega^{p+q}\left(\mathcal{M}\right)\\
\\
 & \left(\hat{\omega}_{p},\hat{\chi}_{q}\right) & \mapsto & \hat{\omega}_{p}\wedge\hat{\omega}_{q}= & {\displaystyle \frac{\left(p+q\right)!}{p!q!}\omega_{[\underline{\hat{m}_{1}}\ldots\underline{\hat{m}_{p}}}\,\chi_{\underline{\hat{n}_{1}}\ldots\underline{.\hat{n}_{q}}]}\,\textrm{d}x^{\hat{m}_{1}}\wedge\ldots}\\
 &  &  & \vphantom{^{X^{X^{X^{X^{X^{X}}}}}}} & \ldots\wedge\textrm{d}x^{\hat{m}_{p}}\wedge\textrm{d}x^{\hat{n}_{1}}\wedge\ldots\wedge\textrm{d}x^{\hat{n}_{q}}.
\end{array}
\end{equation}
In this context, we choose the notation $\left(\hat{\omega}_{p}\right)^{n}=\overset{n\:\textrm{factors}}{\overbrace{\hat{\omega}_{p}\wedge\hat{\omega}_{p}\wedge\ldots\wedge\hat{\omega}_{p}}}$
for exterior products of a $p$-form $\omega_{p}$ with itself. 
\item The exterior derivative $\textrm{d}$ is given by 
\begin{equation}
\begin{array}{ccl}
\textrm{d}:\Omega^{p}\left(\mathcal{M}\right) & \rightarrow & \Omega^{p+1}\left(\mathcal{M}\right)\\
\\
\hat{\omega}_{p} & \mapsto & {\displaystyle \textrm{d}\hat{\omega}_{p}=\frac{1}{p!}\frac{\partial\omega_{\hat{m}_{1}\ldots\hat{m}{}_{p}}}{\partial x^{\hat{n}}}\textrm{d}x^{\hat{n}}\wedge\textrm{d}x^{\hat{m}_{1}}\wedge\ldots\wedge\textrm{d}x^{\hat{m}_{p}}}.
\end{array}
\end{equation}

\item The Hodge star operator\textit{ }$\star$ is defined by
\begin{equation}
\begin{array}{@{}cccl@{}}
\star: & \Omega^{p}\left(\mathcal{M}\right) & \rightarrow & \Omega^{D-p}\left(\mathcal{M}\right)\\
\\
 & \hat{\omega}_{p} & \mapsto & {\displaystyle \star\hat{\omega}_{p}=\frac{1}{\sqrt{g}p!\left(D-p\right)!}\varepsilon_{\hat{m}_{1}\ldots\hat{m}_{p}\hat{m}_{p+1}\ldots\hat{m}_{D}}g^{\hat{m}_{1}\hat{n}_{1}}\ldots g^{\hat{m}_{p}\hat{n}_{p}}\omega_{\hat{n}_{1}\ldots\hat{n}_{p}}}\textrm{d}^{D-p}x.
\end{array}\label{eq:Hodge-Star}
\end{equation}
In particular, one can define a scalar product of two $p$-forms $\hat{\omega}_{p}$
and $\hat{\chi}_{p}$ by
\begin{equation}
\hat{\omega}_{p}\wedge\star\overline{\hat{\chi}_{p}}=\frac{\sqrt{g}}{p!}\omega_{\hat{m}_{1}\ldots\hat{m}_{p}}\overline{\chi_{\hat{n}_{1}\ldots\hat{n}_{p}}}g^{\hat{m}_{1}\hat{n}_{1}}\ldots g^{\hat{m}_{p}\hat{n}_{p}}\textrm{d}^{D}x.\label{eq:Scalar-Product-Differential-Forms}
\end{equation}
On $D-$dimensional Lorentzian manifolds, $\star$ satisfies the bijectivity
condition 
\begin{equation}
\star\star\hat{\omega}_{p}=\left(-1\right)^{p\left(d-p\right)+1}\omega_{p}.
\end{equation}
Using this, one can show that the $b$-twisted Hodge star operator
(\ref{eq:b-twisted-hodge-star}) squares to $-1$, 
\begin{equation}
\star_{b}\star_{b}=-1.
\end{equation}
When splitting a differential $p$-form $\hat{\omega}_{p}=\eta_{p-n}\wedge\rho_{n}$
living in $M^{10}$ into two forms $\eta_{p-n}\in\Omega^{p-n}\left(M^{1,4}\right)$
and $\rho_{n}\in\Omega^{n}\left(CY_{3}\right)$, the Hodge-star operator
splits as 
\begin{equation}
\star\hat{\omega}_{p}=\left(-1\right)^{n\left(p-n\right)}\star\eta_{p-n}\wedge\star\rho_{n}.
\end{equation}
As a consequence, one obtains for the involution operator (\ref{eq:MukaiPairing-Involution})
\begin{equation}
\star\lambda\left(\hat{\omega}_{p}\right)=\star\lambda\left(\eta_{p-n}\right)\wedge\star\lambda\left(\rho_{n}\right).
\end{equation}

\item For differential poly-forms, we define the projectors $\left[\cdot\right]_{n}$
to give as output the $n$-form components of the argument.
\end{itemize}

\section{\label{sec:Overview}Complex and K\"ahler Geometry}

This appendix provides an overview on geometric properties of Calabi-Yau
3-folds and $K3\times T^{2}$ used for the calculations of section~\ref{sec:3}
and section~\ref{sec:4}, respectively. Most of the technical steps are based
on the notions complex and K\"ahler geometry, which shall be discussed
here.

Both $CY_{3}$ and $K3\times T^{2}$ are complex manifolds, allowing
for a standard complex structure $I$ satisfying
\begin{equation}
\begin{array}{lcl}
I^{a}{}_{b}=i\delta^{a}{}_{b}, & \quad & I^{\bar{a}}{}_{\bar{b}}=-i\delta^{\bar{a}}{}_{\bar{b}},\\
\vphantom{^{X^{X^{X^{X^{X^{X}}}}}}}I^{a}{}_{\bar{b}}=0, & \quad & I^{\bar{a}}{}_{b}=0.
\end{array}
\end{equation}
Being also K\"ahler and, thus, Hermitian manifolds, the only non-vanishing
components of their metric $g$ are $g_{a\bar{b}}=\overline{g}_{\bar{a}b}$.
They are related to the K\"ahler form $J$ by 
\begin{equation}
J_{a\bar{b}}=ig_{a\bar{b}},\quad J_{\bar{a}b}=-ig_{\bar{a}b}\label{eq: Appendix B: Relation zwischen J und g}
\end{equation}
and, in real coordinates, 
\begin{equation}
J_{ij}=g_{im}I^{m}{}_{j}.
\end{equation}
For the holomorphic three-form of $CY_{3}$, we employ the normalization
\begin{equation}
\frac{i}{8}\Omega\wedge\star\overline{\Omega}=\frac{1}{3!}J^{3},
\end{equation}
leading to the relations 
\begin{equation}
\begin{array}{rcl}
\Omega_{abc}\overline{\Omega}_{\bar{a}\bar{b}\bar{c}}g^{c\bar{c}} & = & 8\left(g_{a\bar{a}}g_{b\bar{b}}-g_{a\bar{b}}g_{b\bar{a}}\right),\\
\vphantom{^{X^{X^{X^{X^{X^{X}}}}}}}\Omega_{abc}\overline{\Omega}_{\bar{a}\bar{b}\bar{c}}g^{b\bar{b}}g^{c\bar{c}} & = & 16g_{a\bar{a}},\\
\vphantom{^{X^{X^{X^{X^{X^{X}}}}}}}\Omega_{abc}\overline{\Omega}_{\bar{a}\bar{b}\bar{c}}g^{a\bar{a}}g^{b\bar{b}}g^{c\bar{c}} & = & 48.
\end{array}\label{eq: Appendix B: Omega-Relations}
\end{equation}
The same normalization is applied to holomorphic form $\Omega:=\Omega_{K3}\times\Omega_{T^{2}}$
of $K3\times T^{2}$ (with $J:=J_{K3}+J_{T^{2}}$), and one obtains
similarly 
\begin{equation}
\begin{array}{rcl}
\Omega_{gab}\overline{\Omega}_{\bar{g}\bar{a}\bar{b}}g^{g\bar{g}} & = & 8\left(g_{a\bar{a}}g_{b\bar{b}}-g_{a\bar{b}}g_{b\bar{a}}\right),\\
\vphantom{^{X^{X^{X^{X^{X^{X}}}}}}}\Omega_{gab}\overline{\Omega}_{\bar{g}\bar{a}\bar{b}}g^{b\bar{b}} & = & 8g_{g\bar{g}}g_{a\bar{a}},\\
\vphantom{^{X^{X^{X^{X^{X^{X}}}}}}}\Omega_{gab}\overline{\Omega}_{\bar{g}\bar{a}\bar{b}}g^{a\bar{a}}g^{b\bar{b}} & = & 16g_{g\bar{g}},\\
\vphantom{^{X^{X^{X^{X^{X^{X}}}}}}}\Omega_{gbb}\overline{\Omega}_{\bar{g}\bar{b}\bar{b}}g^{g\bar{g}}g^{a\bar{a}}g^{b\bar{b}} & = & 16.
\end{array}
\end{equation}
\newpage{}

\clearpage 
\bibliographystyle{utphys}
\bibliography{references}

\end{document}